%% file: ljets_prd.tex
\documentclass[aps,prd,twocolumn,showpacs,groupedaddress,floatfix]{revtex4}

\usepackage{graphicx}
\usepackage{color}
\usepackage{dcolumn}
\usepackage{bm}
\usepackage{xspace}
\usepackage{epsfig}
\usepackage{psfrag}
\usepackage{rotating}
\usepackage{afterpage}
\usepackage{longtable}
\usepackage{amssymb}   
\usepackage{amsmath}
\usepackage{amsfonts}
\usepackage{multirow}
\usepackage{changebar}
\usepackage{here}
\usepackage{booktabs}

\newcommand{\dzero}     {D0}
\newcommand{\ttbar}     {\mbox{$t\bar{t}$}\xspace}

\newcommand{\ppbar}     {\mbox{$p\bar{p}$}\xspace}

\newcommand{\pythia}    {\sc{pythia}}

\newcommand{\alpgen}    {\sc{alpgen}}

\newcommand{\evtgen}    {\sc{evtgen}}

\newcommand{\geant}     {\sc{geant}}
\newcommand{\met}       {\mbox{$\not\!\!E_T$}\xspace}

\newcommand{\ljets}     {\mbox{$\ell$+jets}}
\newcommand{\ejets}     {\mbox{$e$+jets}}
\newcommand{\mujets}    {\mbox{$\mu$+jets}}

\newcommand{\wplus}     {$W$+jets\xspace}

\begin{document}



\hspace{5.2in} \mbox{FERMILAB-PUB-07/128-E}
\title{ Measurement of the $t\overline{t}$ production cross section in
\ppbar ~collisions at {\mbox{$\sqrt{s}$ =\ 1.96\ TeV}} using kinematic
characteristics of lepton + jets events}

\input{list_of_authors_r2_new.tex}

\date{May 18, 2007}


\begin{abstract}
We present a measurement of the top quark pair production cross section 
in \ppbar ~collisions at \( \sqrt{s}=1.96 \) TeV utilizing 
425 pb$^{-1}$ of data collected with the D0 detector at the Fermilab Tevatron
Collider.  
We consider the final state of the top quark pair containing one high-$p_{T}$ 
electron or muon and at least four jets. 
We exploit specific kinematic features of \ttbar ~events to extract the 
cross section.  
For a top quark mass of 175 GeV, we measure
$
\sigma_{t\overline{t}} = 
6.4^{+1.3}_{-1.2}\:{\rm (stat)}
\pm 0.7 \:{\rm (syst)}                
\pm 0.4\:{\rm(lum)}\:{\rm pb},$
\noindent in good agreement with the standard model prediction.
\end{abstract}

\maketitle

\section{Introduction}
\label{sec:intro}
\input{intro}

\section{\dzero ~detector}
\label{sec:detector}
\input{detector}

\section{Object identification}
\label{sec:pid}
\input{pv}
\input{electron}

\input{muon}
\input{jets}
\input{met}

\section{Data Samples and Monte Carlo simulation}
\label{sec:samples}
\input{trigger}
\input{trigger_meas}
\input{mc}
\input{dataMCcor}

\section{Method overview}
\label{sec:method}
\input{method}

\section{Event selection}
\label{sec:select}
\input{selection}

\section{Backgrounds}
\label{sec:bckg}
\input{bckg_nils}

\section{Kinematical analysis}
\label{sec:topo}
\input{topo_nils}

\section{Cross section extraction}
\label{sec:xs}
\input{xs}


\section{Systematic uncertainties}
\label{sec:syst}
\input{systematics}

\section{Summary}
\input{summary}

\section{Acknowledgements}
\input{acknowledgement_paragraph_r2.tex}


\begin{appendix}
\input{appendix/topovar}

\end{appendix}

\end{document}

%% file: list_of_authors_r2_new.tex
%
\author{V.M.~Abazov$^{35}$}
\author{B.~Abbott$^{75}$}
\author{M.~Abolins$^{65}$}
\author{B.S.~Acharya$^{28}$}
\author{M.~Adams$^{51}$}
\author{T.~Adams$^{49}$}
\author{E.~Aguilo$^{5}$}
\author{S.H.~Ahn$^{30}$}
\author{M.~Ahsan$^{59}$}
\author{G.D.~Alexeev$^{35}$}
% ...

\affiliation{\vspace{0.1 in}(The D\O\ Collaboration)\vspace{0.1 in}}
% ...
\affiliation{$^{5}$University of Alberta, Edmonton, Alberta, Canada,
                Simon Fraser University, Burnaby, British Columbia, Canada,
                York University, Toronto, Ontario, Canada, and
                McGill University, Montreal, Quebec, Canada}
% ...
\affiliation{$^{28}$Tata Institute of Fundamental Research, Mumbai, India}
% ...
\affiliation{$^{30}$Korea Detector Laboratory, Korea University, Seoul, Korea}
% ...
\affiliation{$^{35}$Joint Institute for Nuclear Research, Dubna, Russia}
% ...
\affiliation{$^{49}$Florida State University, Tallahassee, Florida 32306, USA}
% ...
\affiliation{$^{51}$University of Illinois at Chicago,
                Chicago, Illinois 60607, USA}
% ...
\affiliation{$^{59}$Kansas State University, Manhattan, Kansas 66506, USA}
% ...
\affiliation{$^{65}$Michigan State University,
                East Lansing, Michigan 48824, USA}
% ...
\affiliation{$^{75}$University of Oklahoma, Norman, Oklahoma 73019, USA}
% ...

%% file: intro.tex
Since the discovery of the top quark in 1995 by the CDF and D0 
experiments~\cite{topdisc},
the Fermilab Tevatron $p\bar{p}$ Collider with its center-of-mass energy of 
$\sqrt{s}=1.96~\rm{TeV}$ 
is still the only collider where top quarks can be studied.
Within the standard model, top quarks are produced either in pairs via strong 
interactions or as single top events via electroweak interactions 
with a lower expected cross section~\cite{single_top}.
Evidence for the latter production mode has been recently found by the 
D0 collaboration~\cite{single_top_new}.
At the current Tevatron Collider center-of-mass energy, 
top quark pair production is predicted to occur via $q\bar{q}$ annihilation 
or gluon fusion with a ratio of approximately 85:15.

The $t\bar{t}$ pair production cross section was measured in various 
channels during Run~I of the Fermilab Tevatron Collider at a center-of-mass energy of 
$\sqrt{s}=1.8~\rm{TeV}$~\cite{runI}. The precision of these measurements
was severely limited by 
available statistics. The $10\%$ higher collision energy 
of the current Tevatron Collider run leads to a $30\%$ higher 
expected top quark pair production rate;
together with an increased luminosity, the precision on measurements of the top 
quark production and decay properties can therefore be substantially increased. 
The latest theoretical calculations 
\cite{SMtheory_B,SMtheory_K,SMtheory_C} of the $t\bar{t}$ production cross section 
at next-to-next-to-leading order (NNLO) have an 
uncertainty ranging from $9\%$ to $12\%$.
Recent measurements with a data set approximately twice as large as in 
Run I \cite{top_r2_xs_d0,top_r2_xs_cdf} are consistent with these 
predictions within the uncertainties. 

Deviations from the standard model could occur due to the presence of new 
physics, such as
resonant $t\bar{t}$ production~\cite{top_res}, a novel top quark decay 
mechanism, as for example,  
$t \rightarrow H^+ b $~\cite{top_H+} or a similar final state signature 
from a top-like particle~\cite{top_like}. Some of these effects could
cause the inclusive $t\bar{t}$ cross section ($\sigma_{t\bar{t}}$) to be different 
from the standard model prediction. Others could cause differences
in top decay branching fractions, thus leading to $\sigma(t\bar{t})$
measured in different decay channels to disagree with the expectations
computed using the standard model branching fractions.
Therefore measurements of $\sigma_{t\bar{t}}$ in different top quark
decay channels and using different analysis methods complement each other.

In this paper we present a new measurement of the top quark production 
cross section
in the \ljets ~channel, where one of the $W$ bosons decays hadronically, 
and the other one leptonically into an electron ($W \rightarrow e \nu $) 
or a muon ($W \rightarrow \mu \nu$).
$W$ boson decays into a $\tau$ lepton with a subsequent decay of the latter 
into an electron or a muon are included in the signal sample. 
Each of the two decay channels represent approximatively 17\% 
of the total top quark pair production and decay. 
We exploit the kinematic properties of the events to separate $t\bar{t}$ 
signal from $W$+jets background, instead of the often-exploited requirement of 
a final-state separated vertex that is consistent with the $b$ decay.  
This choice makes this measurement less dependent on the assumption
that a top quark decays into a $b$-quark.  

The measurement is based on a data sample taken between August 2002 and 
August 2004 with an integrated luminosity of $425~\rm{pb^{-1}}$, which 
represents approximatively a factor two increase with respect to the 
previously published measurement by the D0 experiment~\cite{top_r2_xs_d0}.

After a short description of the relevant D0 detector parts and 
underlying object identification algorithms,
we describe the data and Monte Carlo samples, 
the event selection, the background determination and the procedure to extract 
the top quark signal. Finally, we discuss the systematic uncertainties
associated with the cross section measurement.

%% file: detector.tex
The D0 detector~\cite{nimpaper} is a nearly hermetic multi-purpose apparatus built 
to investigate $p\bar{p}$ interactions at high 
transverse momentum. The measurements reported here rely on the tracking 
system, the Uranium-Liquid Argon calorimeter, 
the muon spectrometer and the luminosity detectors, which are briefly 
described below. The coordinate system is 
right handed with the $z$ axis along the Tevatron  
proton beam direction, the $y$ axis vertical and the $x$ axis pointing outside of the 
accelerator ring. 
The coordinates are also expressed in terms of the azimuthal 
angle $\varphi$, rapidity $y$ and pseudorapidity $\eta$. The latter are 
defined as functions of the polar angle $\theta$ as 
$y(\theta,\beta) \equiv
{1 \over 2} \ln{[(1+\beta\cos{\theta})/(1-\beta\cos{\theta})]}; 
\eta(\theta) \equiv y(\theta,1)$, where $\beta$ is the ratio of particle momentum
to its energy.  
When the center of the D0  
detector is considered as the origin of the coordinate system,
these coordinates 
are referred to as detector coordinates 
$\varphi_{\rm det}$ and $\eta_{\rm det}$;
when the reconstructed interaction vertex is considered as
the origin of the coordinate sytem, these coordinates are referred to
as physics coordinates $\varphi$ and $\eta$.

The tracking system includes the Silicon Microstrip Tracker (SMT) 
and the Central Fiber Tracker (CFT). 
A superconducting solenoid surrounds the tracking system and provides a 
uniform 
magnetic field of $2~\rm{T}$. The SMT is a system closest to the beam pipe. 
It has six barrels in the 
central region of $|\eta_{\rm det}| <1.5$, each barrel is $12~\rm{cm}$ long and 
capped at high $|z|$ by a 
disk with an external radius of $10.5~\rm{cm}$. Each barrel has four 
silicon readout layers, composed 
of two staggered and overlapping sub-layers. Each small-radius disk is 
composed of twelve double-sided wedge-shaped detectors. Track 
reconstruction in the forward region up to $|\eta_{\rm det}|<3$ is provided by
two units composed of three small and two large radius disks located 
at $|z|=44.8,~49.8,~54.8~\rm{cm}$ and $110,~120~\rm{cm}$ respectively. 
Large radius disks are composed of
48 single-sided wedges with an external radius of $26~\rm{cm}$.
The CFT consists of 8 concentric cylinders and covers the radial space from $20$ to $52~\rm{cm}$. 
The two innermost cylinders are $1.66~\rm{m}$ long, and the outer six cylinders are $2.52~\rm{m}$ long. 
Each cylinder supports two doublets of overlapping scintillating fibers with the diameter of 
$0.84~\rm{mm}$, one doublet being parallel to the beam axis, the other with an 
alternating stereo 
angle of $\pm 3^{\circ}$. Light signals are transfered via clear optical fibers to solid-state 
visible light photon counters (VLPCs) that have a quantum efficiency of about $80\%$.
Tracks are reconstructed combining the hits from both tracking detectors. 

The calorimeter is used to reconstruct jets, electrons, 
photons and missing transverse energy of non-interacting particles such as
neutrinos. 
The D0 Uranium-Liquid Argon calorimeter which surrounds the tracking system 
is divided into the Central 
Calorimeter (CC) up to  $|\eta_{\rm det}| \simeq 1.0$ and two Endcap 
Calorimeters (EC) extending the coverage to 
$|\eta_{\rm det}|\simeq 4.0 $, housed in separate cryostats. Each 
calorimeter consists of an 
electromagnetic section with depleted Uranium absorber plates, a fine hadronic 
section with an 
Uranium-Niobium absorbers and a coarse hadronic section with Copper 
(Stainless Steel) absorbers in
the CC (EC). The calorimeter is compact and highly segmented in the transverse and the 
longitudinal 
directions with about 56,000 channels in total. In $\varphi$, the electromagnetic part 
 is divided 
into 64 modules and the hadronic part into 32 modules. 
The electromagnetic part has a depth of about $20$ radiation
lengths ($X_0$); and with the hadronic sections, the calorimeter has
a total of $7.2$ nuclear interaction length ($\lambda_I$) at $\eta=0$ and 
of $10.3 \lambda_I$ at $|\eta| \simeq 4$.
The inter-cryostat region is equipped with scintillation detectors
(inter-cryostat detectors or ICD) to improve energy resolution. 

The muon system is the outermost part of the D0 detector. It 
consists of three 
layers of tracking detectors used for precise coordinate measurements
and triggering and two layers of 
scintillation counters used for triggering~\cite{muonnimpaper}. Proportional drift 
tubes (PDT) cover the central region ($|\eta_{\rm det}| < 1.0$), and mini drift tubes (MDT) 
extend the coverage to $|\eta_{\rm det}| = 2.0$. One layer of scintillation counters 
in the central region and two layers in the forward region ($1.0 < |\eta_{\rm det}| < 
2.0$) along with two layers of drift tubes (B and C layers) are located outside of a 
$1.8~\rm{T}$ iron toroid while the innermost layers (A) of muon tracking detectors 
and scintillators are located in front of it. 
The support structure underneath the D0 detector allows only for 
partial coverage in this region. 

The luminosity is determined from the rate of inelastic collisions measured by the 
luminosity monitors (LM) 
located in front of the ECs at $z=\pm 140~\rm{cm}$. The LM consists of two arrays of 
24 plastic scintillator 
counters with photomultiplier readout and covers the pseudorapidity range $|\eta_{\rm det}|$ 
between $2.7$ and $4.4$.
The uncertainty on the luminosity measurement is $\pm$ 6.1\% ~\cite{d0lumi} 
and is dominated by the uncertainty on the \ppbar ~inelastic cross section.

%% file: pv.tex
\subsection{Primary vertex}

The primary (or hard scatter) vertex of the event is reconstructed in three  
steps. At the first step, we locate a beam spot position using 
reconstructed tracks with the transverse momentum of $p_T>0.5~\rm{GeV}$. These
tracks should have at least two hits in the SMT detector and the significance
of the distance of closest approach $S_{\mathrm{dca}}=|dca/\sigma_{\mathrm{dca}}|<100$. 
The distance of closest approach $(dca)$ is 
calculated with respect to the center of the detector in the plane 
transverse to the beamline.
At the second step, we impose a more stringent requirement on the tracks,
$S_{\mathrm{dca}}<3$, where $S_{\mathrm{dca}}$ is calculated with respect to the beam spot 
determined in the previous 
step. These tracks are then used to fit the final primary vertices. 
We use information on the position of these tracks along the
beamline to identify tracks belonging to different interactions and build 
clusters of the tracks within 2 cm from each other. All tracks in each cluster
are fitted to a common vertex using the Kalman filter technique~\cite{kalman}. 
Finally, to distinguish the position of the hard scatter interaction from 
the simultaneously produced minimum bias scatters, a minimum bias probability is 
computed for each reconstructed vertex based on the transverse 
momenta and the total number of associated tracks. The primary vertex with the 
lowest minimum bias probability is selected as the hard scatter.

The primary vertex finding algorithm reconstructs vertices in the fiducial region 
of the SMT with an efficiency close to 100\%. 
The position resolution, measured in data as a difference between the reconstructed vertex 
position and the position of the beam spot center, depends on the number of tracks 
fitted to the primary vertex and is around $40~\rm{\mu m}$ in the plane transverse to the 
beam direction. It is dominated by the beam spot size of about $30~\rm{\mu m}$. 

For the analysis, we select events with the primary vertex within the SMT 
fiducial region $|z_{\mathrm{PV}}| \le 60~\rm{cm}$ and at least three tracks attached  
to the vertex. 


%% file: electron.tex
\subsection{Electrons}
\label{sub:em}
The electron identification is based on clusters of calorimeter cells 
found in the CC within $|\eta_{\rm det}| < 1.1$ using a simple cone 
algorithm with a cone size of 
${\cal{R}} = \sqrt {(\Delta\eta)^2+(\Delta\varphi)^2} = 0.2$.
A cluster is considered as a ``loose'' electron if 
({\it i}\,) at least $90\%$ of its reconstructed energy is in the 
electromagnetic part of the calorimeter ($f_{\mathrm{EM}} > 0.9$), 
({\it ii}\,) the cluster is isolated, 
({\it iii}\,) its shower shape is consistent with an electromagnetic shower and 
({\it iv}\,) there is at least one track in a 
$\Delta\eta \times \Delta\varphi$ road of size $0.05 \times 0.05$ around the 
cluster.
The angular coordinates {$\eta, \varphi$} of the electron are taken 
from the parameters of the matched track; its energy is determined
from the calorimeter cluster. 
The isolation criterion $f_{\mathrm{iso}}$ requires the ratio of 
the difference of the total energy within the cone size ${\cal{R}}< 0.4$ 
around
the center of the cluster and the energy deposited in electromagnetic 
layers within the cone size ${\cal{R}}< 0.2 $   
to the reconstructed electron energy not to exceed 15\%. 

The electron shower shape estimator is built from seven observables
characterizing the electron shower shapes, which are
the energy deposits in the first five layers of  the calorimeter, the azimuthal 
extension of the cluster in the finely segmented third layer of the 
calorimeter, and the logarithm of the cluster total energy. 
From these observables a covariance matrix is built, where the
matrix elements are computed from reference Monte Carlo samples
at different cluster energies and pseudorapidities. The covariance parameter
$\chi^2_{H}$ measures the consistency of a given shower 
to be an electromagnetic one. As the observables are not normally 
distributed, $\chi^2_{H}$ does not follow a normal $\chi^2$ distribution
and a cut on $\chi^2_H<50$ is applied for electrons.

To define a ``tight'' electron we combine in a likelihood discriminant 
the variables defined above 
($f_{\mathrm{EM}}$, $\chi^2_{H}$) with  
({\it i}\,) the ratio of the transverse component of
the cluster energy measured in the calorimeter to the transverse momentum 
of the matched track, $E_T^{\mathrm{cal}}/p_T^{\mathrm{track}}$, ({\it ii}\,) the $\chi^2$ probability 
of a track matched to the calorimeter cluster, 
({\it iii}\,) the
 $dca$ of the matched track with respect to the primary vertex, 
({\it iv}\,) the number of tracks within a cone of ${\cal{R}}=0.05$ around 
the matched track and ({\it v}\,) the sum of transverse momenta of the tracks 
inside a cone of ${\cal{R}}<0.4$ around, but excluding the candidate track. 
By construction,
a discriminant value close to unity corresponds to a prompt isolated 
electron.      
We require that tight electrons satisfy the loose criteria and have a   
likelihood discriminant ${\cal {L}}_{\mathrm{em}} > 0.85$. 

The electron energy scale is fixed by comparing the di-electron invariant mass distribution in $Z\rightarrow ee$ events 
selected from the data with the simulated expectation based on a $Z$ boson mass of 
$91.19~\rm{GeV}$~\cite{PDG2006}. 
Additional random smearing of the electron inverse energy is applied to tune the simulated electron energy 
resolution to that observed in the data.


%% file: muon.tex
\subsection{Muons}
\label{muonID}
Muons are identified from tracks reconstructed in the  
layers of the muon system and matched to a track reconstructed in the 
central tracking system taking
advantage of its superior momentum and position resolution.
For this analysis, we accept muons having ({\it i}\,) at least
two wire hits and at least one scintillator hit in both the A-layer inside the toroid and 
the B- and C-layers outside, ({\it ii}\,) three matched reconstructed muon 
track segments from all three muon system tracking layers, ({\it iii}\,)
a good quality matched track in the central tracking system 
($\chi^2/N_{\mathrm{dof}} < 4$) 
and
({\it iv}\,) consistency with originating from the primary interaction vertex. 
The last condition includes the requirements that the timing of the muon, 
determined from associated scintillator hits, has to be within $10~\rm{ns}$ 
of the 
beam interaction time, that the smallest distance along $z$ axis between the 
primary vertex and the muon track must be less than 1~cm   
and $S_{\mathrm{dca}}<3$.

Muons are distinguished as ``loose'' and ``tight'' depending on their isolation with respect to other reconstructed objects in the event.
The loose muon isolation criterion is defined by demanding that a muon 
is separated from a jet by $\Delta {\cal {R}}(\mu, \mathrm{jet}) > 0.5$ 
where $\Delta {\cal {R}}$ is the distance in pseudorapidity-azimuthal angle 
space. 
For a tight muon identification, the muon is additionally required to be isolated 
from
energy depositions in the calorimeter and additional tracks in the tracking system.
The calorimeter isolation requires the sum of the calorimeter cells' transverse 
energies between two cones of 
radius ${\cal{R}}=0.1$ and ${\cal {R}}=0.4$ around the muon track to be 
smaller than $8\%$ of the muon $p_T$. 
The track isolation is based on the sum of the tracks' momenta 
contained in a cone of ${\cal{R}}=0.5$ around the muon track, excluding the muon 
track itself. We require the sum to be less than 
$6\%$ of the muon $p_T$. 

The muon momentum is measured from the matched reconstructed central track. 
Due to the limited acceptance of the SMT some tracks have hits in the CFT part
of the central tracking system only, and therefore their resolution is degraded. 
To improve the momentum 
resolution of such tracks we apply a correction to the inverse track 
transverse momentum. It is based on a fit constraining the track $dca$ to zero with 
respect to the primary vertex in the transverse plane.

The muon momentum scale is fixed by comparing the di-muon invariant mass distribution in $Z\rightarrow \mu\mu$ events 
selected from the data with the simulated expectation based on the $Z$ boson mass.
Additional random smearing of the muon inverse transverse momenta is performed to tune the simulated muon momentum 
resolution to that observed in the data.  


%% file: jets.tex
\subsection{Jets}
\label{jetid}
Jets are reconstructed from calorimeter cells using the iterative, seed-based 
cone algorithm including midpoints ~\cite{jet_algo} with a cone radius of 
${\cal{R_{\mathrm{jet}}}} = \sqrt {(\Delta y)^2+(\Delta\varphi)^2} = 0.5$.
The minimum $p_T$ of a reconstructed jet is required to be 8 GeV before 
any energy corrections are applied.
To remove jets resulting from noise in the calorimeter or created by
electromagnetic particles, further quality criteria are applied:
({\it i}\,) the jet has to 
have between 5\% and 95\% of its reconstructed energy in the electromagnetic 
calorimeter and less than 40\% of its energy in the outermost hadronic section 
of the calorimeter,
({\it ii}\,) the ratio of the highest to the next-to-highest transverse 
momentum cell in a jet has to be less than 10, 
({\it iii}\,) a single calorimeter tower must not contain more than 90\% of the
jet energy and
({\it iv}\,) the jet has to be confirmed by the independent calorimeter trigger  
readout.

Previously reconstructed electrons and photons might also be reconstructed 
as jets in the calorimeter. To 
avoid the resulting double-counting, we reject any jet
overlapping with an electromagnetic object within a cone of ${\cal{R}} < 0.5$ 
fulfilling the electron identification criteria ({\it i}\,)--({\it iii}\,) 
of Sect.~\ref{sub:em} and having $p_T>15$ GeV and $|\eta_{\rm det}|<2.5$.

We correct the $p_T$ of each reconstructed jet to the particle level 
by applying jet energy scale (JES) corrections \cite{massideo}. These 
corrections account for imperfect calorimeter response, the jet energy offset due
to the underlying event, multiple interactions, pile-up effects and noise,  
and the jet energy loss due to showering outside of the fixed-size jet cone.
We make use of transverse momentum  
conservation in a sample of photon + jet events to calibrate the jet energy and determine
the jet energy scale corrections separately for data and simulation. 
Since the jet identification efficiency and energy resolution differ between 
data and simulation, the jet inverse energies are smeared and depending on the 
jet $|\eta_{\rm det}|$ from 1\% to 3\% of the jets 
are removed to reproduce the data.

%% file: met.tex
\subsection{\label{pid_met} Missing $E_T$} 
The presence of a neutrino in the final state can be inferred
from the energy imbalance of an event in the transverse plane. It is
reconstructed from the vector sum of the transverse energies of all
cells surviving various noise suppression algorithms and not 
belonging to a coarse hadronic section of the calorimeter. The latter 
cells are generally
excluded due to their higher noise level. They are however included
if clustered within jets. The vector opposite to this total
visible momentum vector is referred to as raw missing transverse energy vector.

The calorimeter response to electromagnetic particles such as photons, 
electrons or $\pi^0$s is different from that due to hadrons and in 
particular from
that due to jets. In events with both electromagnetic objects and jets,
this imbalance propagates directly into missing transverse energy (\met). 
As a JES correction is derived for all jets satisfying criteria 
({\it i}\,)--({\it iv}\,) of Sect.~\ref{jetid}, it also has to be applied 
to ~\met. In order to do so, the JES
correction (limited to the response part) applied
to jets is subtracted from the \met\ vector. In an equivalent
way the EM correction for electromagnetic objects is applied 
to the \met\ vector. 

Muons are minimum ionizing particles throughout the entire detector.
Hence they will deposit only a small amount of energy in the
calorimeter and their presence can thus fake missing transverse
energy. Therefore we replace the transverse 
energy deposited by muons, satisfying requirements ({\it i}\,)--({\it iii}\,) of
Sect.~\ref{muonID}, in the calorimeter by the 
transverse momentum measured by the tracking system.

%% file: trigger.tex
\subsection{Event trigger}
The D0 trigger is based on a three-level pipeline system. 
The first level consists of hardware and firmware components that make a trigger decision
based on fast signal inputs from the luminosity monitor, the tracking system, 
the calorimeter and the muon system. The second level combines the same 
information to construct simple physics objects, 
whereas the third level is software based
and uses the full event information obtained with a simplified reconstruction.
The accepted event rates are $2~\rm{kHz}, 1~\rm{kHz}$ and $50~\rm{Hz}$ respectively
for level~1 (L1), level~2 (L2) and level~3 (L3). 
For all events used in this analysis the trigger system is required to find at least 
one jet and an electron or muon.

The D0 calorimeter trigger is based on energy deposited in towers of calorimeter cells with a transverse granularity of $\Delta \eta \times
\Delta \varphi = 0.2 \times 0.2$. In addition, towers are segmented longitudinally into electromagnetic (EM) and hadronic
(HAD) sections. The level 1 electron trigger requires a minimum transverse energy 
($E_T$) deposition in the
electromagnetic section of a tower. At level 2, a seed-based cluster 
algorithm sums the energy in neighboring towers and bases the trigger decision on the 
$E_T$ and the electromagnetic fraction ($f_{\mathrm{EM}}$) of a cluster.
At level 3, the electron identification is based on a simple cone algorithm 
with ${\cal R}<0.25$ and the trigger decision is based on the requirements
on $E_T$, $f_{\mathrm{EM}}$ and a shower-shape estimator. 

The level 1 jet trigger is based on the $E_T$ deposited in a full 
calorimeter trigger tower. At level 2, these towers are summed by a seed based cluster algorithm within
a $5 \times 5 $ tower array. The level 3 jet algorithm uses a simple cone algorithm with 
${\cal R }<0.5$ or ${\cal R}<0.7$ and a decision is taken based on the $E_T$ within the cone.

The level 1 muon trigger is based on input from the muon scintillator counters, the muon wire 
chambers and the track trigger system. At level 2, muons are reconstructed from the muon
scintillator and wire chamber information and requirements on the number of muons, their
transverse momentum $p_T$ and position in $\eta$ as well as on their quality can be made. The
quality is based on the number of scintillators and wires hit.
At level 3, muon tracks are fitted using information from the tracking and 
muon systems. This  
refines the selection in $p_T$, $\eta$, and reconstruction quality.

The data used for the measurement presented in this paper were collected 
between August 2002 and August 2004 and correspond to an integrated luminosity of 422
$\pm$ 26 pb$^{-1}$ in the $\mu$+jets and 425 $\pm$ 26 pb$^{-1}$ in the $e$+jets
channel, respectively \cite{d0lumi}. The trigger criteria evolved over this 
period of time to account for the increase in instantaneous luminosity 
while keeping a constant trigger rate. 
The different trigger criteria and the corresponding integrated luminosity 
collected are summarized in Table~\ref{table.trigger} for the 
$e+$jets and the $\mu+$jets data. 

\begin{table*}[htpb]
\footnotesize{
\begin{tabular}{l|c|c|c|c} \hline
\multicolumn{5}{c}{$e$ + jets channel}\\
\hline
Trigger Name & $\int {\mathcal L} dt$ & Level 1 & Level 2 & Level 3 \\
 &(pb$^{-1}$) & & & \\ \hline
EM15\_2JT15 & 128 & 1 EM tower, $E_T>10\;\rm GeV$ & 1$e$, $E_T>10\;\rm GeV$,
$f_{\mathrm{EM}}>0.85$
& 1 tight $e$, $E_T>15\;\rm GeV$ \\
& & 2 jet towers, $p_T>5\;\rm GeV$ & 2 jets, $E_T>10\;\rm GeV$
 & 2 jets, $p_T>15\;\rm GeV$\\ \hline
E1\_SHT15\_2J20         & 244 &
1 EM tower, $E_T>11\;\rm GeV$ & None & 1 tight $e$, $E_T>15\;\rm GeV$ \\
 & & &&   2 jets, $p_T>20\;\rm GeV$\\ \hline
E1\_SHT15\_2J\_J25      & 53 &
1 EM tower, $E_T>11\;\rm GeV$ & 1 EM cluster, $E_T>15\;\rm GeV$
& 1 tight $e$, $E_T>15\;\rm GeV$ \\
 & & &&  2 jets, $p_T>20\;\rm GeV$ \\
 & & & & 1 jet, $p_T>25\;\rm GeV$ \\
\hline
\multicolumn{5}{c}{$\mu$ + jets  channel}\\
\hline
Trigger Name& $\int {\mathcal L} dt$ & Level 1 & Level 2 & Level 3 \\ 
 & (pb$^{-1}$) & & & \\ \hline
MU\_JT20\_L2M0 & 132 &
1$\mu$, $|\eta|<2.0$ & 1$\mu$, $|\eta|<2.0$  & 1 jet, $p_T>20\;\rm GeV$\\
 & &1 jet tower, $p_T>5\;\rm GeV$ &  & \\ \hline
MU\_JT25\_L2M0 & 244 &
1$\mu$, $|\eta|<2.0$  & 1$\mu$, $|\eta|<2.0$ & 1 jet, $p_T>25\;\rm GeV$\\
 & & 1 jet tower, $p_T>3\;\rm GeV $ & 1 jet, $p_T>10\;\rm GeV$ & \\ \hline

MUJ2\_JT25 & 30 &
1$\mu$, $|\eta|<2.0$  & 1$\mu$, $|\eta|<2.0$ & 1 jet, $p_T>25\;\rm GeV$\\
 &  & 1 jet tower, $p_T>5\;\rm GeV$ &
1 jet, $p_T>8\;\rm GeV$ & \\ \hline

MUJ2\_JT25\_LM3 & 16 &
1$\mu$, $|\eta|<2.0$  & 1$\mu$, $|\eta|<2.0$ & 1$\mu$, $|\eta|<2.0$ \\
        &  & 1 jet tower, $p_T>5\;\rm GeV$ &
1 jet, $p_T>8\;\rm GeV$ & 1 jet, $p_T>25\;\rm GeV$ \\ \hline
\end{tabular}
}
\caption{Trigger requirements for different data-taking periods.}
\label{table.trigger}
\end{table*}

\subsection{Trigger efficiency}
\label{sc:treff}
Only a fraction of all produced \ttbar\ events will pass the selection criteria imposed 
by the trigger system.
The trigger efficiency for \ttbar\ events is estimated by folding into
simulated events the per-lepton and per-jet probability
to satisfy the individual trigger conditions at L1, L2 and L3. 
The total probability for an event to satisfy a set of
trigger requirements is obtained assuming that the probability for a
single object, described below, to satisfy a specific trigger condition 
is independent
of the presence of other objects in the event. Under this assumption,
the contributions from the lepton and the jets to the total event
probability factorize, so that
\begin{equation}
P_{\mathrm{event}} = P_{\mathrm{lepton}} \times P_{\mathrm{jet}}.
\end{equation}
Furthermore, under the assumption of independent trigger
objects, the probability $P_{\mathrm{jet}}$ for at least one out of
$N_{\mathrm{jet}}$ jets in the event to fulfill the jet part of the
trigger requirement is given by
\begin{equation}
P_{\mathrm{jet}} = 1 - \prod_{i=1}^{N_{\mathrm{jet}}}( 1- P_{i}),  
\end{equation}
where $P_i$ is the probability for one jet to pass the trigger
conditions.  

The total trigger efficiency is then calculated as the luminosity-weighted 
average of the event probability associated to the trigger
requirements corresponding to each data taking period.

%% file: trigger_meas.tex
\subsection{Trigger efficiency measurement}
\label{app:triggerEffMeas}

The probability for a lepton or a
jet to satisfy a particular trigger requirement is measured in  
samples of events that are unbiased with respect to the trigger requirement 
under study. Reconstructed leptons or jets are identified in the
event offline and the trigger efficiency is determined by measuring the fraction 
of objects satisfying the trigger condition under study. These efficiencies are
generally parameterized as a function of the object $p_T$ and
$\eta_{\mathrm{det}}$.  

We use a sample of {\mbox{$Z \rightarrow e^+e^-$}} 
({\mbox{$Z \rightarrow \mu^+\mu^-$}}) events 
to calculate the fraction of electrons (muons), fulfilling the requirements 
defined in Sect.~\ref{sub:em} and~\ref{muonID}, that pass the
trigger requirement under study. We selected events triggered by a single 
electron (muon) trigger and require the presence of two reconstructed electrons
(muons) fulfilling the tight selection criteria 
defined in
Sect.~\ref{sub:em} (\ref{muonID}) for electrons (muons), respectively.
The invariant mass of the two selected leptons is required to be within
a window around the $Z$ mass,
80\,GeV$<\it M_{\ell\ell}<$100\,GeV. We choose one electron (muon) as a 
``tag'' and require it to have $p_T$ above 20 GeV and to be matched to an  
electron (muon) object at all relevant trigger levels. We use the 
other ``probe'' electron (muon) 
to calculate the efficiency of the trigger criterion studied. If both
leptons fulfill the tag requirements, each of them serves both as a 
tag and as a probe. 

\begin{figure}[hbt]
\centering
\includegraphics[width=0.45\textwidth,clip=]{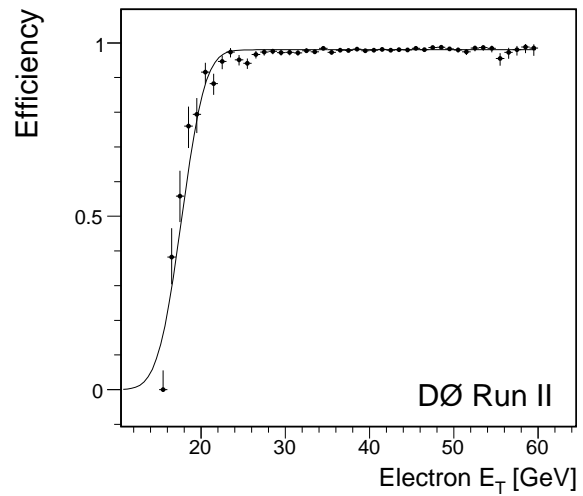}
\caption{\label{fig:emTriggerEff} Electron L3 trigger efficiency
for the last data-taking period and its parameterization as a function of 
the electron $E_T$.}
\end{figure}

\begin{figure}[hbt]
\centering
\includegraphics[width=0.45\textwidth,clip=]{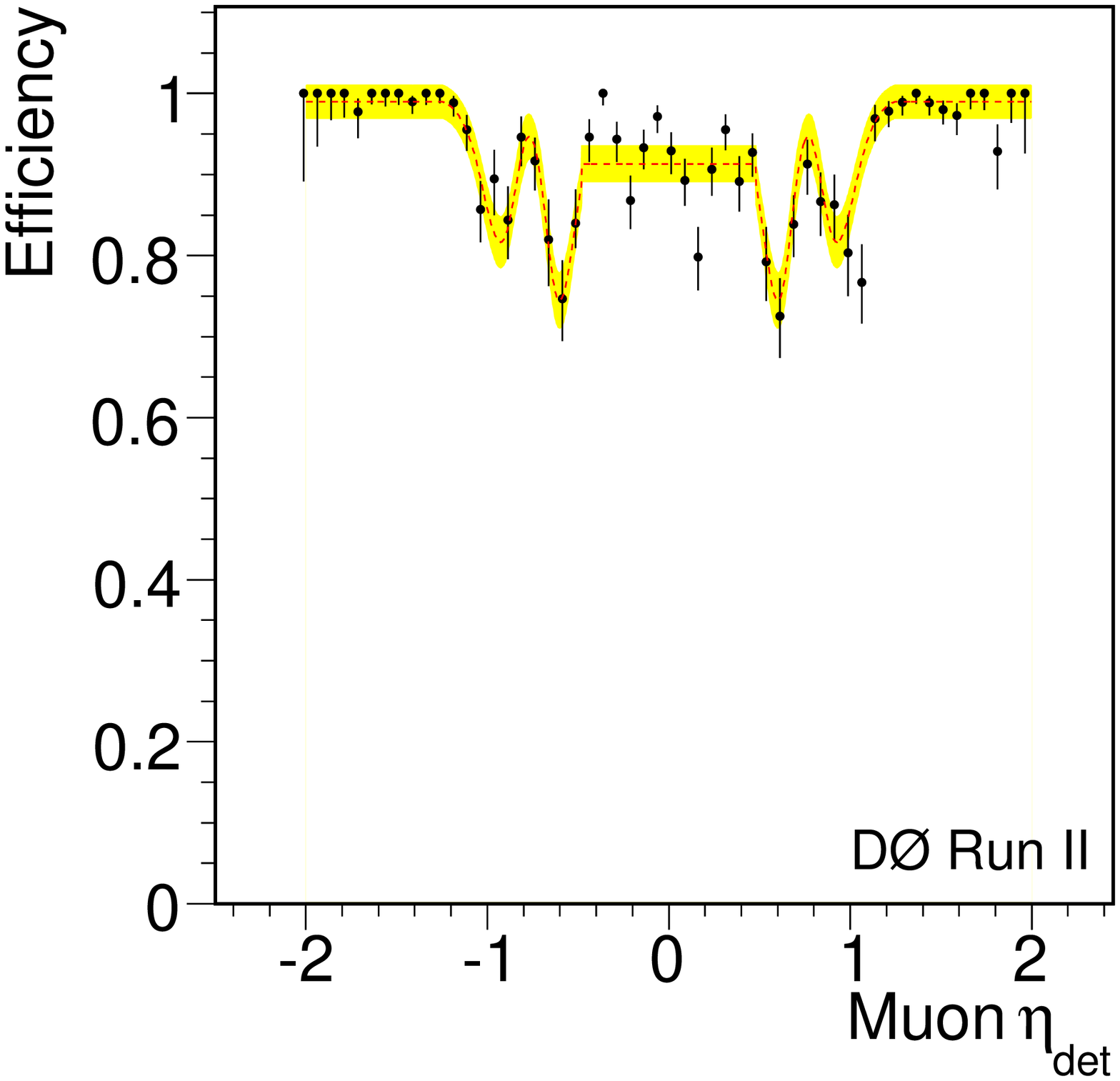}
\includegraphics[width=0.45\textwidth,clip=]{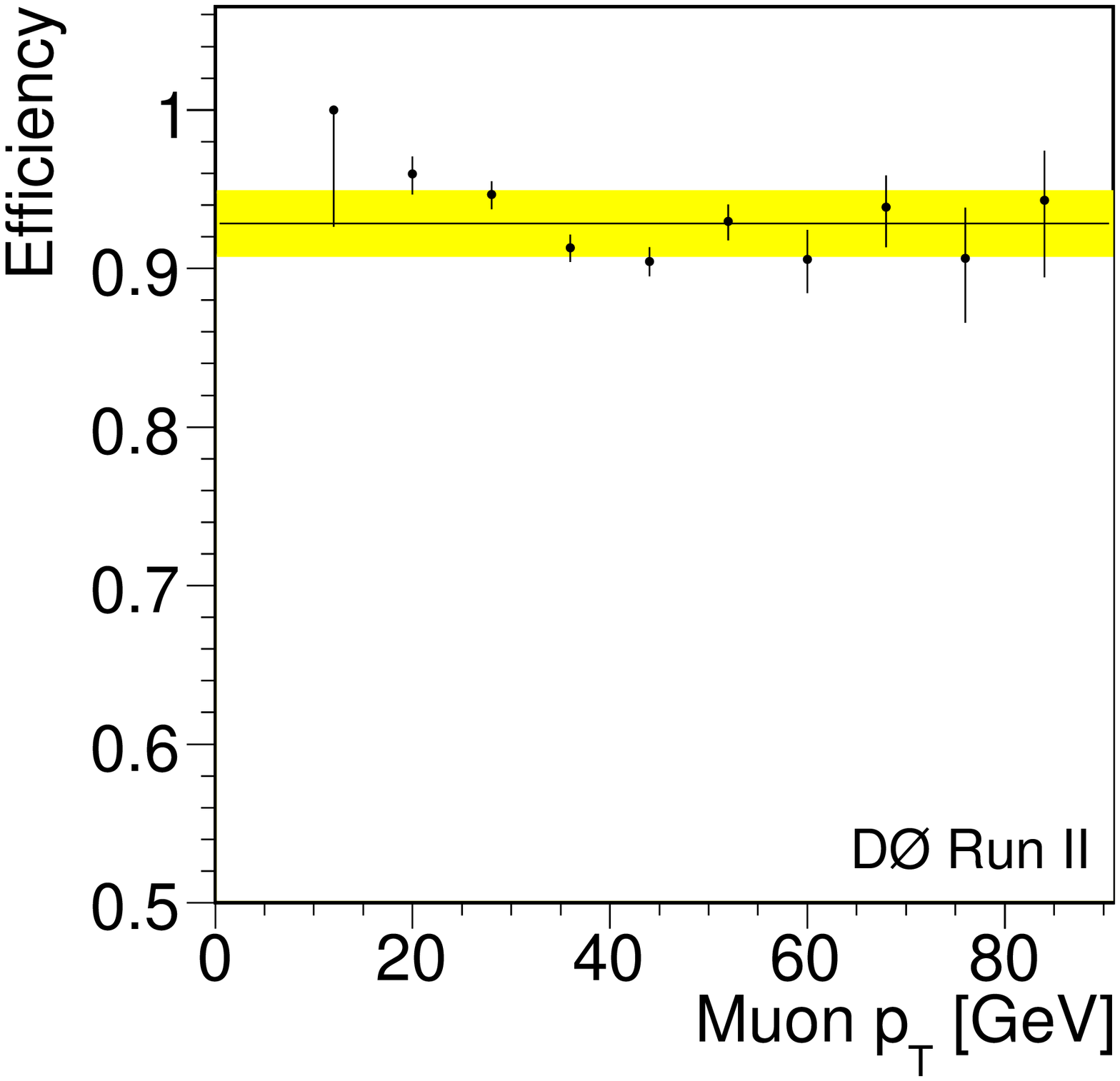}
\caption{\label{fig:muonTriggerEff} Muon trigger efficiencies
for the first and second data-taking periods. The parameterization as a 
function of the muon $\eta_\mathrm{det}$
is shown in the upper plot as the dashed line, the statistical  
error of the fit added in quadrature with the systematic uncertainty 
is given by the band. The lower plot shows
the muon trigger efficiency as a function of the muon $p_T$ and the chosen 
central value along with the uncertainty band. }
\end{figure}

Figure~\ref{fig:emTriggerEff} shows the measured
probability that the electron passes the L3 condition and the 
parameterization used
in the analysis for the last data-taking period.  
Figure~\ref{fig:muonTriggerEff} shows the measured muon trigger
efficiencies for the first and second data-taking periods. The measured 
efficiency is
parameterized as a function of the muon $\eta_\mathrm{det}$ with the 
fit function chosen to be symmetric in $\eta_\mathrm{det}$. 
Both the muon detector 
geometry and the details of offline reconstruction contribute to the 
observed shape of the distribution. We do not 
use the dependence of the trigger efficiency on the muon transverse
momentum in the parameterization. However, due to the spread of efficiencies 
observed, an
overall uncertainty of $\pm$2\,\% is added in quadrature to the
statistical fit uncertainty.  

\begin{figure}[htb]
\leftline{
\includegraphics[width=0.35\textwidth,clip=]{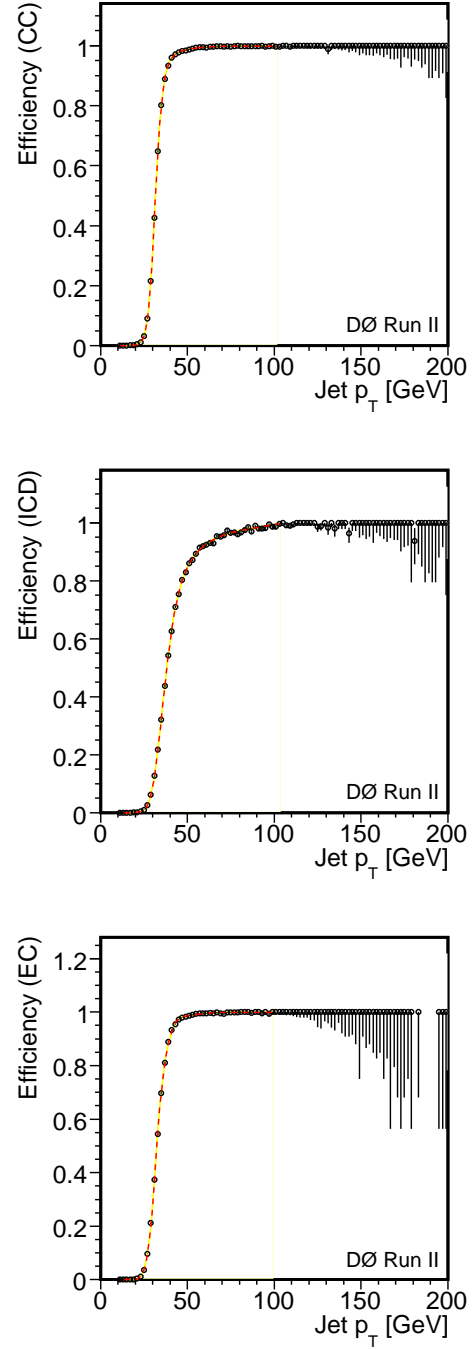}
}
\caption{\label{fig:jetTrigTV9} The trigger efficiency for a jet to pass 
  L1, L2 and L3 trigger requirements for the three
  different calorimeter regions: CC (top), ICD
  (middle) and EC (bottom). 
}
\end{figure}

We measure jet trigger efficiencies in a sample of data events which
fire one of the many muon triggers present in a set of triggers corresponding 
to a data-taking period of interest. The
jet trigger efficiencies are parameterized as a function of jet $p_T$
in three regions of the calorimeter: CC~($|\eta_\mathrm{det}|<0.8$),
ICD~($0.8\le|\eta_\mathrm{det}|<1.5$) and
EC~($|\eta_\mathrm{det}|\ge1.5$). An example of the parameterizations 
obtained for the second data-taking period is shown in
Fig.~\ref{fig:jetTrigTV9}. Systematic uncertainties associated with the method
are evaluated by varying the jet sample selection. The difference between 
the efficiencies derived in different samples is added in
quadrature to the statistical uncertainty of the fits.     

%% file: mc.tex
\subsection{Monte Carlo simulation}
We use Monte Carlo simulated samples to calculate selection efficiencies 
and to simulate kinematic characteristics of the events.  
Top quark signal and $W$+jets and $Z$+jets background processes are generated 
at $\sqrt{s}=1.96~\rm{TeV}$ using {\alpgen}~{\sc 1.3.3}~\cite{alpgen} for the
appropriate matrix element simulation and {\pythia}~{\sc 6.202}~\cite{pythia} 
for subsequent hadronization. We used the {\sc cteq5l}~\cite{cteq} parton
distribution functions for modeling the initial and final state
radiation, decays and hadronization in {\pythia}. The ``tune A''~\cite{tuneA} parameter set is used 
for simulating the underlying event. Minimum bias simulated  
proton-antiproton events are superposed on all simulated events after 
hadronization.  

In the {\ttbar} signal simulation we set the top quark mass to
$175~\rm{GeV}$ and choose the factorization scale for calculation of 
the {\ttbar} process 
to be $Q^2= m_t^2$. We use {\evtgen}~\cite{evtgen} to provide the 
branching fractions and lifetimes for all $b$ and $c$ hadrons. 
The main background consists of $W$+jets and  is simulated at the factorization 
scale
$M_W^2+\sum p_{T_j}^2$ where $M_W$ is the $W$ boson mass and $p_{T_j}$ is  
the transverse momentum of the jet $j$ in the event. 
For $Z$+jets events   
the scale is set to the squared invariant mass of the lepton pair $M_{\ell\ell}^2$. 
We include
virtual photon process (Drell-Yan production) and the
interference between the photon and $Z$~boson in the model. 

Generated events are processed through the {\geant}-based~\cite{geant} 
simulation of the {\dzero} detector and are reconstructed with the 
same program as used for collider data. 

%% file: dataMCcor.tex
\subsection{Calibration of Monte Carlo simulations}
\label{sc:dataMCcor}
We smear (i.e., convolute with a Gaussian) the reconstructed inverse energies 
of electrons and inverse transverse momenta of 
muons and jets in the simulation to improve the agreement with the observed 
momentum resolutions
in data, as already described in Sect.~\ref{sub:em}--\ref{jetid}.
In addition, we correct the simulation for possible 
inaccuracies in describing individual object identification efficiencies. 
We derive correction factors to account for the difference in the 
following efficiencies between data and the simulation: 
({\it i}\,) electron (muon) reconstruction and identification,       
({\it ii}\,) electron (muon) track match,
({\it iii}\,) electron likelihood,
({\it iv}\,) muon isolation,
({\it v}\,) muon track quality and the distance of
closest approach significance (requirements {\it iii} and {\it iv} 
of Sect.~\ref{muonID}, respectively),  
({\it vi}\,) primary vertex selection, and
({\it vii}\,) electron (muon) promptness 
by comparing the efficiencies measured in 
{\mbox{$ Z\rightarrow {\ell^+\ell^-}$}}\ data events to the ones obtained 
from the simulation. Two typical examples of the methods used to 
determine correction factors and their systematic uncertainties are provided 
below.  

To measure the efficiency of electron (muon) reconstruction, we use the same 
tag and probe method as that used in the trigger efficiency calculation.
To avoid bias due to trigger requirements events used for the measurement have to be 
recorded with a single electron (muon) trigger, and we require a tag electron  
(muon) be matched to the electron (muon) trigger object at all trigger levels.  
We repeat the same measurement using simulated 
{\mbox{$ Z\rightarrow {\ell^+\ell^-}$}}\ events and plot the ratios of   
the efficiencies as a function of detector $\eta$, $\phi$ and $p_{T}$ for muons
and, additionally, as a function of the distance to the closest jet in the event for
electrons to probe the dependence of the electron reconstruction on the jet
activity. Since no strong dependence on any of these
quantities is found, we use inclusive factors to correct simulation to data yielding 
$0.98 \pm 0.027 {\rm\; (syst)}$ ($1.00 \pm 0.04 {\rm\; (syst)}$) for the electron
(muon) reconstruction and identification efficiency. 
Systematic uncertainties are assigned based on the spread of
measured ratios in the $\Delta {\cal{R}}(e,\mathrm{jet})$ distribution for electrons and 
$\vert\eta_{\rm det}\vert$ for muons. 
 
\begin{figure}
\centering
\includegraphics[width=0.5\textwidth,clip=]{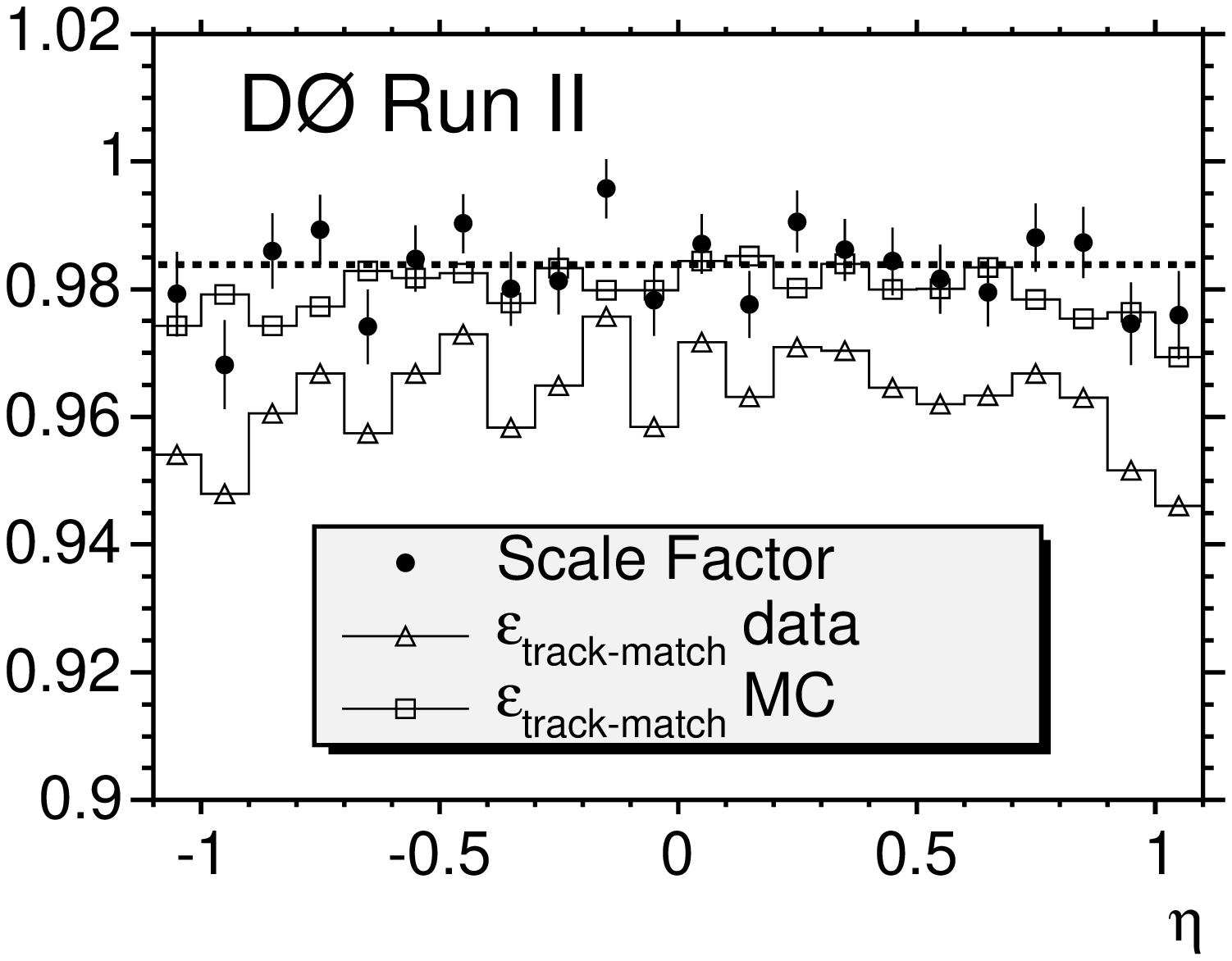}
\includegraphics[width=0.5\textwidth,clip=]{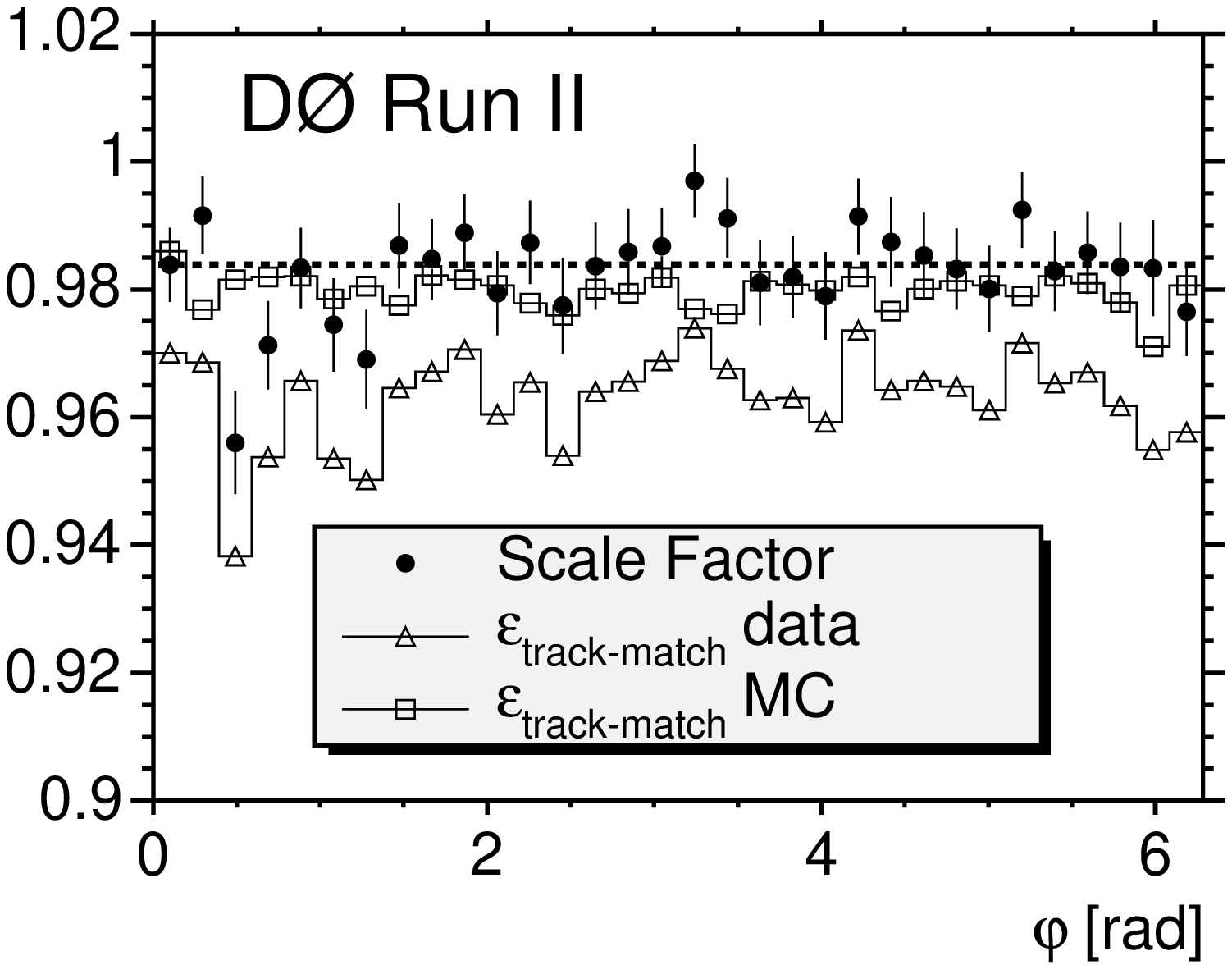}
  \caption{Track match efficiency in {\mbox{$ Z\rightarrow {e^+e^-}$}}\ data 
  and Monte Carlo and their ratio as a function 
  of $\eta$ (top) and $\varphi$ (bottom) in CC.}
  \label{fig:etrack_vs_eta_phi}
\end{figure}

We determine the efficiency of finding a track matched to a electron (muon)
by applying the same tag and probe method to 
{\mbox{$ Z\rightarrow {\ell^+\ell^-}$}}\ events selected with a tight electron
(muon) as a tag and an electromagnetic cluster satisfying criteria 
({\it i}\,)--({\it iii}\,) of Sect.~\ref{sub:em} 
(a muon identified in the muon chambers) as probe for electrons (muons). 
The correction factors obtained by comparing efficiencies in data and
the simulation are found to be $0.983 \pm 0.007 {\rm\; (syst)}$ for 
electrons and $0.99 \pm 0.03 {\rm\; (syst)}$ for muons. Systematic uncertainties 
arise mainly from the minor dependence of the correction factors on the $p_T$, 
$\eta$ and $\phi$ of the leptons. An example of such a dependence 
is shown in Fig.~\ref{fig:etrack_vs_eta_phi} for the electron
track match efficiency.   


%% file: method.tex
The analysis strategy is outlined briefly in the following. First, 
we select events that have the same signature 
as \ttbar ~signal events decaying in the lepton+jets channel, i.e., 
a truly isolated lepton and genuine \met from the $W$ boson decay. 
Multijet events produced by strong interactions are expected
to contain neither isolated leptons nor \met. However, they are present in the selected 
samples due to the imperfect reconstruction in the detector. 
In particular, the selected $\ejets$ sample contains 
contributions from multijet events in which
a jet is misidentified as an electron. Events where   
a muon originating from the semileptonic decay of a heavy quark 
appears isolated contribute to the selected sample in the $\mujets$ channel. 
Significant $\met$ can arise from fluctuations and mismeasurements of the 
muon and jet energies. 
In order to model these effects, we use a dedicated data sample to describe the
kinematic properties of the surviving multijet events. 

The background within the selected samples is dominated by $W$+jets events.    
Its contribution is estimated using Monte Carlo simulations.  
We validate the background model by
comparing observed distributions to the predictions from our model in
samples of events with low jet multiplicities where only a small
signal fraction is expected. For these comparisons we assume a
\ttbar production cross section of 7\,pb as predicted in the SM. 
The Kolmogorov-Smirnov probability for data and simulation to
originate from the same underlying distribution is used as an
estimator of the quality of the background model and generally good agreement is found.

To extract the fraction of \ttbar~events in the samples, we 
select kinematic variables which discriminate between the 
$W$+jets background and the \ttbar~signal, and combine them 
into a discriminant function. The selected variables are required 
to be well described by the background model.     

In a final step, we derive the discriminant function for the observed data, 
the \ttbar~signal and the electroweak and multijet backgrounds. 
A Poisson maximum-likelihood fit of the signal and background discriminant
distributions to that of the data yields the fraction of \ttbar~signal and the electroweak
and QCD multijet backgrounds in the data sample. Finally, the \ttbar~production 
cross section is computed from the number of fitted \ttbar~events.

In contrast to the \ttbar~cross section measurement presented in 
Ref.~\cite{btagPRD}, we do not take advantage of the fact that two
jets are expected to contain displaced vertices due to the 
$b$-quark decays for signal events. 
Our cross section estimation is based
solely on the different kinematic properties of the
signal and background events.

%% file: selection.tex

In both channels, we select events containing one lepton with $p_T>20\;\rm GeV$ 
that passes the tight identification criteria, originates from the primary vertex 
($|\Delta z(\ell,\mathrm{PV})|<1$\,cm), and is matched to trigger objects at 
all relevant levels. We accept muons with 
\mbox{${|\eta_\mathrm{det}|<2.0}$} and electrons with 
$|\eta_\mathrm{det}|<1.1$. This choice of cuts is motivated 
by the acceptance of the \dzero ~muon system and central calorimeter,
respectively. Jets in the event are required to have $|\eta|<2.5$ and  
$p_T > 20\;\rm GeV$ except for the highest $p_T$ jet which  
has to fulfill $p_T>40\;\rm GeV$.    
Events with a second isolated high transverse momentum lepton are studied 
elsewhere ~\cite{dileptonPRD} and explicitly vetoed in the event selection 
to retain orthogonality between analyses.

\begin{figure*}[htb]
\epsfig{ file=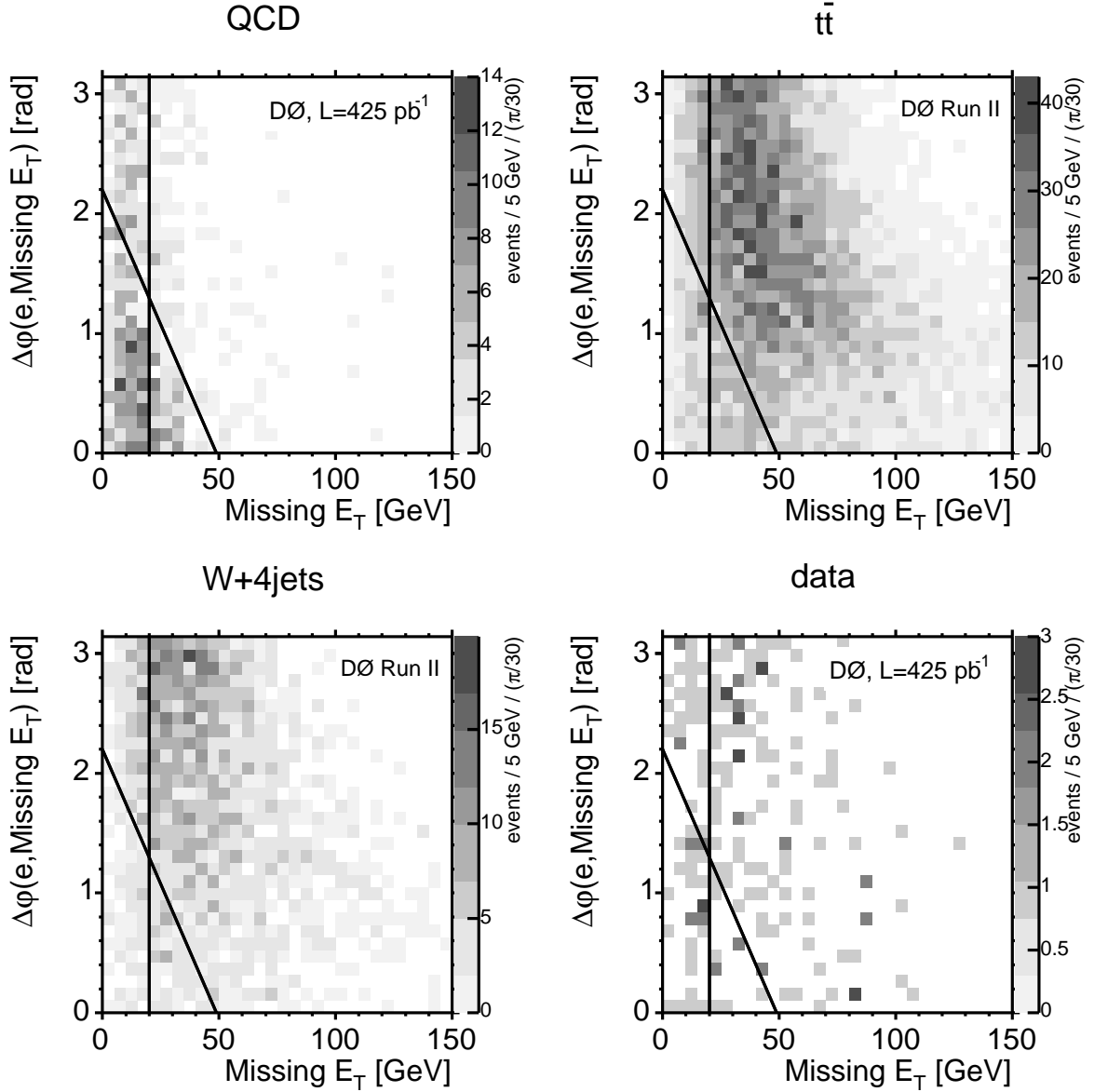, width=16cm}
	\caption{$\Delta\varphi(e, \met)$ versus \met in the multijet QCD data 
	sample, \ttbar Monte Carlo, $W$+jets Monte Carlo and in data. 
	The lines represent the cuts optimized for the fourth 
	inclusive jet multiplicity bin.}
	\label{fig:ejets_presel_dphi}
\end{figure*}

In both channels we require \met $>20$ GeV to reject multijet backgrounds. 
However, a significant fraction of multijet events survive this cut due to the
presence of heavy flavor decays or jet energy mismeasurement. These events
typically have \met either in the direction of the
lepton or back-to-back to it. Figure~\ref{fig:ejets_presel_dphi} illustrates the 
difference in the angular distribution of \met and 
the lepton $(\Delta\varphi(\ell,\met))$
between signal and multijet background events which we exploit 
to further suppress the latter. 
   
We performed a grid search in the $(\met,\Delta\varphi(\ell,\met))$
plane to find cuts that provide the highest product of efficiency and purity 
for \ttbar ~events, where purity is defined as the ratio of the number of signal
\ttbar ~events to the total number of events in the selected sample.  
The optimal cuts are found to be 
$\Delta\varphi(e,$\mbox{$\not\!\!E_T$}$) > 0.7 \pi - 0.045 {\not\!\!E_T} $
and 
$\Delta\varphi(\mu,\met) > 0.6 \pi  \left(1-\met / 
(50\,\mathrm{GeV})\right)$
in \ejets ~and \mujets ~channels respectively in addition to the common \met $>20$ GeV cut. 
  
The $\mu$+jets channel suffers from a significant contribution of
$Z(\mu\mu)$+jets events which pass the selection criteria due to poor 
\met\ resolution in events with four or more jets. A cut on the invariant
dimuon mass of the selected isolated high $p_T$ muon and the additional 
highest $p_T$ muon with relaxed quality requirements is applied at 
70\,GeV $< \it {M_{\mu\mu}} < $~110\,GeV and 
rejects roughly $27$\,\% of the $Z \to \mu\mu$+jet background 
while keeping almost $100$\,\% of the signal in the selected sample. The remaining
$Z\to\mu\mu$+jets background cannot be rejected since no second
muon is reconstructed mainly for reasons of finite acceptance.  
  
The \ttbar\ event selection efficiency is measured using simulated events 
with respect to all \ttbar\ final states 
that contain an electron or a muon originating either directly from 
a $W$ boson or 
indirectly from the $W\rightarrow\tau \nu$ decay. The branching fractions of such  
final states are 17.106\% and 17.036\% \cite{PDG2006} for the $\ejets$ 
and $\mujets$ channels, 
respectively. After applying the correction factors discussed in 
Sect.~\ref{sc:dataMCcor} and the trigger efficiency parameterizations   
(Sect.~\ref{sc:treff}) to the simulated \ttbar\ events, the final \ttbar selection efficiencies yield 
(9.17$\pm$0.09)\% and (9.18$\pm$0.10)\% in the 
$\ejets$ and $\mujets$ channel, respectively. The quoted uncertainties are  
statistical only.

%% file: bckg_nils.tex
\subsection{Multijet background evaluation}
The background within the selected samples is dominated by $W$+jets events, which have
the same final state signature as $\ttbar$ signal events. However, the samples also include
contributions from multijet events in which
a jet is misidentified as an electron ($\ejets$ channel) or in which
a muon originating from the semileptonic decay of a heavy quark
appears isolated ($\mujets$ channel). 
Significant $\met$ can arise from fluctuations and mismeasurements of the
jet energies and the muon momentum in addition to neutrinos originating from 
semi-muonic heavy quark decays.  
These instrumental backgrounds are collectively called ``multijet backgrounds'', and
their contribution is estimated directly from data since Monte Carlo simulations
do not describe them reliably. 

In order to estimate the contribution of the multijet background to the selected data samples 
we define two samples of events in each channel, a ``loose'' and a ``tight'' set where the latter is  
a subset of the former. The loose set (containing $N_{\ell}$ events) corresponds to the 
selected sample described in the previous paragraph, but with only the loose 
lepton requirement applied.
The tight sample (containing $N_t$ events) additionally demands the selected lepton to pass 
the tight criteria and is identical to the selected sample.
The loose sample consists of $N^{s}$ events with a truly isolated lepton 
originating from $W$+jets, $Z$+jets or $\ttbar$ events
and $N^{b}$ multijet background events with a fake isolated lepton: $N_{\ell}=N^{s}+N^{b}$. 
The tight sample consists of  $\varepsilon_{s} N^{s}$ \ttbar\ signal and electroweak background events and
$\varepsilon_{b} N^{b}$ multijet background events, where
$\varepsilon_{s}$ and $\varepsilon_{b}$ are the efficiencies for a loose lepton to also 
fulfill the tight lepton requirements.

Solving the system of linear equations for $N^{b}$ and $N^{s}$ yields:
\begin{eqnarray} 
N^{s} = \frac{N_t-\varepsilon_{b}
N_{\ell}}{\varepsilon_{s} -\varepsilon_{b}} \mbox{\ \ \ and\ \ \ }
N^{b} = \frac{ \varepsilon_{s} N_{\ell}-N_t}{\varepsilon_{s} -\varepsilon_{b}},
\label{eq:matrix2}
\end{eqnarray}
and allows the determination of the size of the multijet background contribution 
in the selected sample. As for the   
\emph{shape} of the multijet background, for a given 
variable it is predicted using a
data sample where the full selection has been applied except for the tight
lepton requirement. Instead, the requirements on the muon
isolation in the $\mu$+jets channel and electron likelihood in the $e$+jets channel 
are inverted, selecting a data sample enriched in events
originating from multijet production processes (``loose$-$tight'' data sample). 
However, truly isolated
leptons from \ttbar and $W/Z$+jets events will leak into this sample. 
The composition of the ``loose$-$tight'' ($N_{\ell -t} = N_\ell - N_t$) 
data sample can be derived from Eq.~\ref{eq:matrix2}: 
\begin{eqnarray} 
N_{\ell -t} = 
\frac{1-\varepsilon_{s}}{\varepsilon_{s}}N_t^{s} + 
N_{\ell -t}^{b}\,,
\label{eq:QCDcor}
\end{eqnarray}
where $N_t^{s} = \varepsilon_{s} N^{s}$
is the number of preselected \ttbar and electroweak background events
as estimated in the following section and $N_{\ell -t}^{b}$ is the
pure multijet contribution to the ``loose$-$tight'' preselected sample.
Using Eq.~\ref{eq:QCDcor}, the contaminations from \ttbar and electroweak
backgrounds are subtracted bin-by-bin from the distribution of the
``loose$-$tight'' preselected data sample in order to predict the shape
of the pure multijet contribution for each individual discriminant input variable under
consideration.

\subsubsection{$\varepsilon_{b}$ determination}
\label{sec:epsQCD}

\begin{figure}
\begin{center}
\vspace{-0.1cm}
\includegraphics[width=0.44\textwidth,clip=]{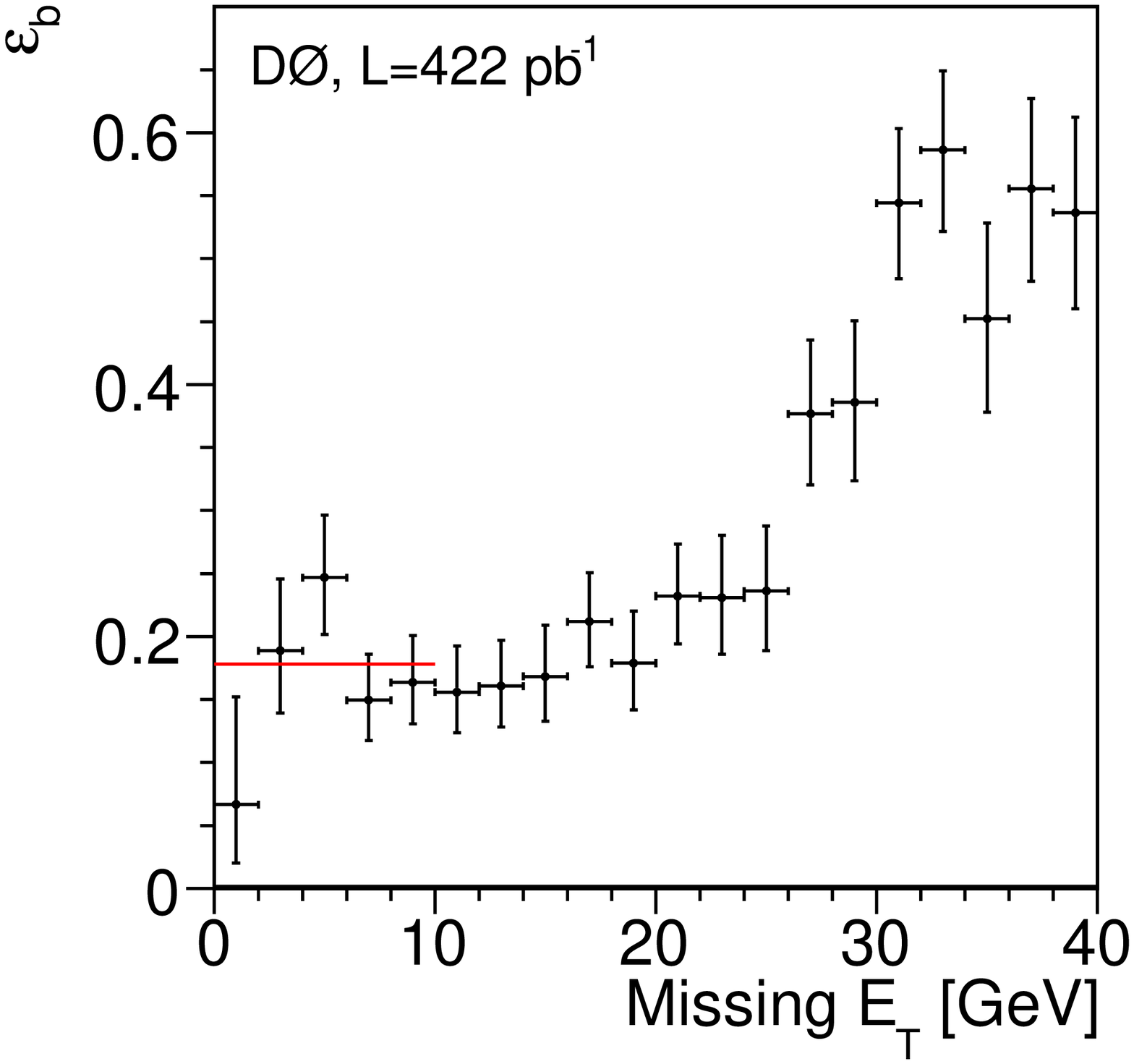}
\vspace{-0.2cm}
\includegraphics[width=0.44\textwidth,clip=]{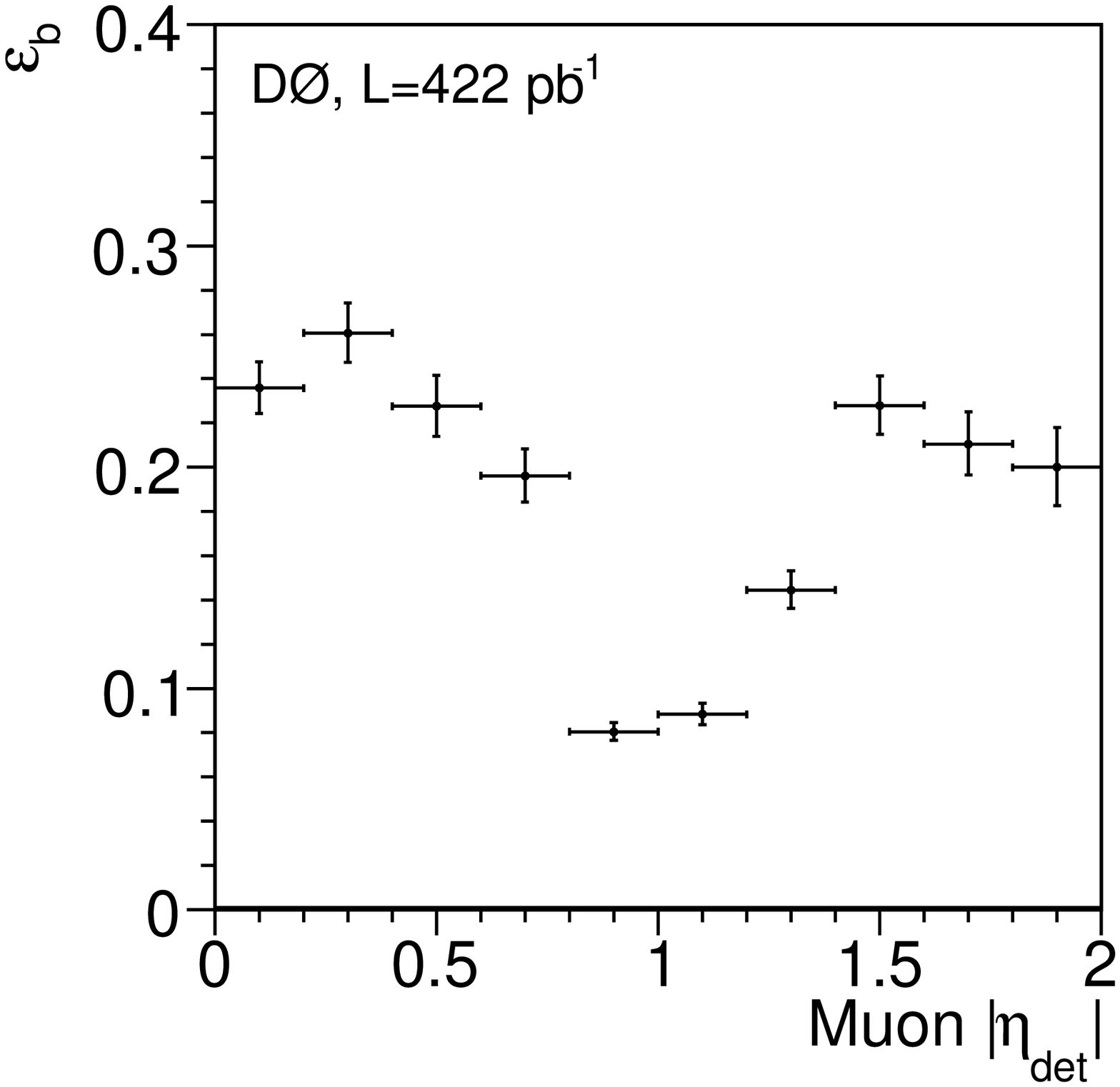}
\hspace{0.5cm}
\vspace{-0.1cm}
\includegraphics[width=0.44\textwidth,clip=]{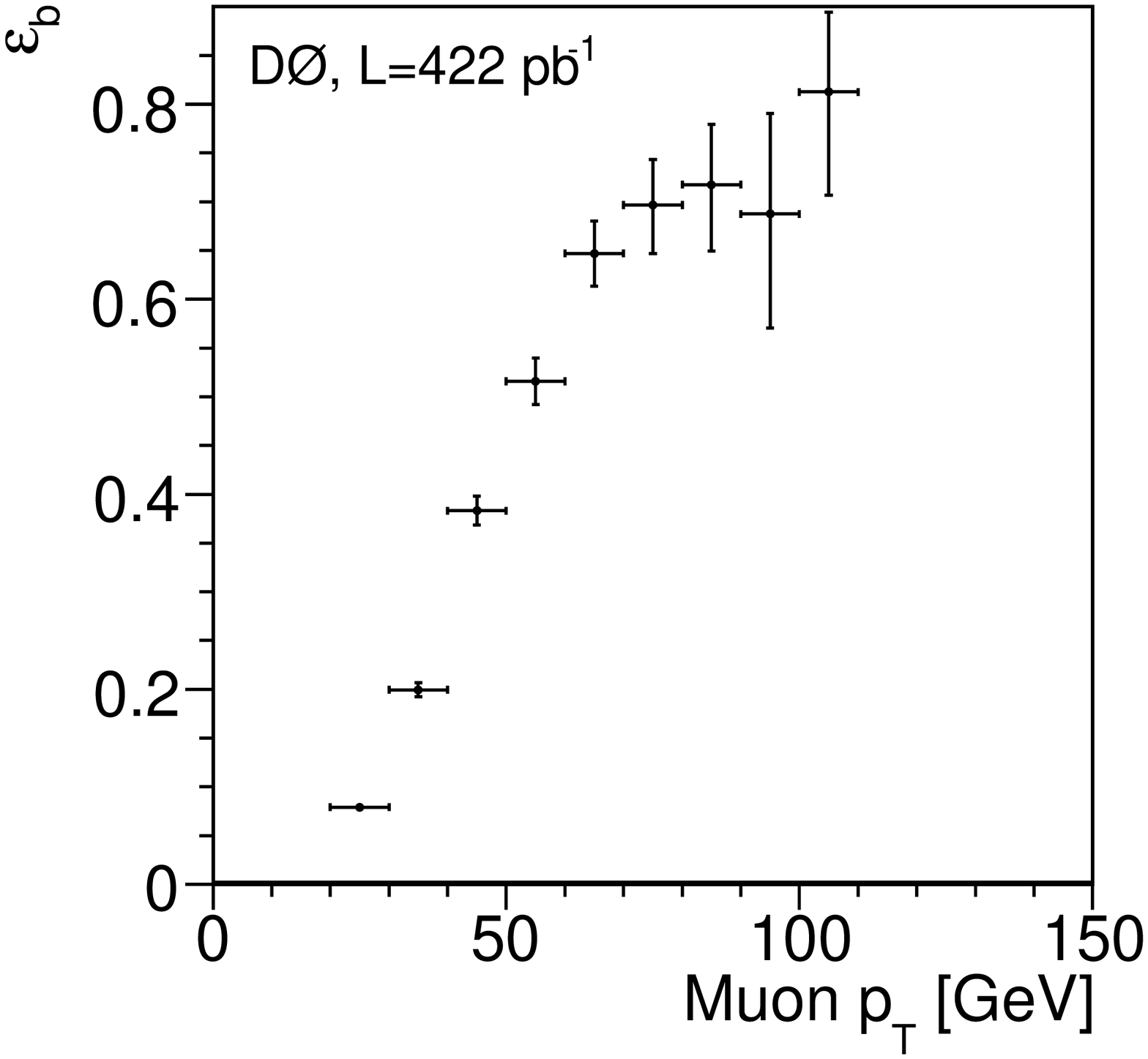}
\end{center}
\vspace{-0.2cm}
\caption{Tight muon isolation efficiency $\varepsilon_{b}$ measured in the 
QCD multijet background dominated data sample as a function of
\met (top), the muon $\eta_{\rm det}$ (middle) and $p_T$
(bottom) in $\mu$+jets channel.}
\label{fig:eQCDmet}
\end{figure}

The rate $\varepsilon_{b}$ at which a lepton in multijet events appears isolated
is measured in a data
sample which passes the same requirements as the selected one but
without applying the \met-related set of cuts discussed in
Sect.~\ref{sec:select}.  
In this data sample, we calculate $\varepsilon_{b}$, the ratio of the number 
of tight events to the number of loose events, as a function of \met. We  
find that it is constant for \met~$<$~10\,GeV, shown in
Fig.~\ref{fig:eQCDmet} (top) for the $\mu$+jets channel, as expected for a 
sample dominated by the multijet events. The value of $\varepsilon_{b}$
given by the constant fit to data in $\met < 10\,\mathrm{GeV}$ region 
is used in the analysis. 

In the muon channel, $\varepsilon_{b}$ does not show significant dependence on 
the jet multiplicity and does not change between different data-taking periods. 
However, rather strong dependences are observed with respect to the muon 
$\eta_\mathrm{det}$ (Fig.~\ref{fig:eQCDmet} middle) and transverse momentum
(Fig.~\ref{fig:eQCDmet} bottom). 
We estimated the effect of these dependences on
the inclusive $\varepsilon_{b}$ by folding them in with the muon 
$\eta_{\rm det}$ and $p_T$ spectra of the selected sample. 
Since the small number of events with four or more reconstructed jets in 
the low \met ~sample does not allow for a precise measurement we determine 
$\varepsilon_{b}$ from the events with three or more jets and assign
systematic uncertainty from the difference between the flat and the muon 
$p_T$-folded measurement:
\begin{equation} 
\varepsilon_{b} = 17.8 \pm 2.0 \;(\mathrm{stat})\ \pm 3.1 {\rm\;(syst)}\%\,.
\end{equation}

\begin{figure*}[htb]
\hspace{-2.0cm}
\leftline{
\epsfig{ file=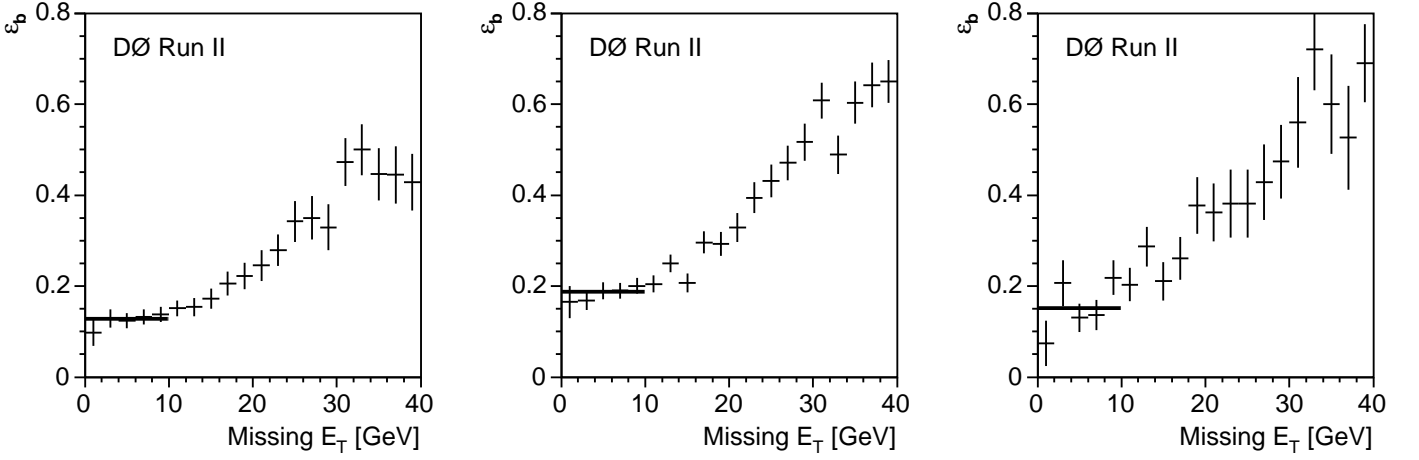, width=19.0cm}
}
\caption{Electron likelihood efficiency in the 
QCD multijet background dominated data sample as a function of \met for 
events with two or more jets, for the first (left), 
second (middle) and third (right) data-taking periods. 
The constant fit to the region of $\met<10\;\rm GeV$ is used to determine 
the value of $\varepsilon_{b}$ used in the analysis.}
\label{fig:eqcd}
\end{figure*}

In the electron channel we find no significant 
dependence of $\varepsilon_{b}$ on the jet multiplicity and electron 
$\eta_{\rm det}$ and $p_T$. However, we observe a statistically 
significant variation of $\varepsilon_{b}$ between different data-taking 
periods. 
In particular, we find a higher value of $\varepsilon_{b}$ for events 
collected during the second data-taking period than during the first one. 
We attribute this increase to the more stringent electron shower shape 
requirements applied at trigger level 3, which improves the 
quality of the fake electrons that enter 
our loose sample, making them more likely to pass the tight criterion. 
Figure~\ref{fig:eqcd} shows the electron isolation efficiency as a function 
of \met ~for events with two or more jets, obtained separately for data  
collected during three data-taking periods. 
A fit to these distributions in the region of $\met < 10\;\rm GeV $ yields 
corresponding $\varepsilon_{b}$. 
In Eq.~\ref{eq:QCDcor} 
we use a luminosity-weighted average $\varepsilon_{b}$ obtained by 
analyzing events with two or more jets:
\begin{equation} 
\varepsilon_{b} = 16.0 \pm 1.2 \;(\mathrm{stat})\ \pm 8.0 {\rm\;(syst)}\%\,.
\end{equation}
The systematic uncertainty of $\varepsilon_{b}$ arises from the small 
observed variation as a function of jet multiplicity and electron $p_T$.   

\subsubsection{$\varepsilon_{\mathrm{s}}$ determination}
\label{sec:epsSIG}
The probability $\varepsilon_{s}$ that a truly isolated lepton (i.e., a lepton originating from 
$W$ boson decays) from a loose sample will survive the tight isolation 
requirements is measured using simulated $W$+jets events with four or more jets 
and corrected with the simulation-to-data scale factor independent of jet 
multiplicity.
  
\begin{figure}
\hspace{-1.0cm}
\leftline{
\includegraphics[width=0.5\textwidth,clip=]{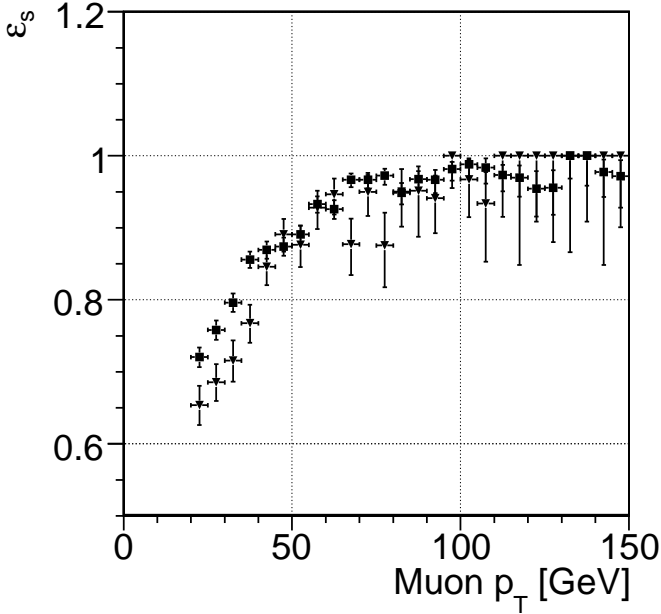}}
\caption{$\varepsilon_{\mathrm{s}}$ as a function of muon $p_{T}$
for simulated $W$+jets~(triangles) and $t\bar{t}$~(squares) events.}
\label{fig:e_sig_muPt}
\end{figure}

In the muon channel, $\varepsilon_{s}$ depends on the muon $p_T$
spectrum, shown in Fig.~\ref{fig:e_sig_muPt}, and hence is slightly
different for $W$+jets and $t\bar{t}$ events. In the signal jet
multiplicity bin (N$_\mathrm{jet} \geq 4$), we add the fraction of
$t\bar{t}$ events corresponding to the expected \ttbar\ cross section of 
7 pb and obtain:
\begin{equation} 
\varepsilon_{\mathrm{s}} = 81.8 \pm 0.7 {\rm\;(stat)}\ ^{+3.3}_{-2.2} {\rm\;(syst)}\%\,,
\end{equation}
where the systematic uncertainty is derived by varying the \ttbar
fraction between 0 and 100\,\%. 
In the electron channel, $\varepsilon_{s}$ determined from the simulated \ttbar
~events agrees with the one obtained from $W$+$\ge$4 jets events and yields: 
\begin{equation} 
\varepsilon_{\mathrm{s}} = 82.0 \pm 0.7 {\rm\;(stat)}\ \pm 1.3
{\rm\;(syst)}\%\,.
\end{equation}
Systematic uncertainty arises from the uncertainty on the simulation-to-data
correction factor.  

\subsection{Expected sample composition}
\label{sec:MMresult}
Equation~\ref{eq:matrix2} is applied separately to events selected in bins of
jet multiplicity
for both the \ejets ~and \mujets ~channels. The yields of 
multijet events $N_t^{b}$ and events with a real isolated lepton $N_t^{s}$ in 
the selected sample with four or more jets are summarized in Table~\ref{MM_eff}.
\begin{table}[b]
\begin{center}
\begin{tabular}{lcccccc}
\hline\hline
channel & $\varepsilon_{s}$(\%) & $\varepsilon_{b}$(\%) & $N_{\ell}$ & $N_{t}$
&$N_{t}^{b}$ & $N_{t}^{s}$ \\\hline
$\mu$+jets & $81.8{\pm} 0.7$ & $17.8{\pm} 2.0$ & $160$ & $100$ & $ 8.6{\pm} 2.0$ &
$91.4{\pm}10.7$ \\
$e$+jets   & $82.0{\pm} 0.7$ & $16.0{\pm} 1.2$ & $242$ & $119$ & $ 19.2{\pm} 2.3$ &
$99.8{\pm}11.6$ \\
\hline\hline
\end{tabular}
\caption{Selected sample composition determined using Eq.\ref{eq:matrix2}. 
Only the statistical uncertainties are quoted where appropriate.}
\label{MM_eff}
\end{center}
\end{table}
Several physics processes contribute to signal-like events $N_t^{s}$ in the
selected sample: \ttbar pair
production decaying into the $\ell$+jets final state, \ttbar pair production 
decaying into two leptons and jets, $t\bar t\to \ell\ell'\nu_{\ell}\nu_{\ell'}b\bar b$, 
where both $W$ bosons decay leptonically, and electroweak background with  
contributions both from $W$+jets and $Z$+jets events. 

We estimate the amount of $Z$+jets background relative to the $W$+jets 
background using the cross sections, branching fractions and  
selection efficiencies determined using simulated events for both processes:
\begin{eqnarray} \label{Eq:Z}
N_{Z+\mathrm{jets}} &=& \frac{\sigma_{Z+\mathrm{jets}}  \cdot   
{\cal B}_{Z\to\mu\mu}}{\sigma_{W+\mathrm{jets}}          \cdot
{\cal B}_{W\to\mu\nu}}                                   \times 
\frac{\varepsilon_{Z+\mathrm{jets}}}{\varepsilon_{W+\mathrm{jets}}}\times
N_{W+\mathrm{jets}}\,. 
\nonumber
\end{eqnarray} 
In the \mujets ~channel, the ratio of the $Z$+jets contribution to the total electroweak 
background in the selected sample is measured to be 7\%. In the \ejets ~channel, 
the $Z$+jets background is negligible. 

The expected contribution from the \ttbar dilepton channel  
is evaluated assuming a standard model cross section of $7$\,pb for \ttbar 
pair production. The fully corrected efficiencies to select \ttbar ~dilepton
events are found to be 0.6\% and 0.5\% in the $\ejets$ and $\mujets$ channel, 
respectively. This results in a 2.0\% (2.3\%) contribution of dilepton events
into the \mujets ~(\ejets) final state.

%% file: topo_nils.tex
The background within the selected samples is dominated by $W$+jets events, 
which have the same final state as $\ttbar$ signal events. To extract the 
fraction of \ttbar ~events in the sample we construct a 
discriminant function that exploits the differences between the kinematic 
properties of the two classes of events: \ttbar ~signal and $W+$jets 
background. All other backgrounds are small and do 
not justify the introduction of an additional event class.

\subsection{Discriminant function}
The discriminant function is built using the method described in 
Ref.~\cite{topmass}, and 
has the following general form:
\begin{eqnarray}
\label{eq:discr0}
{\mathcal D} &=& \frac{S(x_1,x_2,...)}{S(x_1,x_2,...) +
B(x_1,x_2,...)}\;,
\end{eqnarray}
where $x_1,x_2,...$ is a set of input variables and 
$S(x_1,x_2,...)$ and $B(x_1,x_2,...)$ 
are the probability density functions (pdf) for observing a particular set of values 
$(x_1,x_2,...,x_N)$ assuming that the event belongs to the signal or  
background, respectively. 
Neglecting correlations between the input variables, the 
discriminant function can be approximated by:  
\begin{eqnarray}
\label{eq:discr1}
{\mathcal D} &=& \frac{\prod_i s_i(x_i)/b_i(x_i)}{\prod_i s_i(x_i)/b_i(x_i) + 1} \;,
\end{eqnarray}
where $s_i(x_i)$ and $b_i(x_i)$ are the normalized pdf's of each individual variable $i$ 
for \ttbar~signal and $W$+jets background, respectively. 
In the analysis, we express the discriminant as
\begin{eqnarray}
\label{eq:LHdiscr}
{\mathcal D} &=&  \frac{\exp\left(\sum_i 
{(\ln
\frac{s_i(x_i)}{b_i(x_i)})
} 
\right)
}{\exp\left(\sum_i 
{(\ln
\frac{s_i(x_i)}{b_i(x_i)})
}
\right) + 1  
}\\ \nonumber
&=& 
\frac{\exp\left(\sum_i 
{(\ln
\frac{s(x_i)}{b(x_i)})_\mathrm{fitted}^{i}
} 
\right)
}{\exp\left(\sum_i 
{(\ln
\frac{s(x_i)}{b(x_i)})_\mathrm{fitted}^{i}
}
\right) + 1  
}\;.
\end{eqnarray}   
where
$(\ln\frac{s}{b})_\mathrm{fitted}^{i}$ is a fit to the logarithm of the 
ratio of the signal and background pdfs for each kinematic variable $i$.
The application of a fit to the logarithm of the signal to background pdf ratios 
reduces the influence of individual events on the discriminant output.

\subsection{\label{sec:variables}Selection of discriminating variables}
All possible observables with different pdfs for $W$+jets and
\ttbar events have the ability to discriminate between the two. 
As a first step toward the goal of selecting an optimal set of discriminating 
input variables, we 
first evaluate the separation power for a large set of individual variables by estimating 
the expected total uncertainty of the \ttbar cross section when using the 
variable under consideration as sole discriminator. Variables are then 
ranked and selected by increasing
uncertainty. The total expected uncertainty is estimated by adding the
systematic uncertainties related to jet energy scale (JES), jet energy
resolution (JER) and jet reconstruction efficiency (JID) in quadrature to the
statistical uncertainty:
\begin{equation}
\sigma_{\mathrm{tot}} = \sqrt{
	  \left(\frac{\sigma_{\mathrm{stat}}}{\sqrt{2}}\right)^2 +
		  \sigma^2_{\mathrm{JES}}		    +
		  \sigma^2_{\mathrm{JER}}		    +
	  \sigma^2_{\mathrm{JID}}}.		    
\label{eq:fom}
\end{equation}
The optimization is done in \ejets ~and \mujets ~channels separately.
Therefore the statistical uncertainty in Eq.\ref{eq:fom} is reduced by a 
factor of $1/\sqrt{2}$, since the additional data from the complementary 
channel roughly doubles the statistics in the combination.
We select a set of thirteen variables described in Appendix~\ref{app:topovar} as input 
for the second step of the discriminant function optimization. 

\subsection{\label{sec:llhOpti} Optimization of the discriminant function}
The optimization procedure, determining which combination of
topological input variables will form the final 
discriminant function, is performed by estimating the expected
combined statistical and systematic uncertainty of the measured
\ttbar\ cross section.  The expected uncertainty is calculated for all
discrimination functions that can be constructed
from the selected input variables by using all possible subsets of the 13
variables in turn as input.
Pseudo-experiments are performed by drawing pseudo-data discriminant 
output distributions from the output discriminant distributions of
simulated events. The composition of such a pseudo-dataset is taken 
according to the expected sample composition for $N_\mathrm{jet}\geq 4$ 
and $\sigma_{t\bar{t}}=7$\,pb and allowing Poisson fluctuations. 
$3,000$ pseudo-experiments are built for each source of statistical or
systematic uncertainty and discriminant function under consideration.
We select the discriminant function which provides the smallest 
expected total uncertainty, including all sources of systematic 
uncertainties that affect the shape of the discriminant function.

In the \mujets ~channel the discriminant constructed with the following five 
input variables shows the best performance:
({\it i}\,) $H_{T}$, the scalar sum of the $p_T$ of the four leading jets; 
({\it ii}\,) $\sum\eta^2$, the sum of the squared pseudorapidities of the four
leading jets;
({\it iii}\,) $M_T$, transverse mass of the four leading jets;
({\it iv}\,) 
the event centrality ${\mathcal C}$, defined as the ratio 
of the scalar sum of the $p_T$ of the jets to the 
scalar sum of the energy of the jets;
({\it v}\,) the event aplanarity ${\mathcal A}$, constructed from the four-momenta of 
the lepton and the jets. Aplanarity characterizes the event shape and is 
defined, for example, in Ref.~\cite{tensor}. 

\begin{figure}
\begin{center}
\includegraphics[width=0.25\textwidth,clip=]{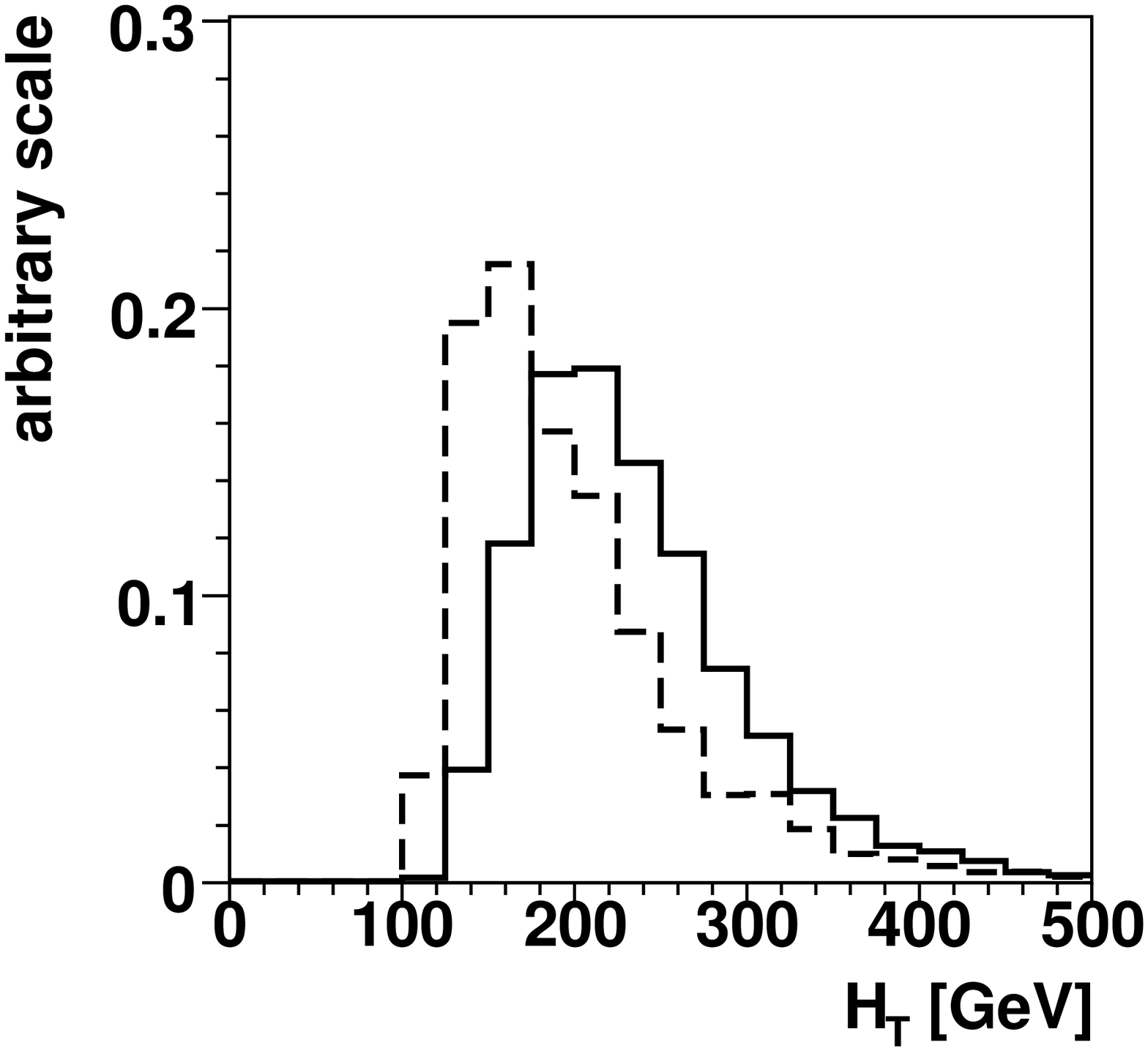} 
\includegraphics[width=0.25\textwidth,clip=]{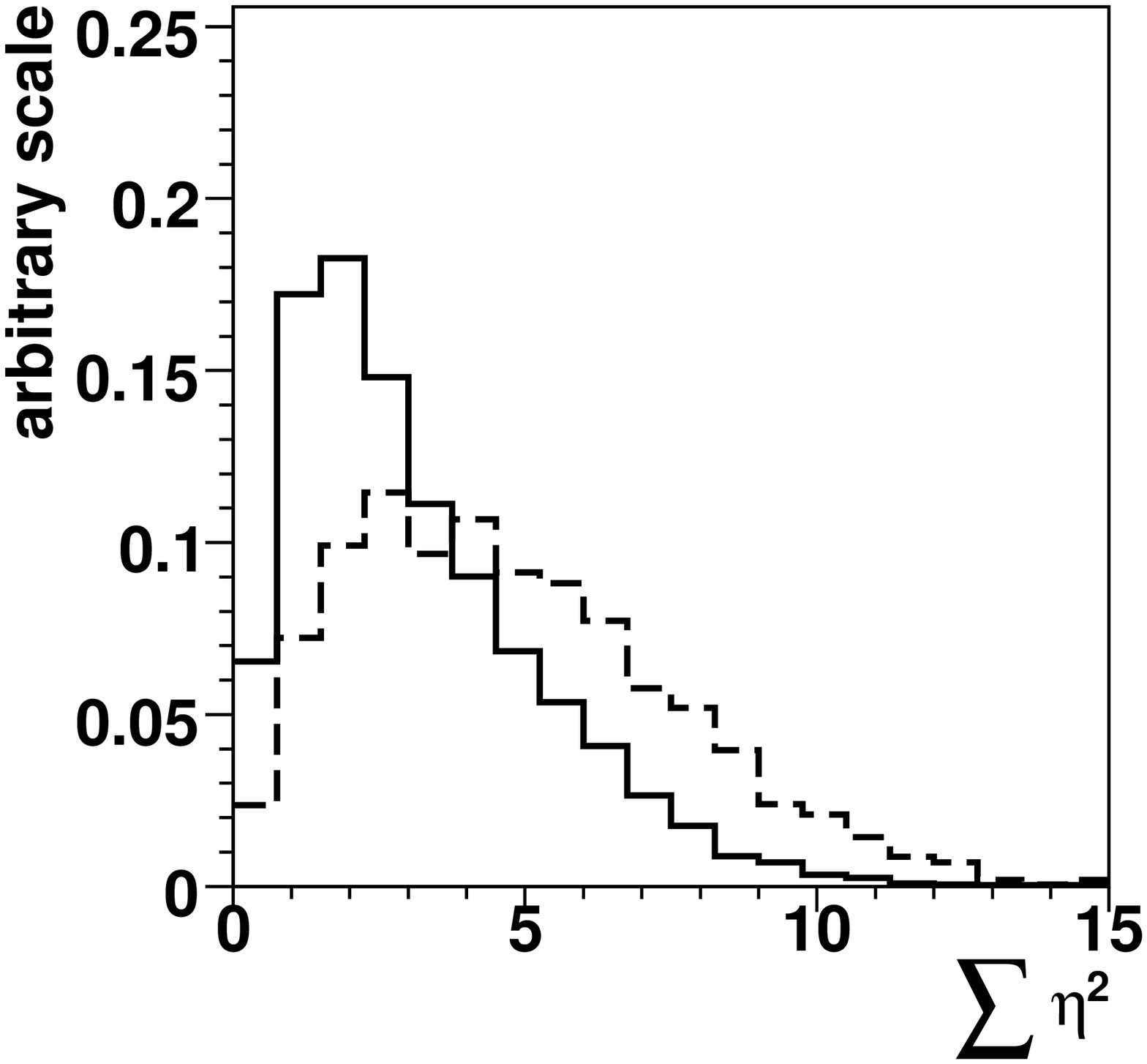} 
\includegraphics[width=0.25\textwidth,clip=]{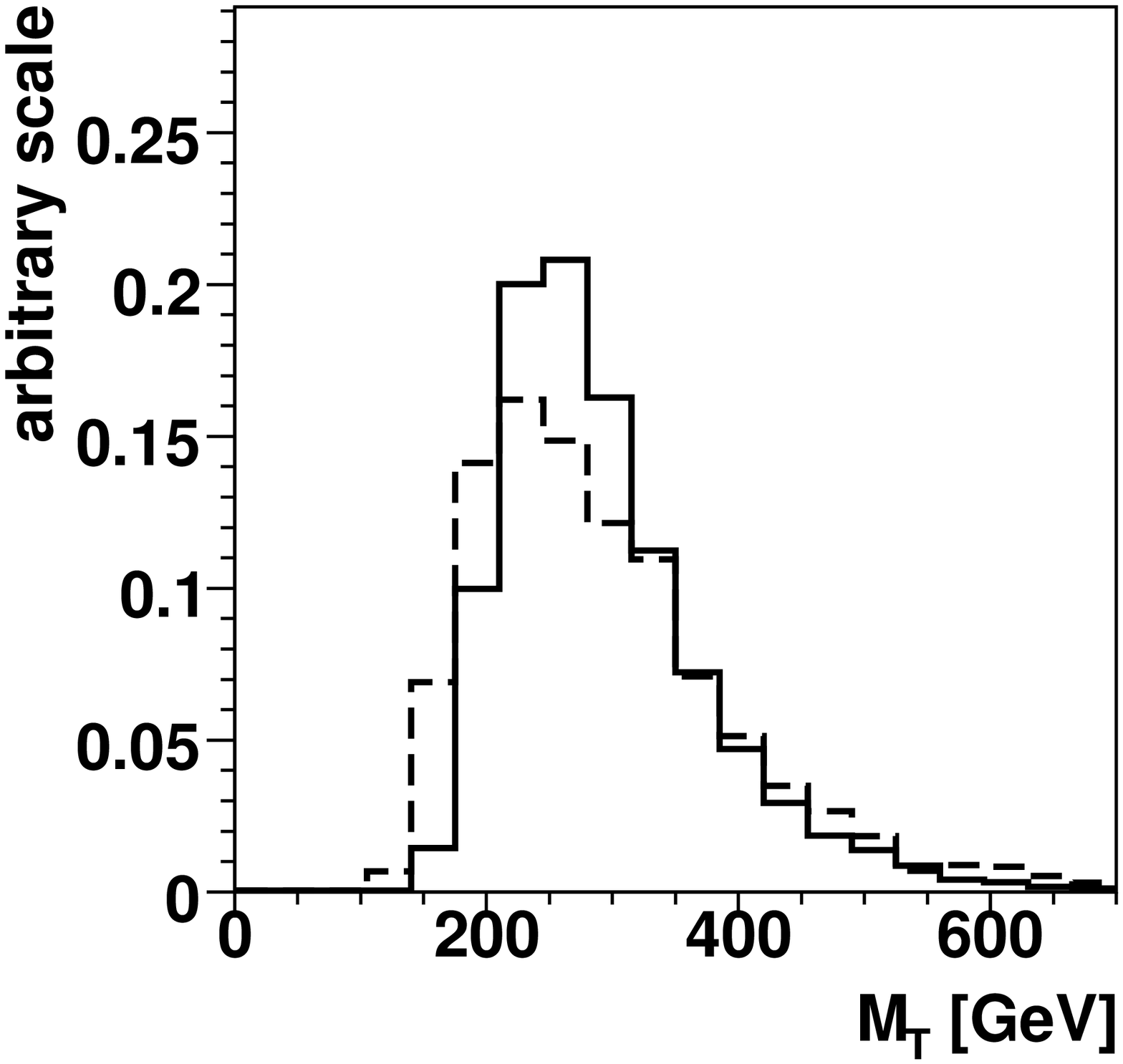}
\includegraphics[width=0.25\textwidth,clip=]{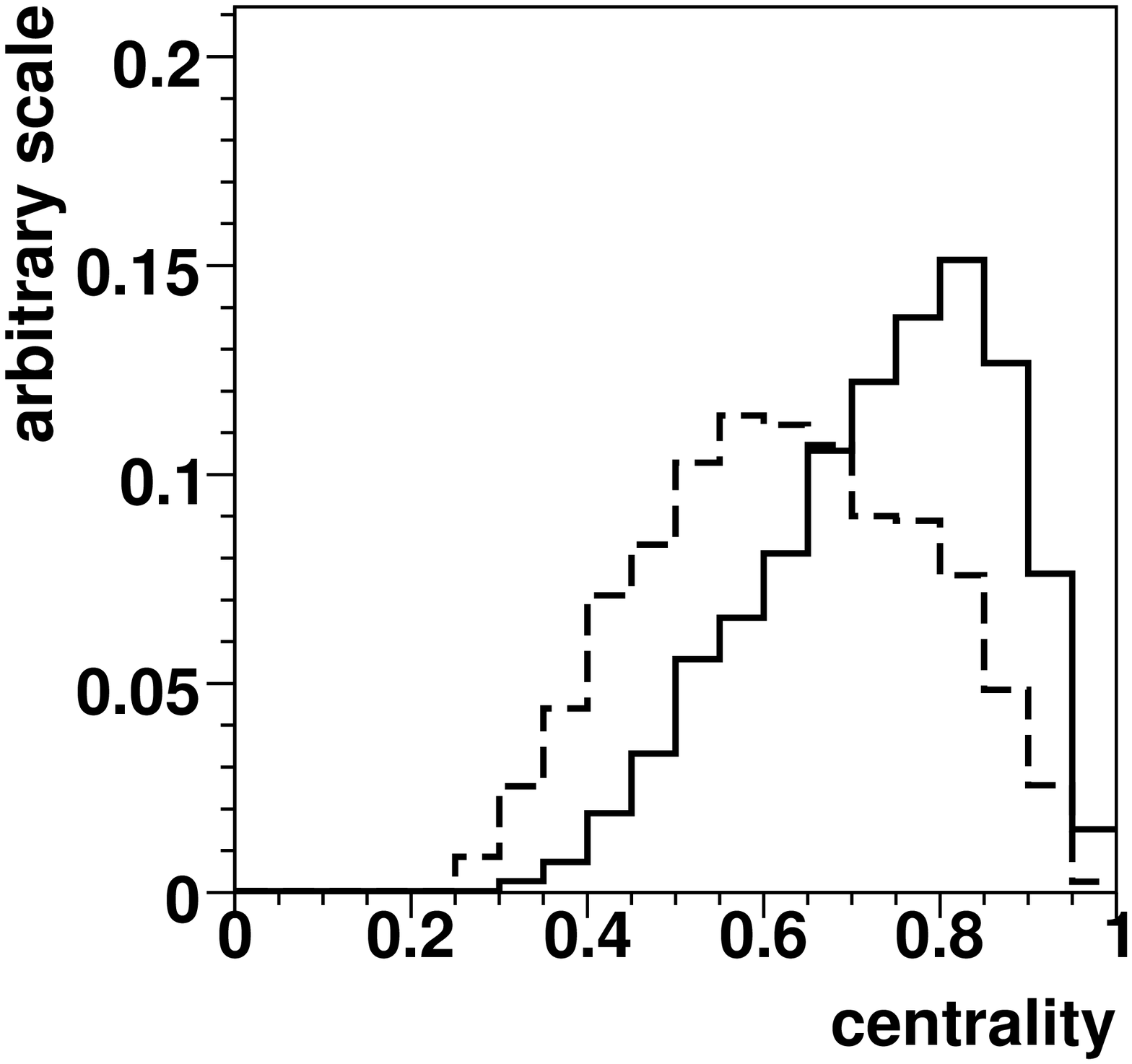} 
\includegraphics[width=0.25\textwidth,clip=]{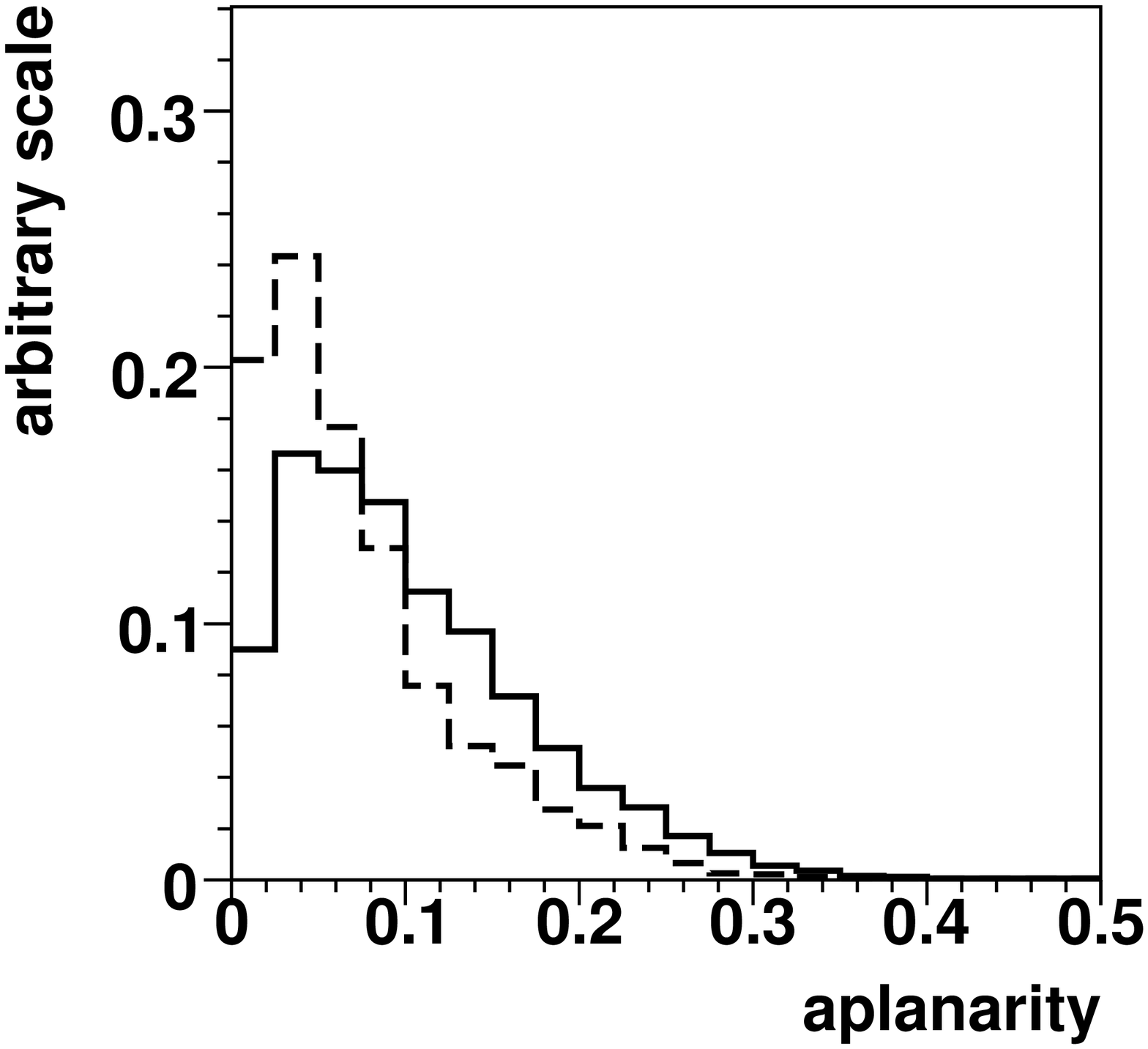} 
\end{center}
\caption{\label{fig:var_mujets}
Distributions of the five variables used as input to the likelihood
discriminant in the \mujets ~channel. 
The \ttbar signal
(solid line) and combined $W/Z$+jets electroweak backgrounds 
(dashed line) are derived from simulations.}
\end{figure}

\begin{figure}
\leftline{\includegraphics[width=0.51\textwidth,clip=]{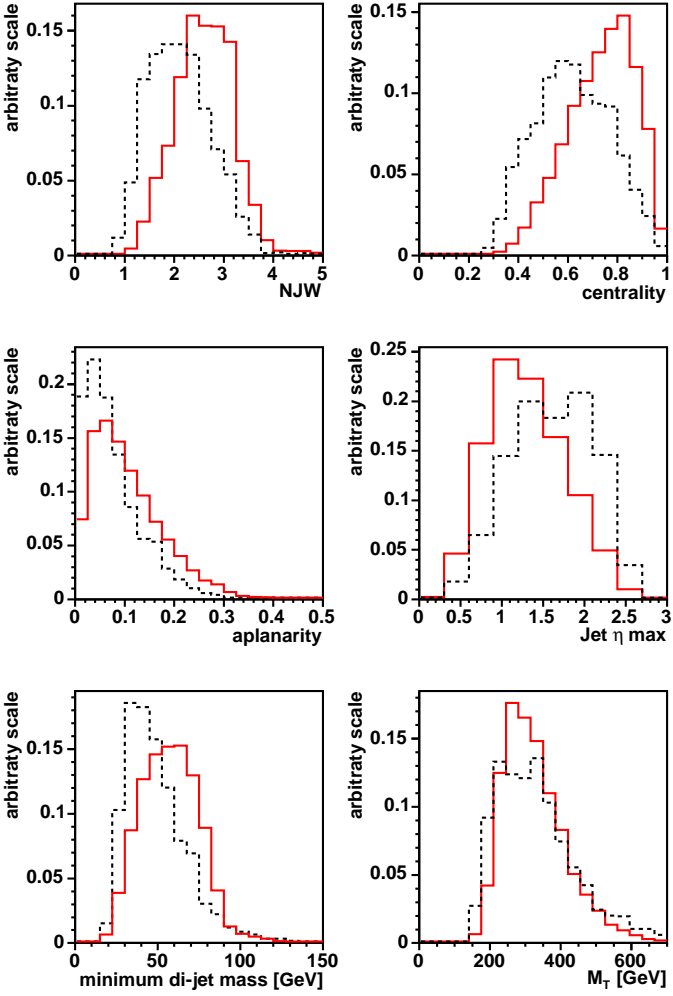}} 
\caption{\label{fig:var_ejets}
Distributions of the six variables used as input to the likelihood
discriminant in the \ejets ~channel. 
The \ttbar signal
(solid line) and $W$+jets background 
(dashed line) are derived from simulations.}
\end{figure}

In the \ejets ~channel the optimal discriminant function is found to be
built from six variables:
({\it i}\,) NJW, the weighted number of jets in the event; 
({\it ii}\,) the event centrality ${\mathcal C}$; 
({\it iii}\,) the event aplanarity ${\mathcal A}$; 
({\it iv}\,) $|\eta_\mathrm{jet}|^\mathrm{max}$, $|\eta|$ of the jet with maximum 
pseudorapidity;
({\it v}\,) minimum of the invariant mass of any two jets in the event; 
({\it vi}\,) $M_T$, transverse mass the four leading jets.
The normalized distributions of the selected kinematic variables for 
\ttbar ~signal and the $W$+jets background are shown in
Fig.~\ref{fig:var_mujets} and Fig.~\ref{fig:var_ejets} for \mujets ~and \ejets  
~channels, respectively. A more detailed explanation of the used variables is
given in Appendix \ref{app:topovar}. Figure~\ref{fig:topovar} demonstrates that 
distributions in data of the kinematic variables selected as input to the 
discriminant 
are well described by the sum of expected \ttbar signal, $W$+jets and
multijet background contributions for events with three jets dominated by the 
background.   

\begin{figure*}[htb]
\begin{center}
\includegraphics[width=0.325\textwidth,clip=]{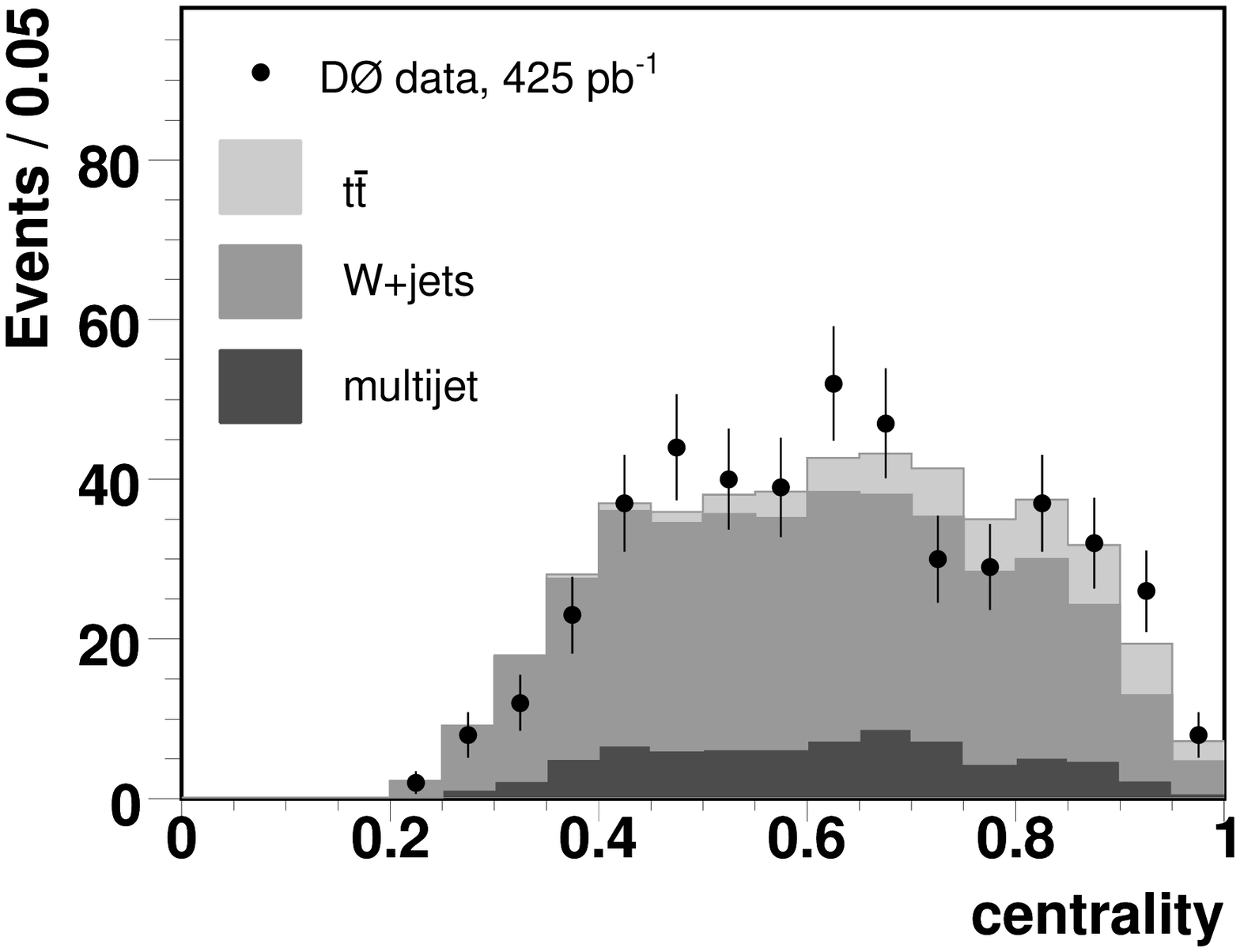} 
\includegraphics[width=0.325\textwidth,clip=]{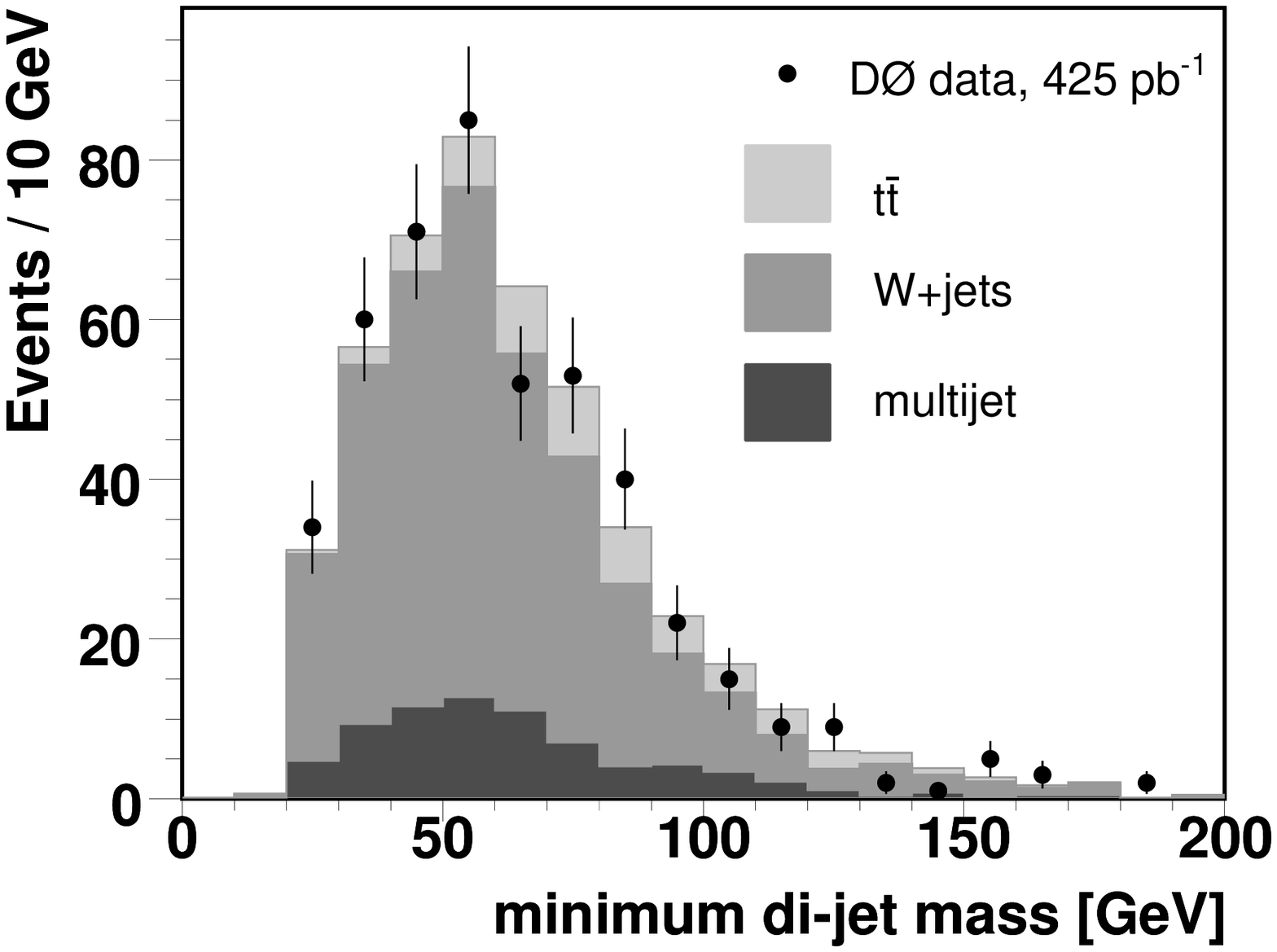} 
\includegraphics[width=0.325\textwidth,clip=]{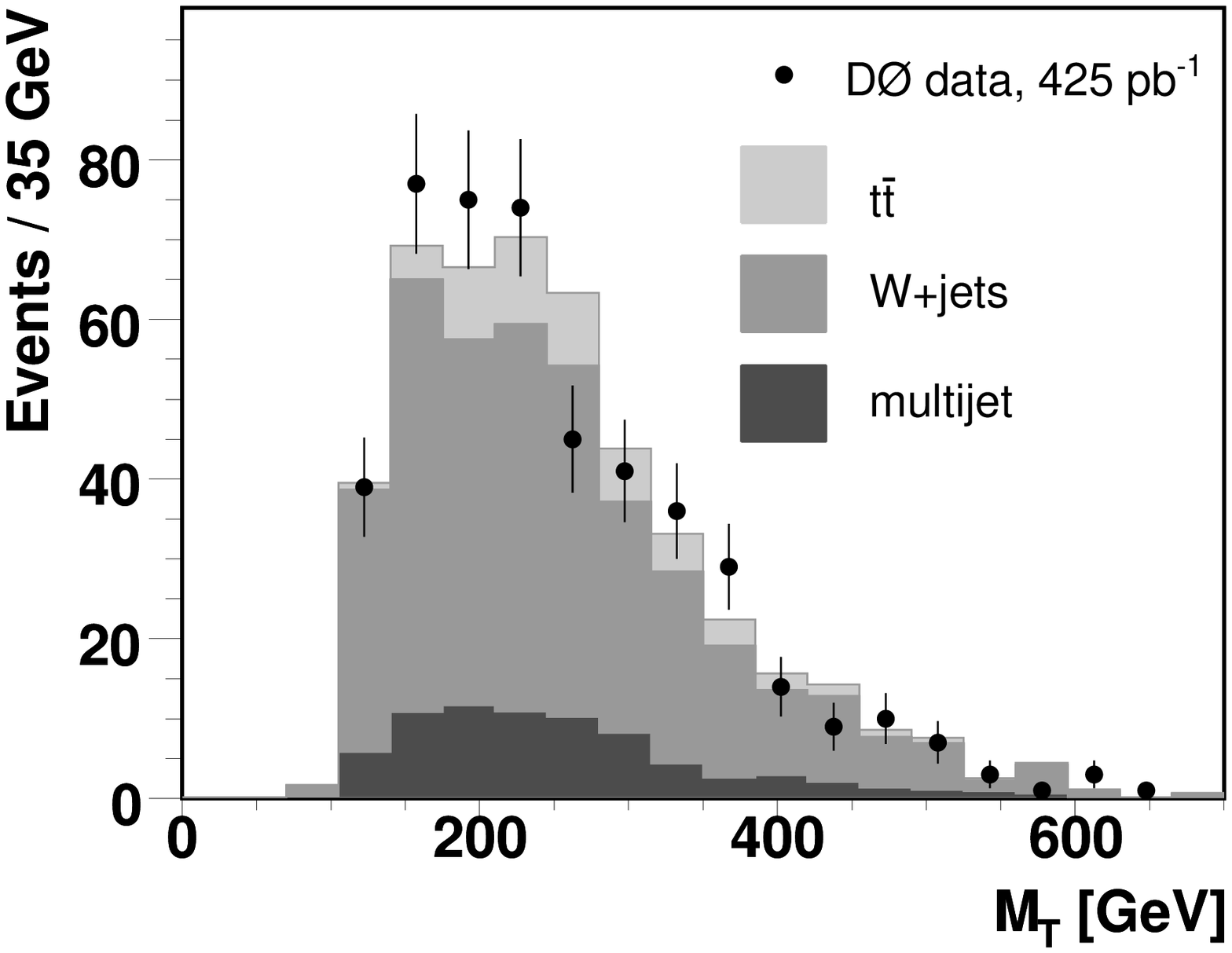} 
\includegraphics[width=0.325\textwidth,clip=]{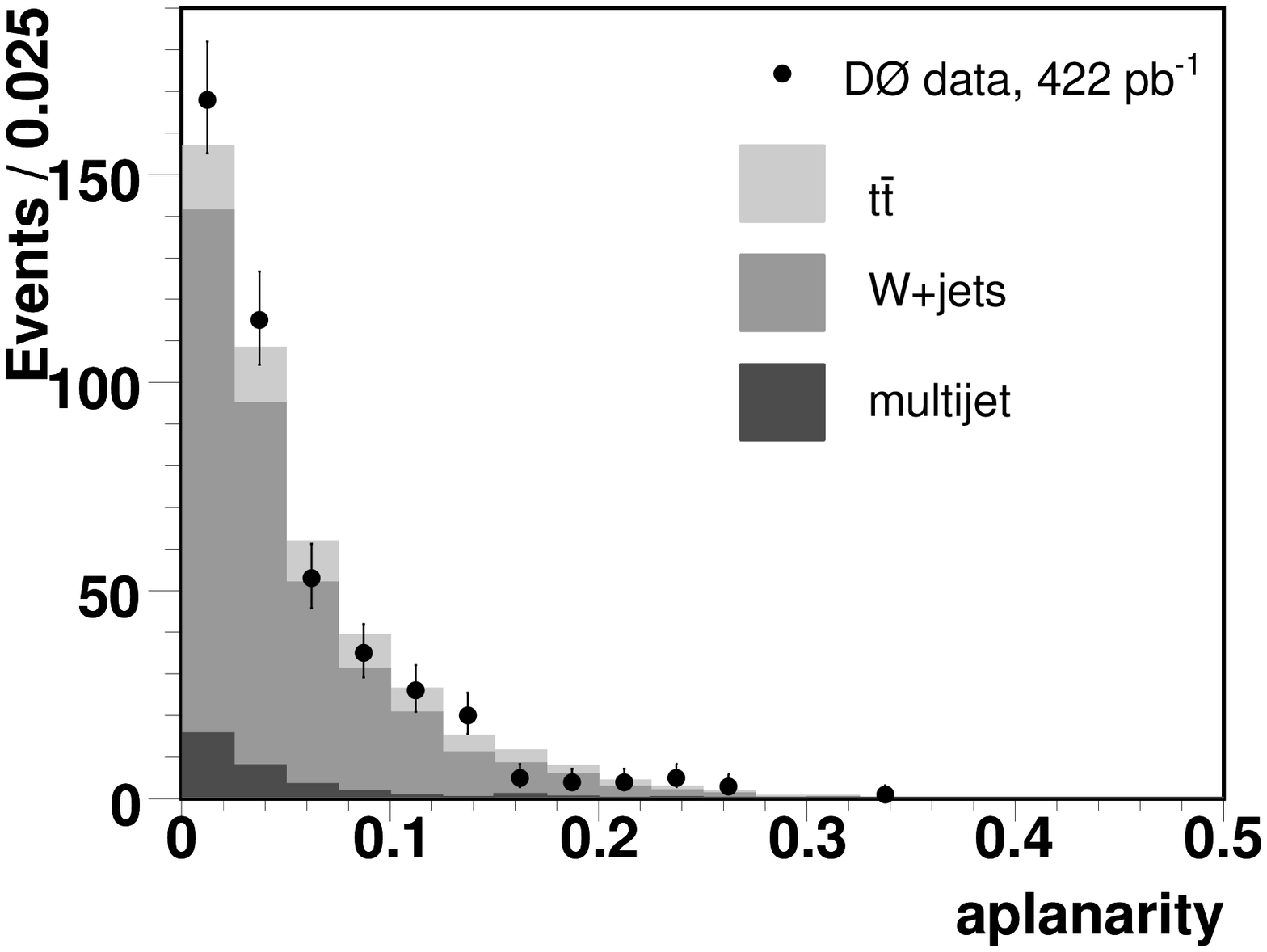} 
\includegraphics[width=0.325\textwidth,clip=]{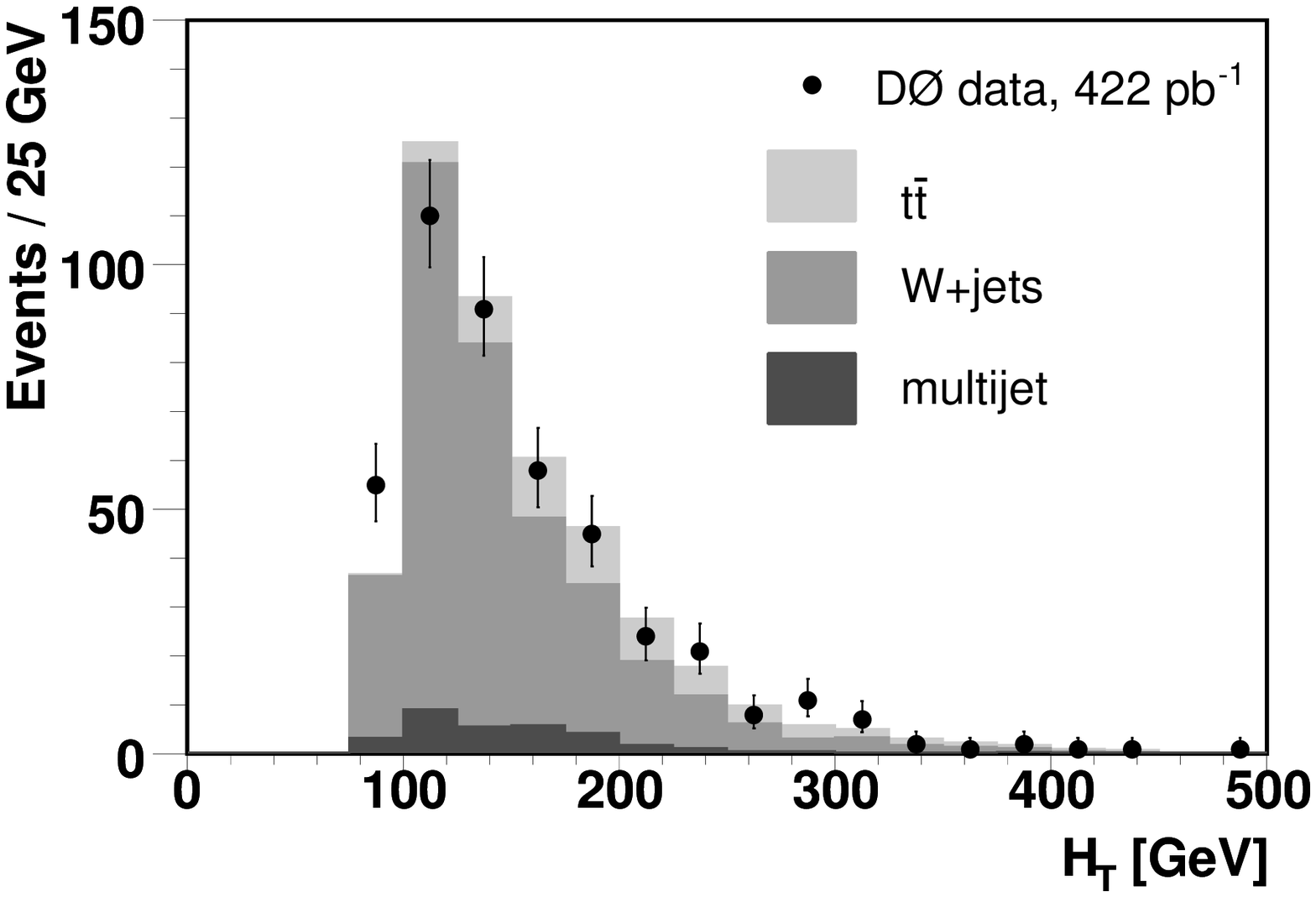} 
\includegraphics[width=0.325\textwidth,clip=]{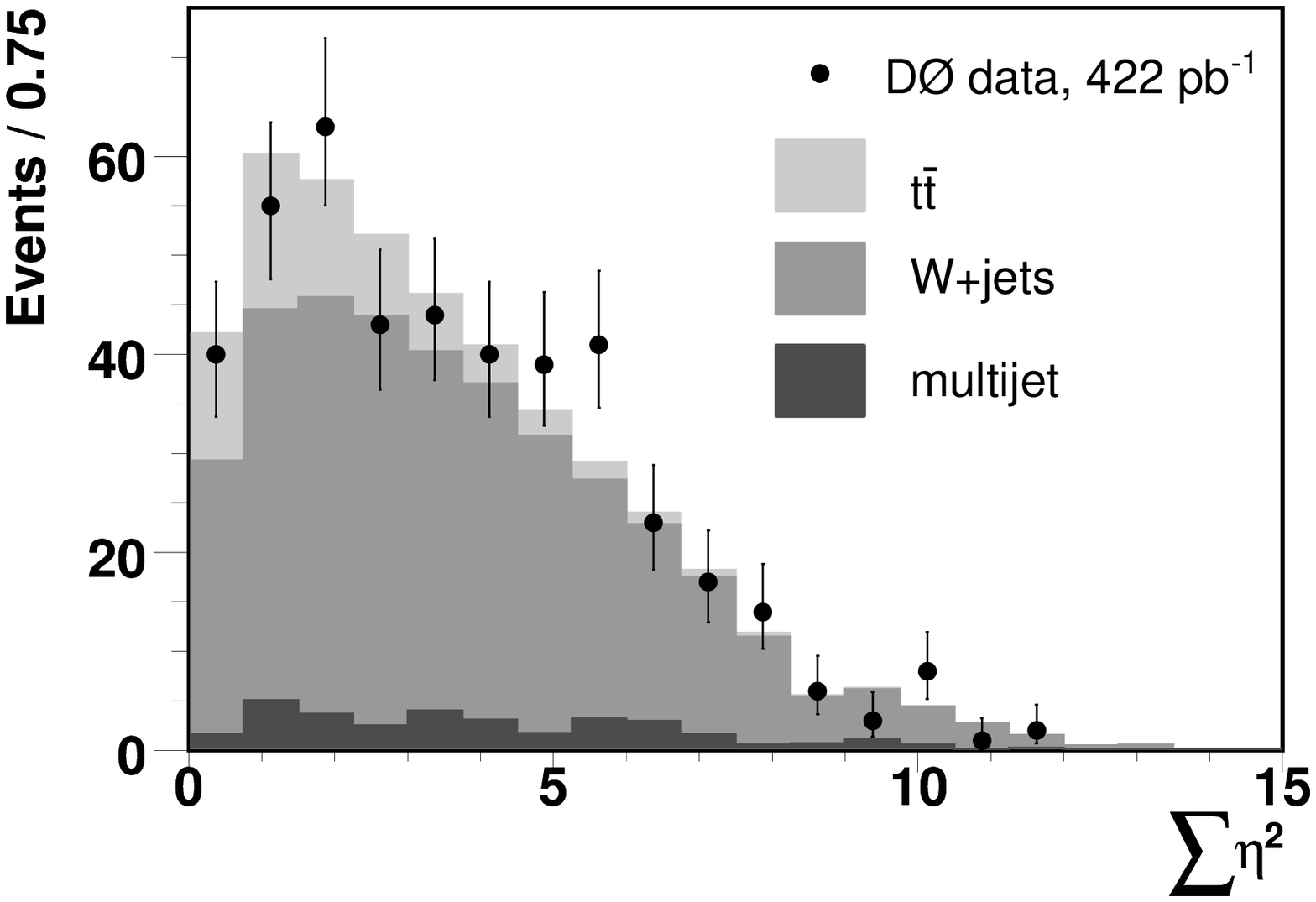}
\end{center}
\caption{\label{fig:topovar}
Distributions of selected variables used as input to the discriminant 
in data overlaid with the predicted background and expected \ttbar signal 
for the events with exactly three jets in the \ejets ~(top row) and 
\mujets ~(bottom row) channels.} 
\end{figure*}
   
\begin{figure}[htb]
\leftline{\includegraphics[width=0.5\textwidth,clip=]{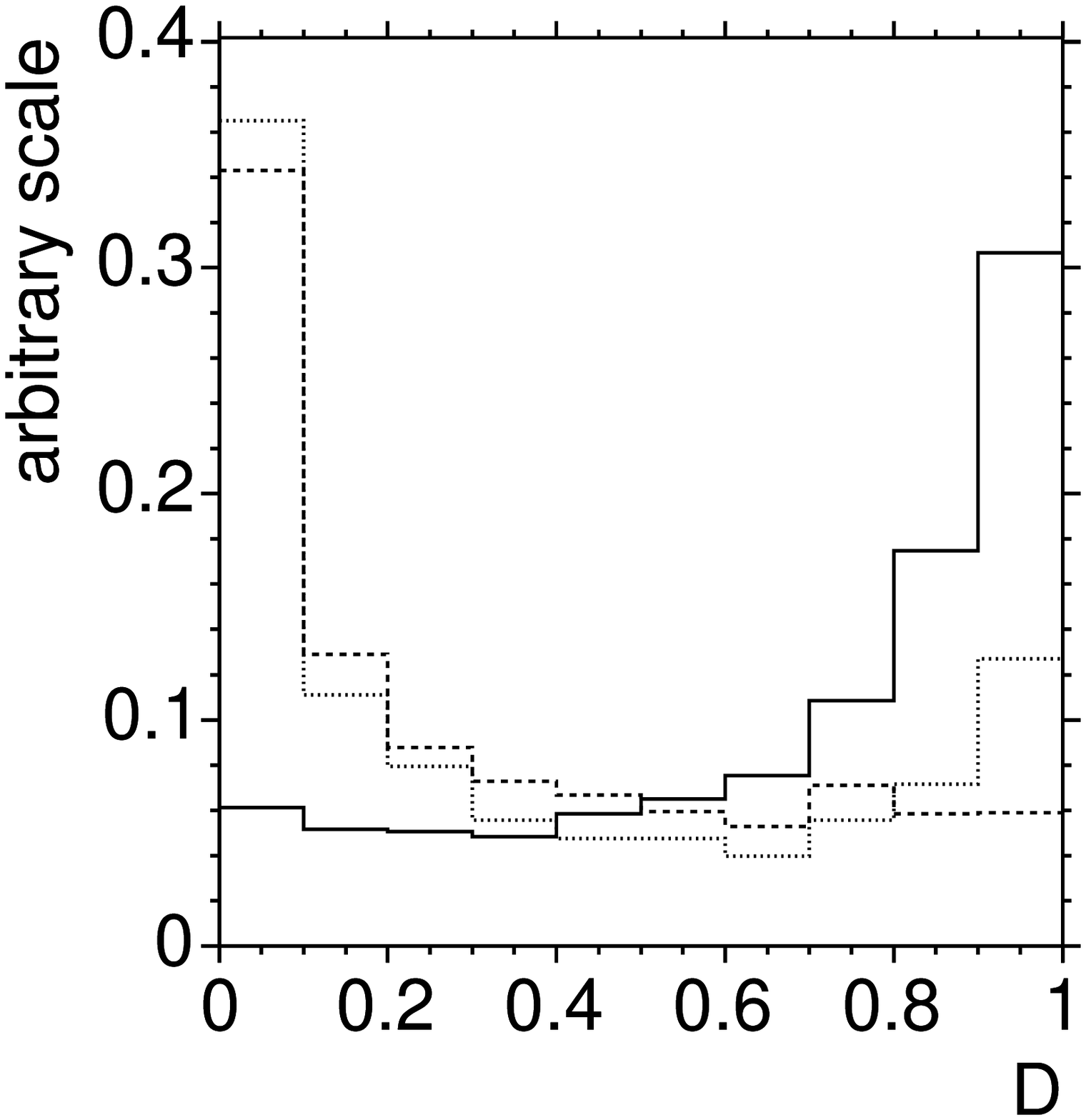}} 
\caption{Discriminant function output distributions for 
the \ttbar signal (solid line), $W$+jets background (dashed
line) and multijet events (dotted line) in the \ejets ~channel. 
\label{fig:ejets_LHD_templates}}
\end{figure}

The discriminant function is built 
according to Eq.~\ref{eq:LHdiscr}, from the fits to the logarithm of the ratio 
of signal (\ttbar) over background (\wplus) based on simulated events. 
Finally, the fully defined discriminant function is 
evaluated for each physics process considered in this analysis. For this
purpose, we use simulated $\ttbar$ events with $\ell$+jets and dilepton final
states, $W/Z$+jets events, and the multijet background data sample 
selected by requiring that the lepton fails the tight selection criterion. 
An example of the 
discriminant distributions for the \ttbar signal and main backgrounds in \ejets
~channel is shown in Fig.~\ref{fig:ejets_LHD_templates}. By  
construction, the discriminant function peaks near zero for the background, 
and near unity for the signal. 

%% file: xs.tex
\subsection{Method}
The number of \ttbar events in the selected data sample is
extracted by performing a maximum likelihood fit to the 
discriminant distribution observed in data using templates for the
\ttbar signal, multijet and $W/Z$+jets ($W$+jets) backgrounds 
in the \mujets ~(\ejets) channel.
The $Z$+jets contribution is added to the $W$+jets discriminant template
according to its fraction determined in Sect.~\ref{sec:MMresult}, 
resulting in a combined electroweak background template in the \mujets ~channel. 
Similarly, the contributions from the
dilepton and $\ell$+jets \ttbar ~signals are combined into a single
\ttbar ~template before fitting by adding the dilepton
contribution to the $\ell$+jets template. Dilepton and $Z$+jets admixtures 
introduce only small corrections to the \ttbar $\ell$+jets and $W$+jets 
template shape, respectively. 

We consider three different contributions to the maximum likelihood fit: 
\ttbar, $W/Z$+jets and multijet, and constrain the relative 
fraction of the latter using Eq.\ref{eq:matrix2}.   
This is realized by defining the following likelihood function:
\begin{eqnarray}
{\it{L}}(N_t^{t\bar{t}},N_t^{ W},N_t^{b}) =
\left[\prod_i {\cal{P}}(n_i^{
o},\mu_i)\right]{\cal{P}}(N_{\ell-t}^{o},N_{\ell-t}) \;,
\label{eq:lhood}
\end{eqnarray}
where ${\cal{P}}(n,\mu)$ denotes the Poisson probability density function
for $n$ observed events given an expectation value of $\mu$ 
and $N_t^{t\bar{t}}$, $N_t^{ W}$, $N_t^{b}$ are the
numbers of \ttbar, ~$W/Z$+jets and multijet events in the selected sample,
respectively. In the first term of Eq.~\ref{eq:lhood}, $i$ runs over all bins 
of the discriminant histogram; $n_i^{ o}$ is the content of bin $i$ 
measured in the selected data
sample; and $\mu_i$ is the expectation for bin $i$, which is a function
of $N_t^{t\bar{t}}$, $N_t^{ W}$ and $N_t^{b}$ as given by:
\begin{eqnarray}
\mu_i(N_t^{t\bar{t}},N_t^{ W},N_t^{b}) = f_i^{t\bar{t}}N_t^{t\bar{t}} + f_i^{
W}N_t^{ W} + f_i^{b}N_t^{b} \;,
\label{eq_mui}
\end{eqnarray}
where $f_i^{t\bar{t}}$, $f_i^{ W}$, $f_i^{b}$ represent
the fractions in bin $i$ of the $t\bar{t}$, $W$+jets and
multijet discriminant templates 
(shown in Fig.~\ref{fig:ejets_LHD_templates} for \ejets ~channel), respectively.
The second term of Eq.~\ref{eq:lhood} effectively implements the constraint
on $N_t^{b}$ via the Poisson probability of the
observed number of events in the ``loose$-$tight'' ($N_{\ell-t}^{o}$) sample,
given the expectation ($N_{\ell-t}$).
The latter can be expressed as: 
\begin{eqnarray}
N_{\ell -t} = \frac{1-\varepsilon_{s}}{\varepsilon_{s}}N_t^{t\bar{t}}
+ \frac{1-\varepsilon_{s}}{\varepsilon_{s}}N_t^{ W}
+ \frac{1-\varepsilon_{b}}{\varepsilon_{b}}N_t^{ b}.
\label{eq_nlmt}
\end{eqnarray}
Thus, the task is to minimize the negative log-likelihood function: 
\begin{eqnarray*}
& &-\log  {\rm{L}} (N_t^{t\bar{t}},N_t^{ W},N_t^{ b})   \simeq   \\
& &\sum_i(-n_i^{ o} \log \mu_i + 
\mu_i) - N_{\ell-t}^{ o}\log N_{\ell -t} + N_{\ell -t} \;, 
\label{eq_final}
\end{eqnarray*}
where any terms independent of the minimization parameters are dropped.
The fitted parameters ($N_t^{t\bar{t}}$, $N_t^{ W}$, $N_t^{b}$) are given by 
their value at the negative log-likelihood function minimum,  
and their uncertainties are obtained by raising the negative log-likelihood
by one-half unit above the minimum while all other parameters of the fit are
allowed to float.
The results of the fits are listed in Table~\ref{tab:nevts} and the 
corresponding correlation coefficients are summarized in Table~\ref{tab:corr}.

\begin{table}[ht]
\newcommand\T{\rule{0pt}{2.6ex}}
\newcommand\B{\rule[-1.2ex]{0pt}{0pt}}
\begin{center}
\begin{tabular}{lccccccc}  \hline \hline 
  channel    & $N_{t}^{t\bar{t}}$ \T \B & $N_{t}^{W}$ & $N_{t}^{b}$ \\ [1pt]\hline
  \\[-7pt]
 \ejets & $67.5^{+13.8}_{-12.9}$ & $32.6^{+14.0}_{-13.0}$ & $19.3 \pm 2.0$ \\[4pt] 
 \mujets & $21.1^{+10.7}_{-9.7}$ & $72.6^{+13.0}_{-12.1}$ & $8.1^{+1.8}_{-1.7}$
 \\ [2pt]\hline \hline
\end{tabular} 
\caption{Fitted number of \ttbar, $W$+jets and multijet background events in the selected  
sample in the \ejets ~and \mujets ~channels. 
$N_{t}^{W}$ includes the $Z$+jets contribution in the \mujets ~sample. }
\label{tab:nevts}
\end{center}
\end{table}

\begin{table}[ht]
\newcommand\T{\rule{0pt}{2.6ex}}
\newcommand\B{\rule[-1.2ex]{0pt}{0pt}}
\begin{center}
\begin{tabular}{lcccccc}  \hline \hline
  & \multicolumn{3}{c} \ejets & \multicolumn{3}{c} \mujets \\ \hline 
  & $N_{t}^{t\bar{t}}$ \T \B & $N_{t}^{W}$ & $N_{t}^{b}$ & $N_{t}^{t\bar{t}}$ & $N_{t}^{W}$ & $N_{t}^{b}$\\ 
\hline 
$N_{t}^{t\bar{t}}$ \T \B         & $+1.00$    &$-0.63$  &$-0.11$ & $+1.00$ &$-0.59$ &$-0.14$ \\
$N_{t}^{W}$          &         & $+1.00$    &$-0.23$  &	  & $+1.00$ &$-0.23$ \\
$N_{t}^{b}$               &         &      &$+1.00$   &	  &	 & $+1.00$ \\ \hline \hline
\end{tabular}
\caption{\label{tab:corr} Matrices of correlation coefficients of the likelihood fit in the \ejets ~and \mujets
~channels.} 
\end{center}
\end{table}

One complication arises due to the fact that the shape of the discriminant 
for the multijet background is obtained from the ``loose$-$tight'' data sample 
which has a small contribution from \wplus and \ttbar ~events (Eq.~\ref{eq_nlmt}).
The contamination of the multijet template is taken into account by using the 
corrected expected number of events in each bin of the discriminant function
\begin{eqnarray*}
& &\mu_i(N_t^{ t\bar{t}},N_t^{ W},N_t^{ b}) =  \\
& &\left(f_i^{ t\bar{t}}N_t^{t\bar{t}} + f_i^{ W}N_t^{
W}\right)\times\left(1-\frac{\varepsilon_{b}}{1-\varepsilon_{b}}\frac{1-\varepsilon_{s}}{\varepsilon_{s}}\right)
+ \\
& & f_i^{ b}\left(N_t^{ b}+\frac{\varepsilon_{b}}{1-\varepsilon_{b}}\frac{1-\varepsilon_{s}}{\varepsilon_{s}}\left(N_t^{
t\bar{t}}+N_t^{ W}\right)\right)\;,
\end{eqnarray*}
in place of the one of Eq.~\ref{eq_mui}.  

\begin{figure}[tbh] 
\epsfig{file=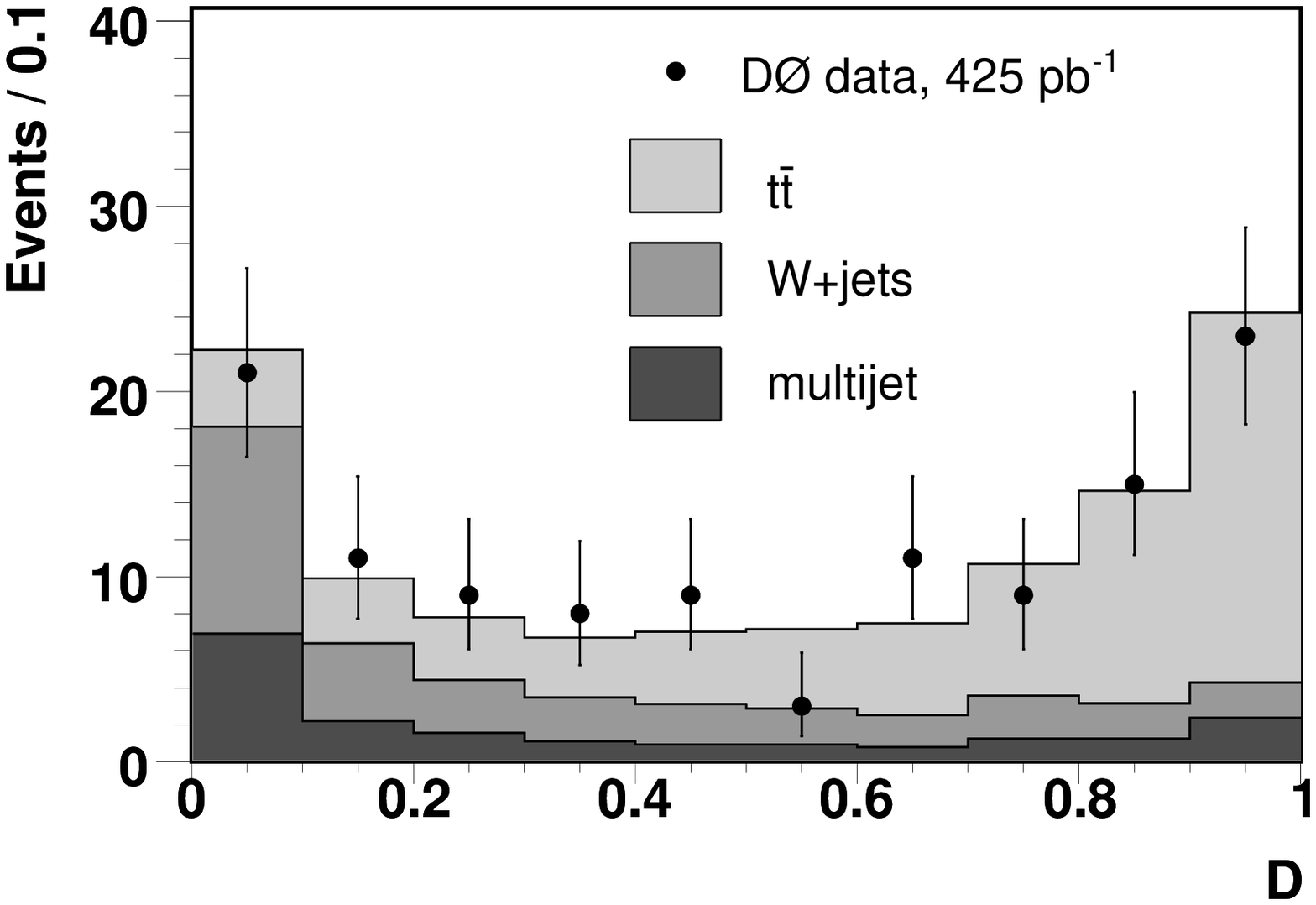, width=9cm, clip=}
\epsfig{file=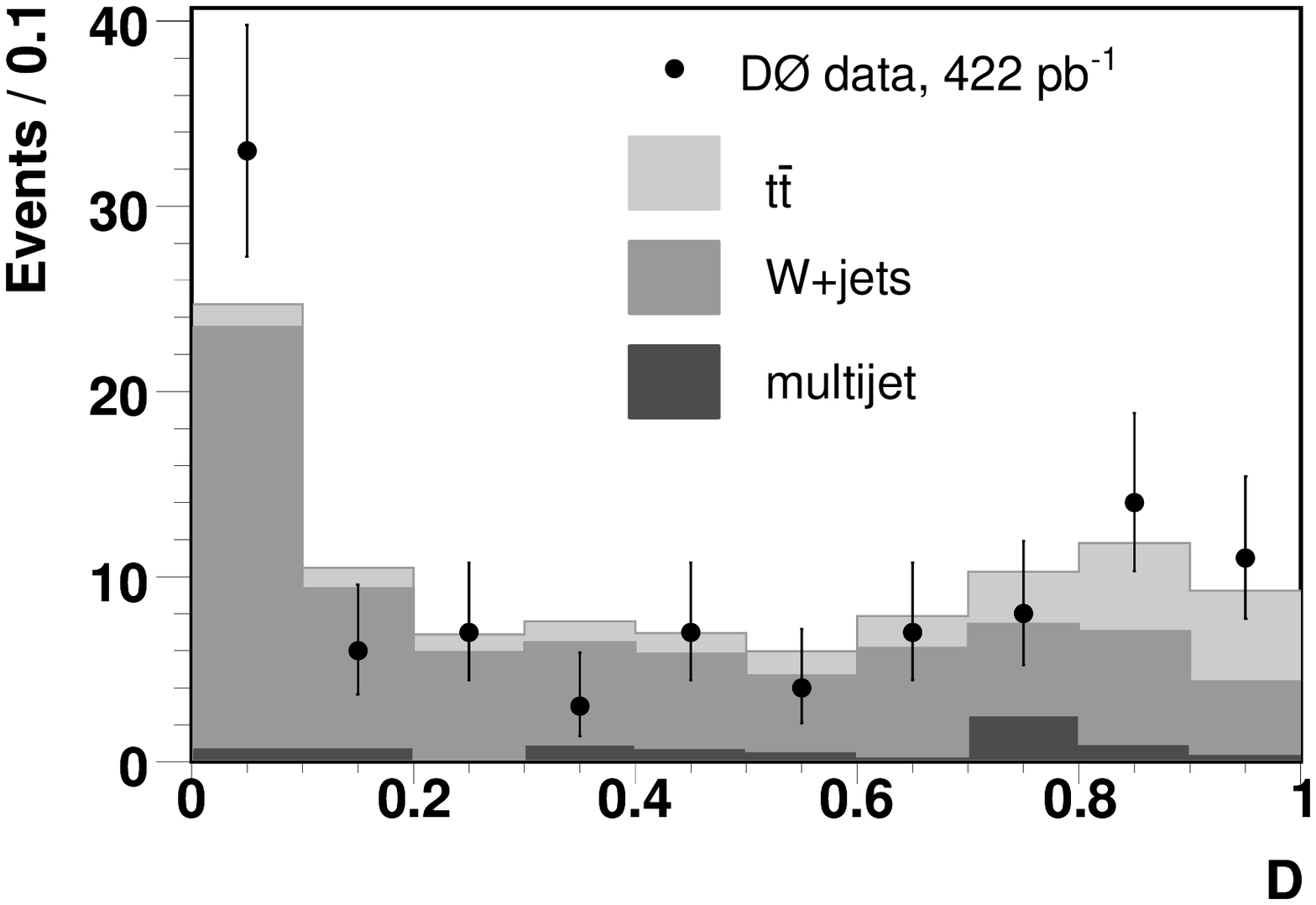, width=9cm}
\caption{\label{fig:topofit}Discriminant distribution for data 
overlaid with 
the result from a fit of $\ttbar$ signal, and $W$+jets and multijet background 
in the \ejets ~(upper plot) and \mujets ~(lower plot) channel.}
\end{figure}

Figure~\ref{fig:topofit} shows the distributions of the discriminant functions 
for data in \ejets ~and \mujets ~channels along with the fitted contributions from the 
$\ttbar$ signal and $W$+jets and multijet backgrounds. 

\subsection{Cross sections in individual channels}

The \ttbar ~production cross section for an individual channel $j$ is  
computed as:
\begin{equation}
\sigma_j = \frac{N_{t}^{t\bar{t}}(j)}{{\varepsilon}_j \; {\cal B}_j \; {\cal L}_j} \;, 
\label{eq_sigma}
\end{equation}
where $N_{t}^{t\bar{t}}(j)$ is the number of fitted \ttbar ~events in channel $j$,
${\cal B}_j$ is the branching fraction for the \ttbar ~final state where the
lepton is allowed to originate either directly from a $W$ boson or from the 
$W\rightarrow\tau\nu$ decay (${\cal B}_{\mathrm{ljets}}$),   
${\cal L}_j$ is the integrated luminosity and $\varepsilon_j$ is the \ttbar ~selection
efficiency. 
The efficiencies $\varepsilon$ are obtained by correcting the
\ttbar ~$\ell$+jets selection efficiencies 
$\varepsilon_{\mathrm{ljets}}$ for the \ttbar\ dilepton final state contribution:
\begin{equation}
\varepsilon = \varepsilon_{\mathrm{ljets}} + \frac{{\cal B}_{\ell\ell}}{{\cal B}_{\mathrm{ljets}}} \;
{\varepsilon_{\ell\ell}} \; ,
\end{equation}
where $\varepsilon_{\ell\ell}$ and ${\cal B}_{\ell\ell}$ are the selection efficiency 
and the branching
fraction for the $t\bar t\to \ell\ell$+jets decay channel.  

The input values for the likelihood fit are summarized in
Table~\ref{tab:ljets}.  
\begin{table}[ht]
\begin{center}
\begin{tabular}{lcccccc}  \hline \hline
  channel    & $N_{l}$ & $N_{t}$ &    ${\cal B}$ & ${\cal L}$ ($\rm pb^{-1}$) &
  $\varepsilon_{\mathrm{ljets}}$ (\%) & $\varepsilon$(\%) \\ \hline
  $e$ + jets & 242     & 119	 & 0.17106 &	  425 		&    9.17	&	   
  9.39     \\ [3pt]
$\mu$ + jets & 160     & 100	 & 0.17036 &	  422 		&    9.18	&	   
  9.36     \\ \hline \hline
\end{tabular} 
\caption{Number of selected events in the loose ($N_{\ell}$) and tight ($N_t$) sample, 
	 branching fraction (${\cal B}$), integrated luminosity (${\cal L}$), selection efficiency
	 for $\ttbar \rightarrow \ell$+jets ($\varepsilon_{\mathrm{ljets}}$) and total selection
	 efficiency ($\varepsilon$) in \ejets ~and \mujets ~channels.}
\label{tab:ljets}
\end{center}
\end{table}

The \ttbar ~production cross sections at $\sqrt{s} = $1.96~TeV for a top quark mass of 175~GeV 
in the $e$+jets and $\mu$+jets channels are measured to be: 
\begin{eqnarray}
{\rm e+jets}: \sigma_{t\bar{t}}  =     9.9^{+2.1}_{-1.9}\:{\rm (stat)}
                                                 \pm 1.0 \:{\rm (syst)}
                                                 \pm 0.6\:{\rm (lum)}\:{\rm pb}; \nonumber \\
{\rm \mu +jets}: \sigma_{t\bar{t}} =     3.1^{+1.6}_{-1.5}\:{\rm (stat)}  
                                                \pm 0.4\:{\rm (syst)}
                                                \pm 0.2\:{\rm (lum)}\:{\rm pb}. \nonumber 
\end{eqnarray}

We estimate the probability to observe the cross
sections measured in the $\mu$+jets and $e$+jets channels by 
generating pseudo-measurements using the expected
statistical and systematic uncertainties evaluated at an assumed true cross
section of 7\,pb. We find that the correlation between two measurements 
is small, 3.3\%, since the dominant uncertainties are of a statistical nature and
therefore uncorrelated between the channels. The probability to observe the 
\ttbar ~cross section above 9.9\,pb (below 3.1\,pb) in one of the channels given 
the true cross section of 7\,pb is 9\% (4\%). 

We estimate the consistency of the cross sections observed  
in the $e$+jets and $\mu$+jets channels by generating pseudo-experiments 
that incorporate all the correlations between the different sources of 
systematic uncertainties  
assuming the measured cross sections in the individual channels. 
From the distribution of the differences between the \ejets ~and $\mu$+jets  
cross sections observed in each pseudo-experiment, 
we conclude that the cross sections agree within 2.4 standard deviations.
The difference 
observed between the measured cross sections is attributed to a statistical 
fluctuation.

    
\subsection{Combined cross section}
\begin{figure}[tbh] 
\epsfig{file=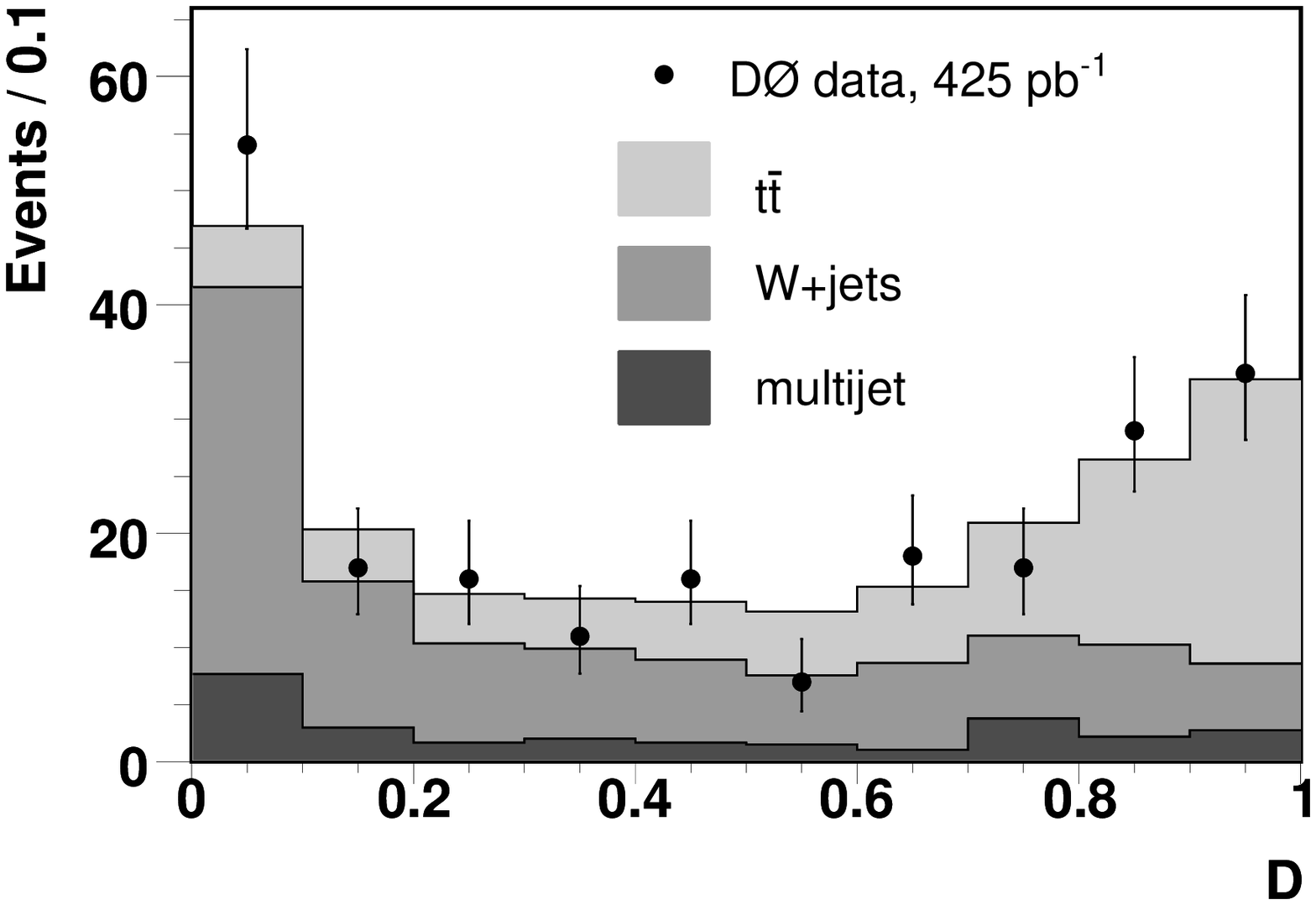, width=9cm}
\caption{\label{fig:topofitComb}Discriminant distribution for data 
overlaid with 
the result from a fit of $\ttbar$ signal, and $W$+jets and multijet background 
in the combined $\ljets$ channel.}
\end{figure}

The combined cross section in the lepton+jets channel is estimated 
by minimizing the sum of the 
negative log-likelihood functions of each individual channel. 
A total of five parameters
are simultaneously fitted: $\sigma_{t\bar{t}}$ (common to both lepton channels) and 
$N_{t}^{W}(j)$ and $N_{t}^{ b}(j)$ separately for each channel. 
Figure~\ref{fig:topofitComb} shows the distribution of the discriminant function 
for data along with the fitted contributions from 
$\ttbar$ signal, $W$+jets, and multijet background events which are found to be
$40\%$, $48\%$ and $12\%$, respectively.
The combined cross section for a top quark mass of $175~\rm{GeV}$ is:   
\begin{eqnarray}
\sigma_{t\bar{t}}(175\ \rm{GeV})      =     6.4^{+1.3}_{-1.2}\:{\rm (stat)}
                                                 \pm 0.7\:{\rm (syst)}
                                                 \pm 0.4\:{\rm (lum)}\:{\rm pb}. \nonumber
\end{eqnarray}


\begin{figure}[h]
\hspace{-1.5 cm}
\leftline
{
\includegraphics[width=0.5\textwidth,clip=]{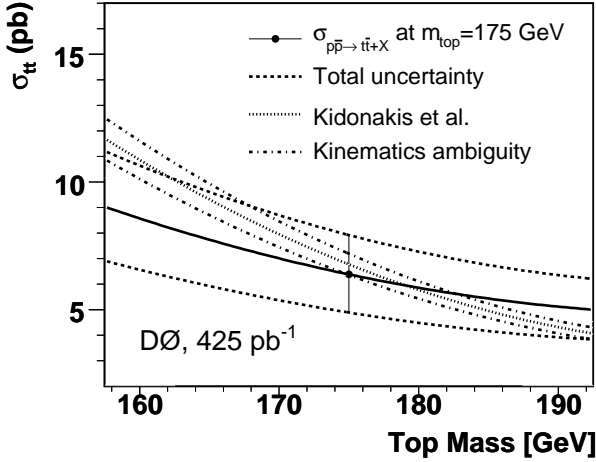}
}
\caption{The combined $t\bar{t}$ production cross section in the lepton+jets channel as a function 
         of top quark mass compared to the theoretical calculations
	 \cite{SMtheory_K}.}
\label{fig:xsec_lj_mass} 
\end{figure}

We have studied the dependence of the measured cross section 
on the top quark mass by using the samples of 
simulated \ttbar ~events with different top quark masses to evaluate signal 
efficiencies and discriminant function outputs and repeating complete analysis. 
Figure~\ref{fig:xsec_lj_mass} shows the dependence of the combined cross section 
in the lepton+jets channel on the top quark mass. Solid line represents the fit to the 
measured cross sections for various masses of the top quark.   
For 170~GeV$<m_{\rm top}<$180~GeV the cross section changes as a function of $m_{\rm top}$ as:
\begin{equation}   
\sigma_{t\bar{t}}(m_{\rm top}) = \sigma_{t\bar{t}}-0.1 \frac {\rm pb}{\rm GeV} \times (m_{\rm top}-175~{\rm GeV}).
\end{equation}    

\begin{figure} 
\hspace{-1.0 cm}
\includegraphics[width=0.5\textwidth,clip=]{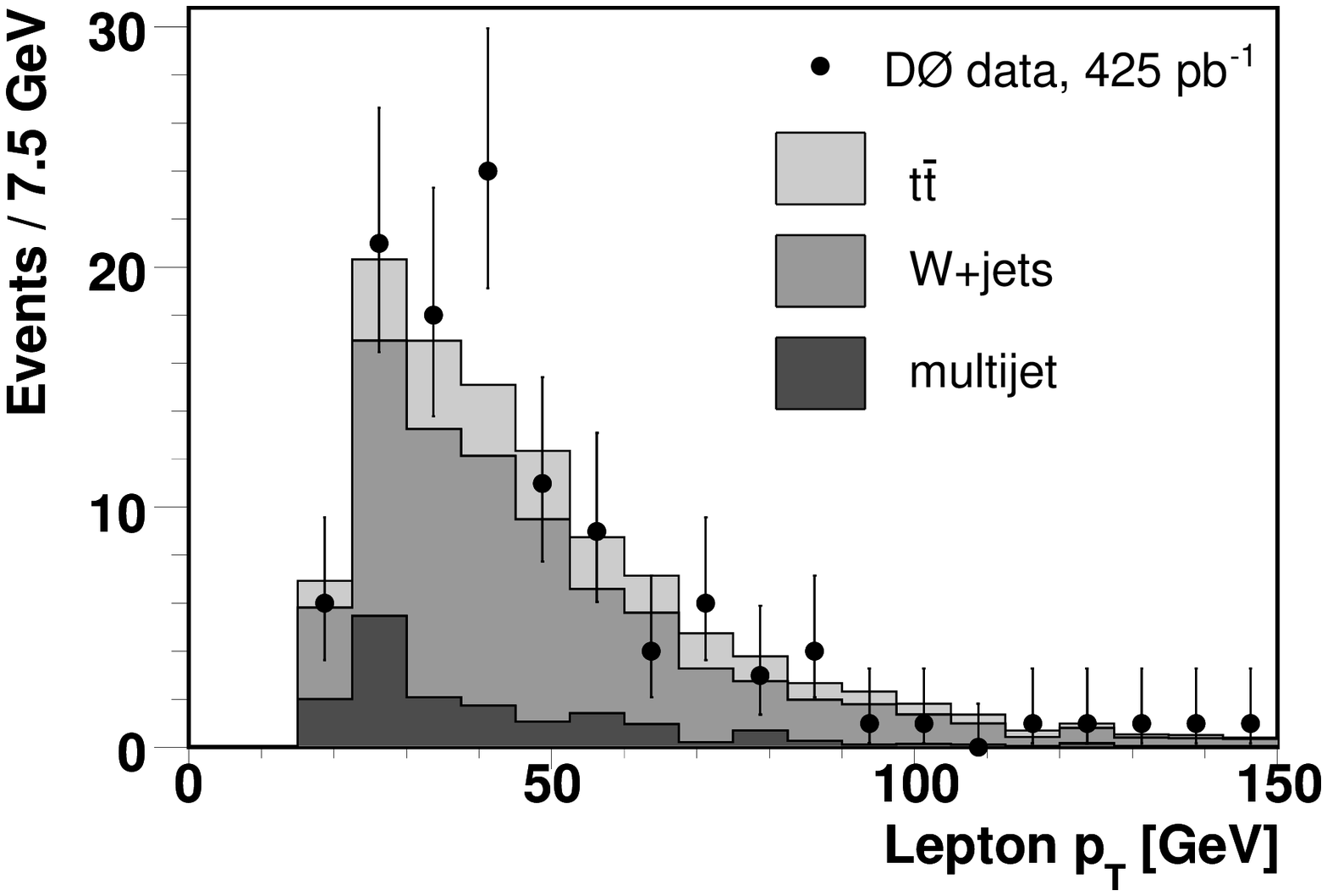}
\vspace{0.5 cm}
\hspace{-1.0 cm}
\includegraphics[width=0.5\textwidth,clip=]{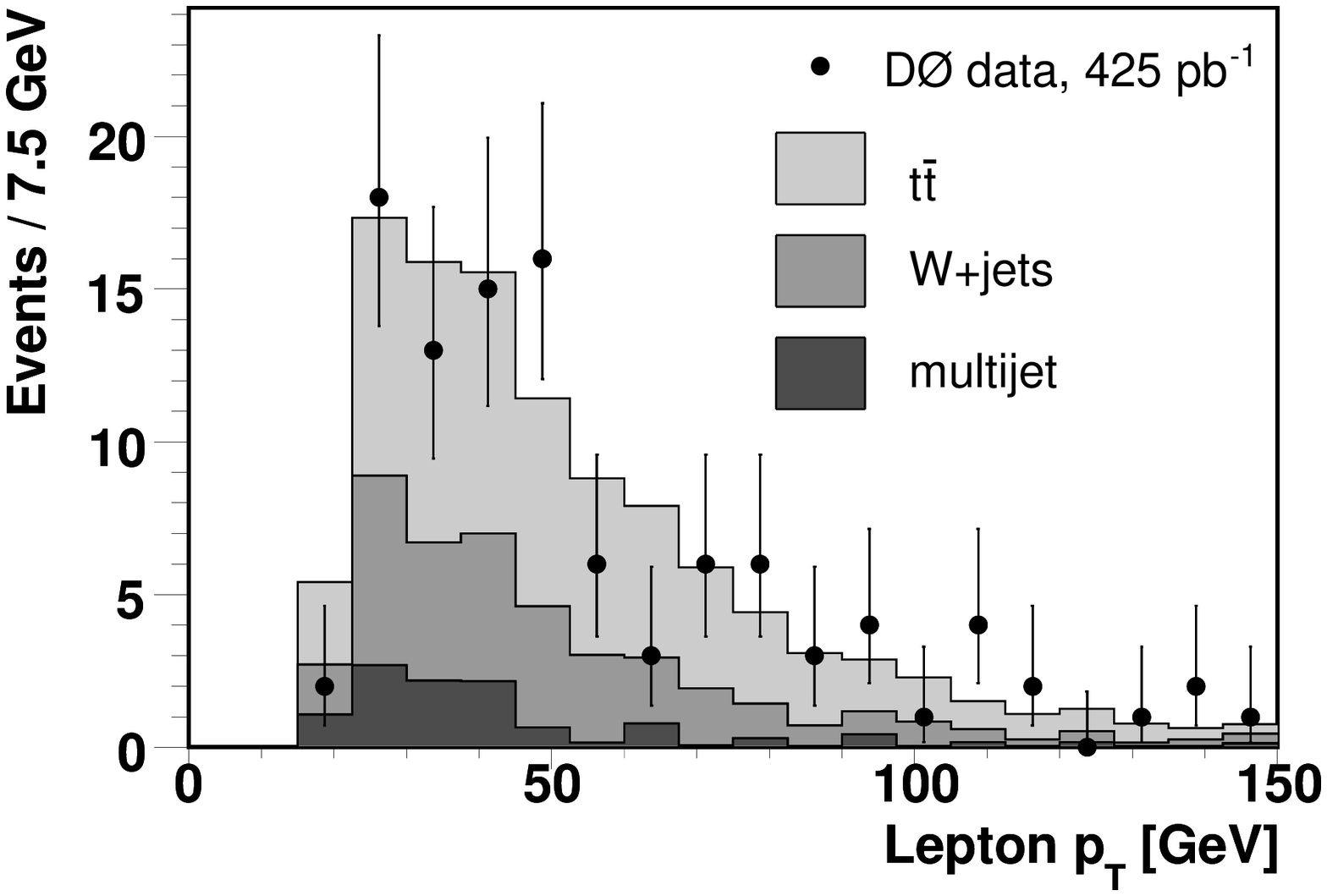}
\caption{\label{fig:lepton_pt} Lepton $p_T$ distribution for $\ljets$ events 
in data with discriminant below 0.5 (upper plot) and discriminant above 0.5 
(lower plot), 
overlaid with the result from a fit of $\ttbar$ signal, and $W$+jets and multijet background.}
\end{figure}

\begin{figure}[tbh] 
\hspace{-1.0 cm}
\includegraphics[width=0.5\textwidth,clip=]{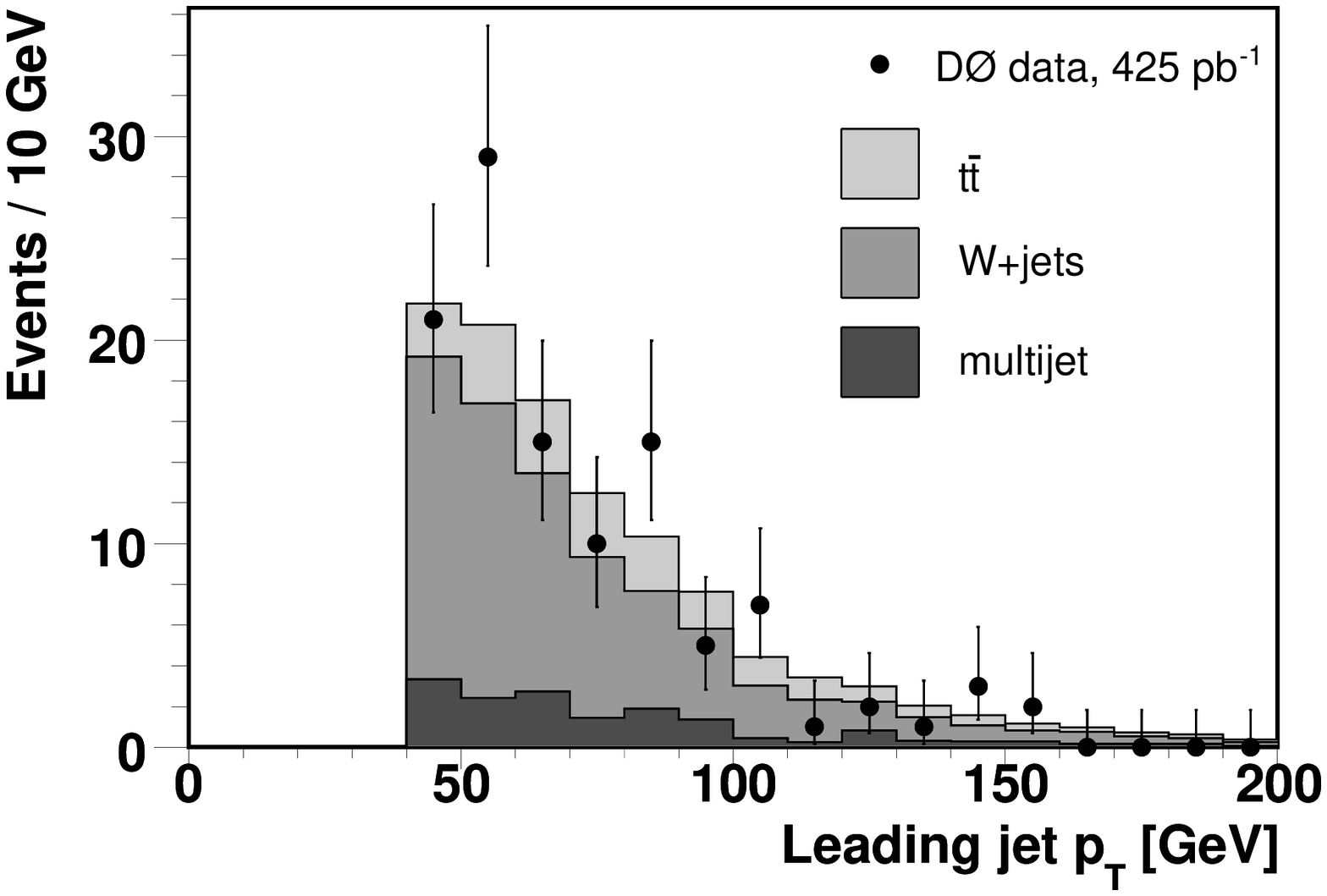}
\vspace{0.5 cm}
\hspace{-1.0 cm}
\includegraphics[width=0.5\textwidth,clip=]{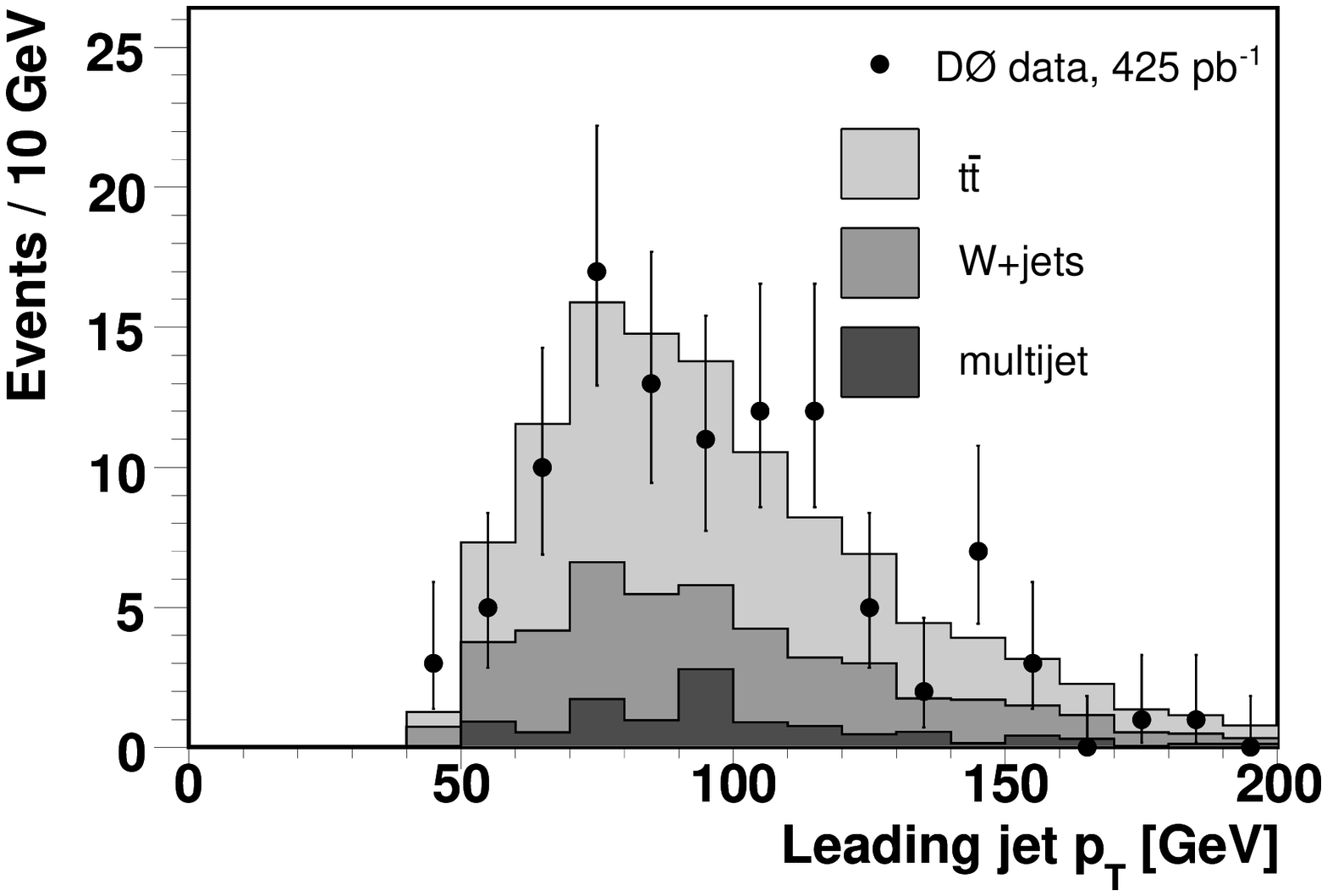}
\caption{\label{fig:jet_pt} Leading jet $p_T$ distribution for $\ljets$ events 
in data with discriminant below 0.5 (upper plot) and discriminant above 0.5 
(lower plot), overlaid with the result from a fit of $\ttbar$ signal, 
and $W$+jets and multijet background.}
\end{figure}

The kinematic distributions observed in
lepton+jets events are well described by the sum 
of $\ttbar$ signal, $W/Z$+jets, and multijet background contributions.
An example of this agreement is 
illustrated in Figs.~\ref{fig:lepton_pt} and \ref{fig:jet_pt} for events 
selected requiring $\mathcal D < 0.5$,  
i.e.,~dominated by background, and events in the $\ttbar$ signal region with
$\mathcal D > 0.5$.
The two variables shown are the lepton $p_T$ and the highest jet $p_T$ in the event and are
not used as input to the discriminant function.  


%% file: systematics.tex
The systematic uncertainty on the \ttbar ~production 
cross section in an individual channel $j$ for each
independent source of systematic uncertainty $i$ is determined by 
varying the source by one standard
deviation up and down and propagating the variation into both the
fitted number of $\ttbar$~events and the signal efficiency
resulting in a new value of the cross section in channel $j$: 
\begin{equation} \label{eq:xsecSyst}
\sigma_j^{i} = \sigma_j\pm\Delta\sigma_j^i = 
\frac {N_{t}^{t\bar{t}}(j)\pm\Delta N_{t}^{t\bar{t}}(j)^i} 
{({\varepsilon}_j\pm\Delta{\varepsilon}_j^i) \; {\cal B}_j \; {\cal L}_j} \;. 
\end{equation}
Variations due to uncertainty sources which modify simultaneously both the 
selection efficiency and the fitted number of $\ttbar$~events are treated as fully 
correlated. 

The variation of the fitted number of $\ttbar$~events due to each individual 
source $i$ is estimated by generating
$10,000$ pseudo-experiments from simulated events. The sample
composition of each pseudo-dataset is the same as the measured sample
composition in data but allowing for Poisson fluctuations in the
number of events from a specific contribution. The discriminant distribution 
for each pseudo-dataset is
fitted in order to extract $N_{t\bar{t}}$ once with the default 
discriminant function templates for $\ttbar$, $W/Z$+jets and multijet 
background and once with the varied ones. The relative
difference between the two results is histogrammed. The  
relative systematic uncertainty is extracted from the histogram by
performing a fit to a Gaussian distribution around the most probable 
value and using the mean of the fit as an estimator for the relative
uncertainty on the fitted number of $\ttbar$~events from source $i$.

Positive (negative) variations of the cross section $\Delta\sigma_j^i$ 
from each individual source of systematics with respect to the central 
value $\sigma_j$ (Eq.\ref{eq:xsecSyst})  
are summed quadratically to obtain total positive (negative) 
systematic uncertainty. 
In addition, a systematic uncertainty of $\pm$6.1\% from the luminosity
measurement is assigned \cite{d0lumi}.
By construction, this 
method does not allow the systematic uncertainties to affect the 
central value of the cross section $\sigma_j$. 

The systematic uncertainty on the combined cross section is estimated following the same  
procedure as described above taking into account the correlations between individual
sources of systematic uncertainties between the channels. 
The systematic uncertainties are classified as either uncorrelated 
(usually of statistical
origin in either Monte Carlo simulation or data) or fully correlated between 
the channels. In particular, we consider the systematic uncertainties coming 
from the primary vertex reconstruction, jet energy calibration, jet identification, 
jet trigger, $W$ background model, and branching fraction to be fully
correlated. Uncertainties associated with the lepton identification, 
lepton trigger, multijet background evaluation, and the limited 
statistics of Monte Carlo samples are taken as uncorrelated in the cross 
section combination.   


Table~\ref{tab:sysxs_ljets} summarizes the contributions
from the various sources of systematic uncertainties to the total systematic
uncertainty on the cross sections in the $e$+jets, $\mu$+jets and combined 
$\ell$+jets channels. The jet energy scale uncertainty dominates, 
followed by the uncertainty of the luminosity measurement. These two 
represent 80\% of the total systematic uncertainty of the combined cross section.   

\begin{table}[h]
\begin{tabular}{lccc}
\hline\hline
  Source                 & $\ejets$  &   $\mujets$   & $\ell+$jets     \\
\hline
Primary Vertex        &  $+0.30-0.28$  & $+ 0.12-0.10$   &  $+ 0.24-0.21$  \\   
Lepton ID             &  $\pm 0.32$	 & $+ 0.17-0.16$   &  $\pm 0.22$   \\
Jet Energy Scale      &  $+ 0.70-0.72$  & $+ 0.05-0.16$   &  $\pm 0.47$  \\   
Jet ID                &  $+ 0.08-0.14$  & $+ 0.11-0.02$   &  $+ 0.03-0.08$  \\  
Trigger 	      &  $+ 0.05-0.21$  & $+ 0.09-0.08$   &  $+ 0.10-0.20$ \\
$W$ bckg model        &  $+ 0.11-0.21$  & $+ 0.13-0.11$   &  $+ 0.12-0.18$  \\  
Multijet bckg         &  $\pm 0.04$	 & $+ 0.13-0.14$   &  $+ 0.05-0.06$   \\
MC statistics	      &  $\pm 0.48$	 & $\pm 0.31 $    &  $\pm 0.33$   \\
$\cal{B}$             &  $+ 0.20-0.19$  & $\pm 0.06 $    &  $\pm 0.14$   \\ \hline  
Subtotal	      &  $+ 0.99-1.03$	 & $\pm 0.44$	  &  $+ 0.70-0.72$ \\ \hline
Luminosity            &  $\pm 0.64$	 & $\pm 0.20 $    &  $\pm 0.42$   \\ \hline
Total		      &  $+ 1.18-1.21$	 & $\pm 0.45$	  &  $+ 0.82-0.83$ \\
\hline \hline
\end{tabular}
\caption{\label{tab:sysxs_ljets} Summary of systematic
uncertainties of the cross section $\Delta \sigma_{t\bar{t}}$ (pb).}
\end{table}

%% file: summary.tex
We have measured the \ttbar ~production cross section in the $\ell$+jets final state by combining 
the measurements performed in the individual $e$+jets and $\mu$+jets channels. The combined cross section  
for a top quark mass of 175 GeV is 
\begin{eqnarray}
{\rm \ell+jets}: \sigma_{t\bar{t}}      =     6.4^{+1.3}_{-1.2}\:{\rm (stat)}
                                                 \pm 0.7\:{\rm (syst)}
                                                 \pm 0.4\:{\rm (lum)}\:{\rm pb}. \nonumber
\end{eqnarray}
The result is in a good agreement with the theoretical predictions of 
$6.7^{+0.7}_{-0.9}$\,pb (\cite{SMtheory_C}) and $6.8 \pm 0.6 $\,pb (\cite{SMtheory_K}) 
based on the full NLO matrix elements 
and the resummation of the leading and next-to-leading soft logarithms.

%% file: acknowledgement_paragraph_r2.tex
%
We thank the staffs at Fermilab and collaborating institutions, 
and acknowledge support from the 
DOE and NSF (USA);
CEA and CNRS/IN2P3 (France);
FASI, Rosatom and RFBR (Russia);
CAPES, CNPq, FAPERJ, FAPESP and FUNDUNESP (Brazil);
DAE and DST (India);
Colciencias (Colombia);
CONACyT (Mexico);
KRF and KOSEF (Korea);
CONICET and UBACyT (Argentina);
FOM (The Netherlands);
Science and Technology Facilities Council (United Kingdom);
MSMT and GACR (Czech Republic);
CRC Program, CFI, NSERC and WestGrid Project (Canada);
BMBF and DFG (Germany);
SFI (Ireland);
The Swedish Research Council (Sweden);
CAS and CNSF (China);
Alexander von Humboldt Foundation;
and the Marie Curie Program.
%

%% file: appendix/topovar.tex
\section{Kinematic variables for discriminant optimization}
\label{app:topovar}

We select a set of thirteen variables as input for the discriminant function
optimization. These variables are designed to address different aspects 
of the \ttbar ~signal and $W$+jets background kinematics: event energy, 
shape, location of the jets in the detector, properties of soft
non-leading jets, etc. 
$W$+jets background tends to have a lower event transverse energy, less
energetic jets and smaller total invariant mass than \ttbar ~events. 
Since the \ttbar ~system is produced nearly at rest at the Tevatron and  
therefore is expected to have a much smaller boost in the beam direction
than $W$+jets, the jets from a \ttbar ~event are more central. The 
\ttbar ~event topology is also different from $W$+jets due to the different
production mechanisms.     

We select the following thirteen variables for the discriminant function
optimization:   
\begin{itemize}
\item $H_{T}$, the scalar sum of the $p_T$ of the four leading jets;
\item $H_{T}'$, $H_{T}$ divided by the scalar sum of the absolute values 
of $p_z$ of the jets, the lepton and the \met;
\item $M_T$, transverse mass of the four leading jets;
\item $M_\mathrm{event}$, invariant mass of up to four leading jets, 
the \met\ and the lepton in the event;
\item Event centrality ${\mathcal C}$, defined as the ratio 
of the scalar sum of the $p_T$ of the jets to the 
scalar sum of the energy of the jets;
\item Event aplanarity ${\mathcal A}=\frac{3}{2}\lambda_{3}$ and 
sphericity ${\mathcal S}=\frac{3}{2}(\lambda_{2}+\lambda_{3})$,  
derived from the normalized momentum tensor, defined by ${\mathcal M}_{ij} =
\frac{\Sigma_o p^o_ip^o_j}{\Sigma_o|\vec{p^o}|^2}$, where $\vec{p^o}$ 
is the momentum vector of jet $o$, $i$ and $j$ are Cartesian
coordinates, and the eigenvalues $\lambda_{k}$ of ${\mathcal M}$ are ordered such that
$\lambda_{1}\ge\lambda_{2}\ge\lambda_{3}$ with
$\lambda_{1}+\lambda_{2}+\lambda_{3}=1$; 
\item $\Delta\varphi(\ell,\met)$, angle between the muon and the \met direction perpendicular to the
beam axis;
\item $|\eta_{jet}|^\mathrm{max}$, 
$|\eta|$ of the jet with maximum pseudorapidity;
\item $\sum\eta^2$, sum of the squared pseudorapidities of up to four 
jets;
\item NJW, built from the transverse momenta of up to four leading jets, it 
corresponds to the jet multiplicity above a given jet $p_T$ threshold, over the
range between 10 GeV and 55 GeV, weighted by the threshold, and is sensitive to
the additional radiation in the event and to the $p_T$ spectrum of the jets in
the event \cite{tkachev};


\item $M_\mathrm{inv}^{123}$, Sum of invariant masses of the three dijet 
pairs formed from the three leading jets in the event;
\item $M_\mathrm{dijet}^{min}$, the minimum of the invariant mass of any 
two jets in the event.  
\end{itemize}

Variables that characterize event energy scale 
($H_{T}$, $H_{T}'$, $M_T$, $M_\mathrm{event}$) show the best 
discrimination power, but they are sensitive to the jet energy calibration,
which is one of the dominant sources of systematic uncertainty on the \ttbar
~cross section. A combination of variables belonging to different classes
provides the best total uncertainty on the cross section.

%% file: ljets_prd.bbl
\begin{thebibliography}{99}
%
\bibitem[*]{alton}
Visitor from Augustana College, Sioux Falls, SD, USA.
\bibitem[\P]{burdin}
Visitor from The University of Liverpool, Liverpool, UK.
\bibitem[\S]{podesta-lerma}
Visitor from ICN-UNAM, Mexico City, Mexico.
\bibitem[\ddag]{voutilainen}
Visitor from Helsinki Institute of Physics, Helsinki, Finland.
\bibitem[\#]{wenger}
Visitor from Universit{\"a}t Z{\"u}rich, Z{\"u}rich, Switzerland.
%
\vskip 0.25cm

\bibitem{topdisc}
CDF Collaboration, F.\ Abe {\sl et al.}, Phys.\ Rev.\ Lett. {\bf 74}, 2626 (1995);
D0 Collaboration, S.\ Abachi {\sl et al.}, Phys.\ Rev.\ Lett. {\bf 74}, 2632 (1995).
%
\bibitem{single_top} 
D0 Collaboration,  V.\ Abazov {\sl et al.}, Phys.\ Lett.\ B {\bf 622}, 265 (2005); 
CDF Collaboration, D.\ Acosta {\sl et al.}, Phys.\ Rev.\ D {\bf 71}, 012005 (2005).
%
\bibitem{single_top_new}
D0 Collaboration, V.\ Abazov {\sl et al.}, Phys.\ Rev.\ Lett. {\bf 98}, 181802 (2007).
%
\bibitem{runI} 
CDF Collaboration, T.\ Affolder {\sl et al.}, Phys.\ Rev.\ D {\bf 64}, 032002 (2001);
D0 Collaboration, V.\ Abazov {\sl et al.}, Phys.\ Rev.\ D {\bf 67}, 012004 (2003). 
%
\bibitem{SMtheory_B} R.\ Bonciani {\sl et al.}, Nucl.\ Phys.\ {\bf B529}, 424 (1998).
\bibitem{SMtheory_K} N.\ Kidonakis and R.\ Vogt, Phys.\ Rev.\ D {\bf 68}, 114014 (2003).
\bibitem{SMtheory_C} M.\ Cacciari {\sl et al.}, J.\ High Energy Phys.\ {\bf 404}, 68 (2004).

\bibitem{top_r2_xs_d0}
D0 Collaboration. V.\ Abazov {\sl et al.}, Phys.\ Lett.\ B {\bf 626}, 35 (2005);
D0 Collaboration. V.\ Abazov {\sl et al.}, Phys.\ Lett.\ B {\bf 626}, 45 (2005);
D0 Collaboration. V.\ Abazov {\sl et al.}, Phys.\ Lett.\ B {\bf 626}, 55 (2005).

\bibitem{top_r2_xs_cdf}
CDF Collaboration, D.\ Acosta {\sl et al.}, Phys.\ Rev.\ D {\bf 71}, 052003
(2005);
CDF Collaboration, D.\ Acosta {\sl et al.}, Phys.\ Rev.\ D {\bf 72}, 032002
(2005);
CDF Collaboration, D.\ Acosta {\sl et al.}, Phys.\ Rev.\ D {\bf 71}, 072005
(2005);
CDF Collaboration, D.\ Abulencia {\sl et al.}, Phys.\ Rev.\ Lett. {\bf 97}, 082004
(2006).

\bibitem{top_res}  
D0 Collaboration, B.\ Abbott {\sl et al.}, Phys.\ Rev.\ Lett. {\bf 82}, 4975 
(1999).

\bibitem{top_H+} J.\ F.\ Gunion {\sl et al.}, {\it The Higgs Hunters Guide} 
Addison-Wesley, Redwood City, California, p 200 (1990);
D0 Collaboration, V.\ Abazov  {\sl et al.}, Phys.\ Rev.\ Lett. {\bf 92}, 221801 
(2004);
CDF Collaboration, T.\ Affolder {\sl et al.}, Phys.\ Rev.\ Lett. {\bf 96}, 042003 
(2005).

\bibitem{top_like} 
H.P.~Nilles, Phys. Rep. {\bf 110}, 1 (1984); 
H.\ J.\ He, N.\ Poleonsky, S.\ Su, Phys.\ Rev.\ D {\bf 64}, 053004 (2001); 
V.\ A.\ Novikov, I.\ B.\ Okun, A.\ N.\ Rozanov, M.\ I.\ Vyosotsky, 
Phys.\ Lett.\ B {\bf 529}, 111 (2002); 
D.\ Choudhury, T.\ M.\ P.\ Tait and C.\ E.\ M.\ Wagner, Phys.\ Rev.\ D {\bf 65}, 
053002 (2002); 
H.\ Cheng and I.\ Low, J.\ High Energy Phys.\ {\bf 309}, 51 (2003).


\bibitem{nimpaper}
D0 Collaboration, V.\ Abazov {\sl et al.},
Nucl.\ Instrum.\ and\ Methods A {\bf 565}, 463 (2006).

\bibitem{muonnimpaper}
D0 Collaboration, V.\ Abazov {\sl et al.},
Nucl.\ Instrum.\ and\ Methods A {\bf 552}, 372 (2005).
                                                                                                
\bibitem{d0lumi} T.\ Andeen {\sl et al.}, FERMILAB-TM-2365. 

\bibitem{kalman} R.\ E.\ Kalman, Transactions of the ASME-Journal of Basic
Engineering, Series D, {\bf 82}, 35 (1960).   

\bibitem{PDG2006} 
W.\ M.\ Yao {\sl et al.}, Journal of Physics G {\bf 33}, 1 (2006).

\bibitem{jet_algo} 
G.\ C.\ Blazey {\sl et al.}, in 
{\sl Proceedings of the Workshop: QCD and Weak Boson Physics in Run II,} 
edited by U.\ Baur, R.\ K.\ Ellis, and D.\ Zeppenfeld, 
Fermilab-Pub-00/297 (2000).

\bibitem{massideo}
D0 Collaboration, V.\ Abazov {\sl et al.}, hep-ex/0702018, 
to be published in Phys. Rev. D.

\bibitem{alpgen} 
M.L.~Mangano {\sl et al.}, J.\ High Energy Phys.\ {\bf 07}, 001 (2003).

\bibitem{pythia}
T.\ Sj\"{o}strand {\sl et al.}, Comp. Phys. Commun. {\bf 135}, 238 (2001).

\bibitem{cteq}
D.~Stump {\sl et al.}, J.\ High Energy Phys.\ {\bf 10}, 046 (2003).

\bibitem{tuneA} S.~Alekhin {\sl et al.}, in {\sl The QCD/SM Working Group: 
Summary Report,} hep-ph/0204316 (2002).

\bibitem{evtgen}
D.\ J.\ Lange, Nucl.\ Instrum.\ and\ Methods A {\bf 462}, 152 (2001).


\bibitem{geant} 
R.\ Brun and F.\ Carminati, CERN program library long writeup W5013 (1993).

\bibitem{btagPRD} 
D0 Collaboration, V.\ Abazov {\sl et al.}, 
Phys.\ Rev.\ D {\bf 74}, 112004 (2006).

\bibitem{dileptonPRD} 
D0 Collaboration, V.\ Abazov {\sl et al.}, in preparation.


\bibitem{topmass} 
D0 Collaboration, B.\ Abbott {\sl et al.}, Phys.\ Rev.\ D {\bf 58}, 052001 (1998).

\bibitem{tensor} 
V.\ Barger, J.\ Ohnemus and R.J.N.~Phillips, Phys.\ Rev.\ D {\bf 48}, 3953 (1993).

\bibitem{tkachev} 
F.\ Tkachev, Int. J. Mod. Phys. A {\bf 12}, 5411 (1997) and private communication.
   

\end{thebibliography}
